\documentclass[review]{elsarticle}

\usepackage{graphicx}
\usepackage{amssymb}

\usepackage[margin=2.5cm]{geometry}
\usepackage{graphicx}
\usepackage{dcolumn}
\usepackage{bm}
\usepackage{braket}
\usepackage{graphicx}
\usepackage{epstopdf}
\usepackage{mwe}    
\usepackage[T1]{fontenc}
\usepackage{caption}
\usepackage{subcaption} 
\usepackage{amssymb}
\usepackage{amsmath}
\usepackage{booktabs}
\usepackage{tabularx}
\usepackage{multirow}
\usepackage{floatrow}
\usepackage[font=normalsize]{caption}
\usepackage{appendix}
\usepackage{xcolor}
\newcommand{\tensor}[1]{\bm{\mathsf{#1}}} 
\newcommand{\ms}{\scriptscriptstyle}
\newcommand{\K}{k}
\newcommand{\subscript}[1]{_{\ms #1}}
\newcommand{\eqp}{eq\prime}
\newcommand{\sg}{\sigma}

\newcommand{\Ks}[1]{k\subscript{#1}}
\newcommand{\Kps}[1]{k^{'}\subscript{#1}}
\newcommand{\Kts}[1]{\tilde{k}\subscript{#1}}
\newcommand{\Ktps}[1]{\tilde{k}^{'}\subscript{#1}}
\newcommand{\dKps}[1]{\Ktps{#1}}
\newcommand{\dKs}[1]{\Kts{#1}}
\newcommand{\ux}{u_x}
\newcommand{\uy}{u_y}
\newcommand{\uz}{u_z}
\newcommand{\uxx}{u_{x}^{\ms2}}
\newcommand{\uyy}{u_{y}^{\ms2}}
\newcommand{\uzz}{u_{z}^{\ms2}}

\newcommand{\f}[1]{f\subscript{#1}}
\newcommand{\ft}[1]{\tilde{f}\subscript{#1}}
\newcommand{\df}[1]{\ft{#1}}

\journal{Journal Name}

\begin{document}

\begin{frontmatter}


\title{Three-Dimensional Central Moment Lattice Boltzmann Method on a Cuboid Lattice for Anisotropic and Inhomogeneous Flows}

\author{Eman  Yahia}
\ead{Eman.Yahia@ucdenver.edu}
\author{William Schupbach}
\ead{William.Schupbach@ucdenver.edu}
\author{Kannan N. Premnath}
\ead{Kannan.Premnath@ucdenver.edu}

\address{\normalsize Department of Mechanical Engineering\\ College of Engineering, Design and Computing\\ University of Colorado Denver\\ 1200 Larimer street, Denver, Colorado 80217 , $U$.S.A}


\begin{abstract}
Lattice Boltzmann (LB) methods are usually developed on cubic lattices that discretize the configuration space using uniform grids. For efficient computations of anisotropic and inhomogeneous flows, it would be beneficial to develop LB algorithms involving the collision-and-stream steps based on orthorhombic cuboid lattices. We present a new 3D central moment LB scheme based on a cuboid D3Q27 lattice. It involves two free parameters representing the ratios of the characteristic particle speeds along the two directions with respect to those in the remaining direction and are referred to as the grid aspect ratios. Unlike the existing LB schemes for cuboid lattices which are based on orthogonalized raw moments, we construct the collision step based on the relaxation of central moments and avoid the orthogonalization of moment basis, which lead to a more robust formulation. Moreover, prior cuboid LB algorithms prescribe the mappings between the distribution functions and raw moments before and after collision by using a moment basis designed to separate the trace of the second order moments (related to bulk viscosity) from its other components (related to shear viscosity), which lead to cumbersome relations for the transformations. By contrast, in our approach, the bulk and shear viscosity effects associated with the viscous stress tensor are naturally segregated only within the collision step and not for such mappings, while the grid aspect ratios are introduced via simpler pre- and post-collision diagonal scaling matrices in the above mappings. These lead to a compact approach, which can be interpreted based on special matrices. It also results in a modular 3D LB scheme on the cuboid lattice, which allows the existing cubic lattice implementations to be readily extended to those based on the more general cuboid lattices. In order to maintain the isotropy of the viscous stress tensor of the 3D Navier-Stokes equations using the cuboid lattice, corrections for eliminating the truncation errors resulting from the grid anisotropy as well as those from the aliasing effects are derived using a Chapman-Enskog analysis. Such local corrections, which involve the diagonal components of the velocity gradient tensor and parameterized by two grid aspect ratios, augment the second order moment equilibria in the collision step. We present a numerical study validating the accuracy of our approach for various benchmark problems at different grid aspect ratios. In addition, we show that our 3D cuboid central moment LB method is numerically more robust than its corresponding raw moment formulation. Finally, we demonstrate the effectiveness of the 3D cuboid central moment LB scheme for the simulations of anisotropic and inhomogeneous flows and show significant savings in memory storage,  and computational cost when used in lieu of that based on the cubic lattice.
\begin{keyword}
Lattice Boltzmann method\sep Cuboid lattice\sep Central moments \sep Multiple relaxation times \sep D3Q27 lattice \sep Chapman-Enskog analysis \sep Three dimensional flow benchmarks \sep Anisotropic flows
\end{keyword}
\end{abstract}




\end{frontmatter}

\section{Introduction} \label{sec:1}
The lattice Boltzmann (LB) methods, which arise as minimally discretized numerical schemes of the Boltzmann transport equation -- a cornerstone formulation in kinetic theory, has attracted much attention in the recent decades~\cite{mcnamara1988use,higuera1989boltzmann,benzi1992lattice,he1997theory,lallemand2021lattice}. As a mesoscopic approach, it has enriched the variety of computational fluid dynamics (CFD) techniques that are being developed and has been applied to a wide range of fluid flows successfully~\cite{chen1998lattice,yu2003viscous,aidun2010lattice,sharma2020current}. It involves the evolution of the distribution of the particle populations by a collision step, which is then followed by their lock-step advection along discrete directions, referred to as the streaming step. The former is often modeled by a relaxation of the distributions (e.g.,~\cite{qian1992lattice}) or their moments (e.g.,~\cite{d2002multiple}) to equilibria. The discrete particle velocity directions are referred to as the lattice, which are usually designed to obey the associated physical symmetry and isotropy of the fluid flow being simulated. The lattice is usually referred to by the notation $DdQq$, where $d$ represents the number of spatial dimensions and $q$ denotes the number of discrete particle directions. The macroscopic fields such as the fluid velocity are obtained from the lower order moments of the distribution functions, while the higher order kinetic moments can be constructed to evolve so as to facilitate the numerical robustness of the method. The linear advection and locally nonlinear collision of the LB method, along with its ability to naturally represent the physics of complex fluids and flows based on kinetic models, are among the important assets of this approach.

Anisotropic and inhomogeneous fluid motions involve relatively large spatial variations in the characteristic features of the fluid flow in one or more directions when compared to that in the other directions. These include shear flows, flows around boundary layers and in compacted geometrically disordered media, where the characteristic length scales or spatial gradients in the flow are dominant in certain directions relative to others. Such multidimensional flows governed by the Navier-Stokes equations can be more efficiently computed using methods which use grids that naturally conform with the direction-specific non-homogeneities inherent to the problem of interest. The LB methods, on the other hand, are usually constructed using uniform lattice grids, such as using a square lattice in two-dimensions (2D) and a cubic lattice in three-dimensions (3D) in order to satisfy their symmetry and isotropy requirements. One approach to overcome this issue is to employ nonuniform grids with the LB method augmented using interpolations (e.g.,~\cite{he1996some,lu2002large}), which however introduce considerable additional numerical dissipation compromising the accuracy of the approach~\cite{lallemand2000theory}. Alternatively, conventional discretizations, such as finite difference, finite volume or finite element approaches (e.g.,~\cite{peng1998lattice,li2004least}) could be used which however entail additional complexity and overhead to the LB schemes. It is thus desirable to use the standard LB discretization along the particle characteristics that preserves the lock-step or perfect streaming with relatively low attendant numerical dissipation while allowing the use of different particle speeds along different directions. Such types of LB schemes are associated with the use of rectangular lattice grids in 2D and cuboid lattice grids in 3D.

Starting from the initial work of Koelman~\cite{koelman1991simple} much focus has been given to the construction of the LB methods on rectangular grids during the last two decades using different collision models with some necessary modifications to the algorithm to satisfy the Navier-Stokes (NS) equations. In the simpler rectangular LB versions, the single-relaxation-time (SRT) model has been modified by including additional particle velocities and whose equilibria were constructed via solving a quadrature problem~\cite{hegele2013rectangular}, or by extending the equilibrium distribution functions with additional corrections to restore the isotropy effects~\cite{peng2016lattice}, or by including some counteracting forcing terms~\cite{wang2019simulating}. On the other hand, by exploiting the additional degrees of freedom existing in the multiple-relaxation-time (MRT) collision model based on raw moments, rectangular LB schemes were constructed in the following four different ways: (i) by coupling between various relaxation parameters and the grid aspect ratio via a linear stability analysis~\cite{bouzidi2001lattice}, (ii) by keeping the matrix of the transformation between the moment space and the velocity space of the distribution functions independent of the grid aspect ratio~\cite{zhou2012mrt}, (iii) by using an additional adjustable parameter that determines the relative orientation in the energy-normal stress subspace via an inverse design analysis based on the Chapman-Enskog expansion~\cite{zong2016designing}, and (iv) by extending the equilibrium moments to include the stress components for restoring the isotropy of the recovered macroscopic equations~\cite{peng2016lattice}. It should be pointed out that Zong \emph{et al}~\cite{zong2016designing} demonstrated that the emergent continuum limit equations of some of the earlier rectangular LB formulations~\cite{koelman1991simple,bouzidi2001lattice,zhou2012mrt} were not fully consistent with the NS equations. Moreover, although the latter rectangular LB schemes~\cite{zong2016designing,peng2016lattice} yielded physically correct formulations, they require cumbersome implementations involving the specification of many free parameters and complicated expressions for the transport coefficients and mapping matrices dependent on the grid aspect ratio, and, like the other consistent schemes~\cite{hegele2013rectangular,wang2019simulating}, are subject to major stability issues at lower grid aspect ratios when simulating flows at relatively large characteristic flow velocities or low viscosities.

The rationale for the limitations of the prior rectangular LB schemes were clarified in our recent work~\cite{yahiaAPSDFD2017,yahiaAPSDFD2018,yahia2021central}. These include their choice of the orthogonal moment basis, construction of the discrete equilibria involving only the lower order velocity terms and without correcting for the non-Galilean invariant cubic velocity errors arising from aliasing effects, and the use of collision models based on raw moments. Moreover, it has been pointed out in an earlier work for standard lattices (square or cubic)~\cite{dubois2015stability,geier2015cumulant} that the use of central moments and avoiding the orthogonalization of the moment basis leads to significant stability improvements. The construction of the collision step in the LB formulations using central moments based on the peculiar velocity~\cite{geier2006cascaded,premnath2011three} naturally maintains the Galilean invariance of the moments independently supported by the lattice and its advantages have been demonstrated for a variety of fluid dynamical problems (see e.g.,~\cite{ning2016numerical,chavez2018improving,HAJABDOLLAHI2018838,hajabdollahi2019cascaded,hajabdollahi2021central,adam2021cascaded}). Based on these considerations, we proposed a 2D central moment rectangular LB scheme recently and demonstrated its superior numerical features for simulating flows at higher Reynolds numbers using relatively small grid aspect ratio when compared to the other existing LB methods based on the rectangular lattice~\cite{yahiaAPSDFD2017,yahiaAPSDFD2018,yahia2021central}.

Since anisotropic and inhomogeneous flows in situations of practical interest are often 3D in nature, it would be beneficial to develop LB algorithms on cuboid lattices. Their exist few prior studies in this regard. For example, Hegele \emph{et al}~\cite{hegele2013rectangular} presented a SRT-LB scheme using a D3Q23 lattice that involves the use of two additional particle velocities in each of the two Cartesian directions embedded to the D3Q19 lattice and constructed the equilibrium distribution functions by solving a quadrature problem to ensure the desired isotropy. It evaluated the accuracy of the resulting approach for simulating cylindrical waves by using grid sizes with only small deviations from the cubic lattice. Later, Jiang and Zhang~\cite{jiang2014orthorhombic} presented a cuboid LB scheme using a D3Q19 lattice for simulations of porous media. However, due to severe numerical stability restrictions, it was able to use grid aspect ratios with minor variations from unity. More recently, Wang \emph{et al}~\cite{wang2019lattice} presented a raw moment-based MRT-LB scheme on the D3Q19 cuboid lattice which utilized an orthogonal moment basis. The equilibria and the transport coefficients were constructed via an inverse design analysis to satisfy the Navier-Stokes equations and involved the specification of many free parameters entailing a cumbersome implementation of the method. The resulting numerical approach was validated for some benchmark flow problems at moderate grid aspect ratios. Here, it should be pointed out that the choice of the equilibria plays a crucial role in maintaining the physical isotropy of the fluid flow equations~\cite{bauer2020truncation}. This can be naturally constructed by matching with the continuous Maxwell distribution as done in central moment LB formulations rather than involving complicated fitting of parameters. Moreover, as mentioned above, the use of central moments is expected to provide significant improvements to the state-of-the-art in LBM based on non-cubic grids. However, a 3D central moment LBM based on a cuboid lattice does not currently exist in the literature, which can be constructed via an extension and significant modification of our recent work on the rectangular lattice~\cite{yahia2021central} and is the main objective of this present paper.

In this study, we will present a new 3D cuboid central moment LB method on a D3Q27 lattice and is referred to as the 3DCCM-LBM in what follows, which is parameterized by two grid aspect ratios representing the ratios of the characteristic particle speeds along the two directions with respect to those in the remaining direction. The D3Q27 lattice is chosen since it provides greater possible isotropy and accuracy when compared to its D3Q15 and D3Q19 lattice subsets~\cite{silva2014truncation}; however, the equilibria corrections needed for the D3Q27 cuboid lattice to satisfy the NS equations (see below for more details) are applicable for these lattice subsets as well. In contrast to the prior approaches, our cuboid LB scheme will be constructed by using a natural moment basis that avoids orthogonalization and the discrete central moment equilibria are specified by matching with those corresponding to the continuous Maxwell distribution function. Moreover, the prior LB algorithms on stretched lattice grids based on the various collision models discussed above, including our recent 2D rectangular central moment LB scheme~\cite{yahia2021central}, prescribe the transformations between the distribution functions and raw moments before and after collision by using a moment basis that separates the trace of the second order moments (related to bulk viscosity) from its other components (related to shear viscosity). The use of this strategy in the context of the 3D formulations based on the cuboid lattice would result quite complicated transformation matrices dependent on the grid aspect ratios. This is obviated in our present approach by naturally separating the bulk and shear viscosity effects associated with the viscous stress only within the collision step and not for such transformations, and the effect of the grid aspect ratios are introduced via much simpler pre- and post-collision diagonal scaling matrices in such mappings. We will show the resulting compact approach can be naturally interpreted based on special matrices. The truncation errors resulting from the grid anisotropy associated with the cuboid lattice as well as those due to the aliasing effects will be eliminated by deriving the necessary correction terms from a Chapman-Enskog analysis, which will be augmented to the second order moment equilibria. The resulting 3DCCM-LBM is modular in construction in that the existing 3D central moment (and also its special case involving raw moment) based algorithms developed for the cubic lattices can be readily extended to cuboid lattices by using the corrections to the moment equilibria derived in this work and introducing the pre- and post-collision diagonal scaling matrices based on the grid aspect ratios. Our cuboid LB approach will be first validated against the analytical solutions and/or numerical results for some standard benchmark flow problems. The advantages of this scheme in efficiently simulating an example inhomogeneous and anisotropic flow problem with significant savings in memory storage and computational effort will then be demonstrated. Moreover, we will also show numerical stability improvements in the use of our 3D central moment scheme when compared to that based on raw moments for computation of shear flows at relatively large flow velocities and/or low viscosities.

The organization of this paper is as follows. Section~\ref{sec:2} presents a Chapman-Enskog multiscale analysis of a 3D LB equation based on the non-orthogonal moment basis using a D3Q27 cuboid lattice. Since the viscous stress tensor in the NS equations is based on the second order non-equilibrium raw moments that are the same as the corresponding second order non-equilibrium central moments, it suffices to present our analysis based on the simpler raw moment formulation; and as part of this, we will identify the correction terms involving velocity gradients and grid aspect ratios for eliminating the grid anisotropy and non-Galilean invariant cubic velocity related truncation errors and show consistency with the 3D NS equations. Formulas for computing the local strain rate tensor based on non-equilibrium moments and parameterized by the grid aspect ratios will also be derived in this section. In Sec.~\ref{sec:3}, we will discuss the complete details of the implementation aspects of our 3DCCM-LBM. Additional algorithmic details in this regard are given in various appendices (see~\ref{sec:appendix1}~--~\ref{sec:appendix4}). Formulas for the momentum-augmented bounce-back scheme for the D3Q27 cuboid lattice that are dependent on the grid aspect ratios for simulating shear flows due to moving boundaries are derived and summarized in Sec.~\ref{sec:4}. Section~\ref{sec:5} presents a numerical validation study of the 3DCCM-LBM using different grid aspect ratios for a variety of benchmark flow problems. This is followed by a demonstration of the computational effectiveness of using cuboid grids (via 3DCCM-LBM) in lieu of utilizing cubic grids for simulating inhomogeneous and anisotropic flows in Sec.~\ref{sec:6} and the numerical stability improvements achieved with using central moments rather than raw moments on the D3Q27 cuboid lattice at different grid aspect ratios in Sec.~\ref{sec:7}. Finally, the main conclusions of this work are given in Sec.~\ref{sec:8}.

\section{Chapman-Enskog Analysis on a D3Q27 Cuboid Lattice: Isotropy Corrections, Macroscopic Flow Equations, and Local Formulas for the Strain Rate Tensor} \label{sec:2}
\subsection{Cuboid lattice parameters, moment basis, and definitions of central moments and raw moments}
\begin{figure}[H]
\centering
 \includegraphics[width=0.5\textwidth] {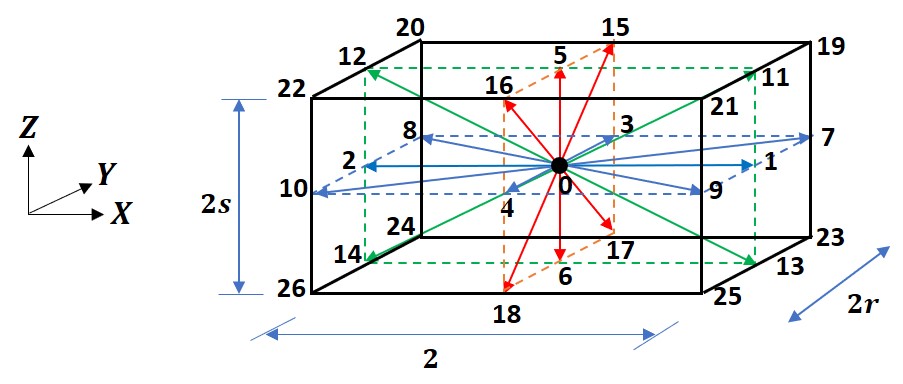}
 \caption{Three-dimensional, twenty seven particle velocities (D3Q27) cuboid lattice.}
 \label{fig:1c}
\end{figure}

The cuboid grid based on the three dimensional, twenty seven velocities (D3Q27) lattice used in deriving the formulation in what follows is shown in Fig.~\ref{fig:1c}.
The components of the particle velocities $\mathbf{e}=\left( \mathbf{e}_x, \mathbf{e}_y, \mathbf{e}_z\right)$ in the $y$ and $z$ directions are related to the grid aspect ratios $r$ and $s$, respectively, which is defined below. In other words, these two free parameters represent the ratios of the characteristic particle speeds along the $y$ and $z$ directions with respect to those in the $x$ direction, and formalize the flexibility accorded by the cuboid lattice. Thus, if $\Delta x$, $\Delta y$ and $\Delta z$ are the space steps in the $x$, $y$ and $z$ directions, respectively, for evolving over a time step $\Delta t$ that would then fix the particle speeds in the respective directions (i.e., $c_x =\Delta x/\Delta t$, $c_y =\Delta y/\Delta t$ and $c_z =\Delta z/\Delta t$), we can then define the grid aspect ratios as $r=\Delta y/ \Delta x$, and $s=\Delta z/ \Delta x$. As is standard in LB formulations, we work with the usual lattice units for simplicity in the following, i.e., the reference space step in the $x$ direction and the time step are taken to be unity: $\Delta x=1$ and $\Delta t=1$. Thus, $\Delta y= r$ and $\Delta z=s$. We now clarify the notations used in Fig.~\ref{fig:1c}, where the magnitude of each of the particle velocity is the distance between the head and tail of an arrow representing that direction; in lattice units, with a unit time step, the Cartesian components of the length of each lattice direction or the streaming distance are bounded by $(1,r,s)$. Now, since in Fig.~\ref{fig:1c}, for every particle velocity direction, its opposite counterpart is also depicted, the total distance for such a pair encompasses a length with components $(2,2r,2s)$ as shown.
Based on these considerations, we can then list the Cartesian components of the particle velocities for the D3Q27 cuboid lattice as follows:
\begin{subequations}
\begin{eqnarray}
\ket{e_{x}} &=& (0,1,-1,0,0,0,0,1,-1,1,-1,1,-1,1,-1,0,0,0,0,1,-1,1,-1,1,-1,1,-1)^\dag, \label{eq:1a}\\
\ket{e_{y}} &=& (0,0,0,r,-r,0,0,r,r,-r,-r,0,0,0,0,r,-r,r,-r,r,r,-r,-r,r,r,-r,-r)^\dag,\label{eq:1b}\\
\ket{e_{z}} &=& (0,0,0,0,0,s,-s,0,0,0,0,s,s,-s,-s,s,s,-s,-s,s,s,s,s,-s,-s,-s,-s)^\dag,\label{eq:1c}
\end{eqnarray}
\end{subequations}
where $\ket{\cdot}$ denotes a column vector based on the standard `ket' notation and $\dag$ refers to taking the transpose of any array. We will also need the following 27-dimensional vector with unit elements in what follows:
\begin{eqnarray}
\ket{1}  = (1,1,1,1,1,1,1,1,1,1,1,1,1,1,1,1,1,1,1,1,1,1,1,1,1,1,1,1)^\dag. \label{eq:2}
\end{eqnarray}
We then define a linearly independent set of non-orthogonal basis vectors for the D3Q27 cuboid lattice using a combination of the monomials of the type $\ket{e_x^m e_y^n e_z^p}$ for integer exponents $m$,$n$ and $p$~\cite{premnath2011three} as follows:
\begin{equation} \label{eq:4}
\tensor{T}= \Big[\ket{T_{0}},\ket{T_{1}},\ket{T_{2}},\ldots,\ket{T_{26}} \Big]^{\dag},
 \end{equation}
where~\cite{premnath2011three}
\begin{eqnarray}\label{eq:5}
\begin{matrix}
 \quad\ket{T_0}=\ket{1}, & \ket{T_9}=\ket{e_x^2 + e_y^2+e_z^2} & \ket{T_{18}},=\ket{e_x^2e_y^2 +e_x^2e_z^2 -e_y^2e_z^2}, \\
 \quad\ket{T_1}=\ket{e_x}, & \;\;\;\;\ket{T_{10}}=\ket{{e_x}e_y^2 +{e_x}e_z^2}, & \ket{T_{19}}=\ket{e_x^2e_y^2 - e_x^2e_z^2 }, \\
 \quad\ket{T_2}=\ket{e_y}, & \;\;\;\;\ket{T_{11}}=\ket{e_x^2{e_y} +{e_y}e_z^2}, & \ket{T_{20}}=\ket{e_x^2{e_y}{e_z}}, \\
 \quad\ket{T_3}=\ket{e_z}, & \;\;\;\;\ket{T_{12}}=\ket{e_x^2{e_z} +e_y^2{e_z}}, & \ket{T_{21}}=\ket{{e_x}e_y^2{e_z}}, \\
 \;\;\;\;\ket{T_4}=\ket{e_x e_y}, & \;\;\;\;\ket{T_{13}}=\ket{{e_x}e_y^2 -{e_x}e_z^2}, & \ket{T_{22}}=\ket{{e_x}{e_y}e_z^2}, \\
 \;\;\;\;\ket{T_5}=\ket{e_x e_z}, & \;\;\;\;\ket{T_{14}}=\ket{e_x^2{e_y} -{e_y}e_z^2}, & \;\;\ket{T_{23}}=\ket{{e_x}e_y^2 e_z^2},\\
 \;\;\;\;\ket{T_6}=\ket{e_y e_z}, & \;\;\;\;\ket{T_{15}}=\ket{e_x^2{e_z} -e_y^2{e_z}}, & \;\;\ket{T_{24}}=\ket{e_x^2{e_y}e_z^2}, \\
 \ket{T_7}=\ket{e_x^2 -e_y^2}, & \ket{T_{16}}=\ket{{e_x}{e_y}{e_z}}, & \quad\ket{T_{25}},=\ket{e_x^2 e_y^2{e_z}}, \\
 \ket{T_8}=\ket{e_x^2 -e_z^2}, & \ket{T_{17}}=\ket{e_x^2e_y^2 +e_x^2e_z^2 +e_y^2e_z^2}, & \;\;\;\;\ket{T_{26}}=\ket{e_x^2e_y^2e_z^2}.
\end{matrix}
\end{eqnarray}
As discussed in the introduction section, we will not orthogonalize the above set of basis vectors further to maintain simplicity and robustness and retain them in their natural forms in the following derivation based on the 3D cuboid lattice analogous to our rectangular central moment LB formulation~\cite{yahia2021central}. It should be noted that while the above basis vectors segregate the trace of the second order components $\ket{e_x^2 + e_y^2+e_z^2}$ from the other second order components in order to enable the independent specification of the bulk viscosity from shear viscosity, as mentioned earlier, in the algorithmic implementation of the central moment approach discussed in the next section (see Sec.~\ref{sec:3}), these operations will be confined only within the collision step and not for performing any mappings between distribution functions and moments. These considerations are crucial in avoiding cumbersome transformations and corrections for eliminating any truncation errors in order to develop an efficient algorithm on a cuboid lattice.

Then, we list the vectors of the distribution functions $\mathbf{f}$, their equilibria $\mathbf{{f}}^{eq}$ and the source terms $\mathbf{{S}}$ accounting for the effect of any body force with components $\bm{F}=(F_x, F_y, F_z)$ experienced by the motion of the fluid with density $\rho$ and velocity $\bm{u}=(u_x,u_y,u_z)$ for the D3Q27 cuboid lattice as $\mathbf{f}=\left(f_{0},f_{1},f_{2},\ldots,f_{26}\right)^{\dag}, \quad \mathbf{{f}}^{eq}=\left({f}_{0}^{eq},{f}_{1}^{eq},{f}_{2}^{eq},\ldots,{f}_{26}^{eq}\right)^{\dag},\quad \mathbf{{S}}=\left({S}_{0},{S}_{1},{S}_{2},\ldots,{S}_{26}\right)^{\dag}$. In anticipation of the developments in Sec.~\ref{sec:3}, we first define the central moments of the distribution functions $f_\alpha$ and their equilibria $f^{eq}$, as well as the source term $S_\alpha$ of order $(m+n+p)$ using the weights as the peculiar velocity components, i.e., the components of the particle velocity shifted by those of the fluid velocity, as follows:
\begin{eqnarray}
\begin{pmatrix}
 k_{mnp}\\
 k^{eq}_{mnp}\\
 \sigma_{mnp}
 \end{pmatrix}
  &=& \sum_{\alpha=0}^{26}
 \begin{pmatrix}
 f_{\alpha}\\
 f_{\alpha}^{eq}\\
 S_{\alpha}
 \end{pmatrix}
 (e_{\alpha x} -u_x)^m  (e_{\alpha y} -u_y)^n (e_{\alpha z} -u_z)^p.\label{eq:3a}
\end{eqnarray}
Also, we define the bare or raw moments of the above three quantities of order $(m+n+p)$ by using just the particle velocity components as the weights, which are given by
\begin{eqnarray}
 \begin{pmatrix}
 k^{\prime}_{mnp},\\
 k_{mnp}^{eq\prime},\\
 \sigma^{\prime}_{mnp}.
 \end{pmatrix}
  &=& \sum_{\alpha=0}^{26}
 \begin{pmatrix}
 f_{\alpha}\\
 f_{\alpha}^{eq}\\
 S_{\alpha}
 \end{pmatrix}
 e_{\alpha x}^m  e_{\alpha y}^n e_{\alpha z}^p.\label{eq:3b}
\end{eqnarray}
Note that here and in the following any quantity with a prime notation ($\prime$) is used to distinguish it as a raw moment (as opposed to a central moment). For convenience, we now enumerate these raw moments as
\begin{align*}
\mathbf{n}=\;\;\Big( &k_{000}^\prime,\; k_{100}^\prime,\; k_{010}^\prime,\; k_{001}^\prime, \;k_{110}^\prime, \;k_{101}^\prime ,\; k_{011}^\prime,\; (k_{200}^\prime-k_{020}^\prime),\; (k_{200}^\prime-k_{002}^\prime),
\;(k_{200}^\prime+k_{020}^\prime+k_{002}^\prime),\\  & (k_{120}^\prime+k_{102}^\prime), \;(k_{210}^\prime+k_{012}^\prime),\;
(k_{201}^\prime+k_{021}^\prime),\; (k_{120}^\prime-k_{102}^\prime),\;(k_{210}^\prime-k_{012}^\prime),\; (k_{201}^\prime-k_{021}^\prime),\;k_{111}^\prime,\\
&(k_{220}^\prime+k_{202}^\prime+k_{022}^\prime),\;
(k_{220}^\prime+k_{202}^\prime-k_{022}^\prime),\; (k_{220}^\prime-k_{202}^\prime),\;k_{211}^\prime,\;k_{121}^\prime,\;k_{112}^\prime,
\;k_{122}^\prime,\;k_{212}^\prime,\;k_{221}^\prime,\;k_{222}^\prime\Big)^{\dag},
\end{align*}
\begin{align*}
\mathbf{n}^{eq}=\;\;\Big( &\K_{000}^{\eqp},\; \K_{100}^{\eqp},\; \K_{010}^{\eqp},\; \K_{001}^{\eqp}, \;\K_{110}^{\eqp},\; \K_{101}^{\eqp} , \;\K_{011}^{\eqp}, \;(\K_{200}^{\eqp}-\K_{020}^{\eqp}),\;(\K_{200}^{\eqp}-\K_{002}^{\eqp}),\;
(\K_{200}^{\eqp}+\K_{020}^{\eqp}+\K_{002}^{\eqp}),\\  & \;(\K_{120}^{\eqp}+\K_{102}^{\eqp}), \; (\K_{210}^{\eqp}+\K_{012}^{\eqp}),\;(\K_{201}^{\eqp}+\K_{021}^{\eqp}),\; (\K_{120}^{\eqp}-\K_{102}^{\eqp}),\;(\K_{210}^{\eqp}-\K_{012}^{\eqp}),\; (\K_{201}^{\eqp}-\K_{021}^{\eqp}),\;\K_{111}^{\eqp},\\
&\;(\K_{220}^{\eqp}+\K_{202}^{\eqp}+\K_{022}^{\eqp}),\;
(\K_{220}^{\eqp}+\K_{202}^{\eqp}-\K_{022}^{\eqp}), \; (\K_{220}^{\eqp}-\K_{202}^{\eqp}),\;\K_{211}^{\eqp},\;\K_{121}^{\eqp},\;\K_{112}^{\eqp},
\;\K_{122}^{\eqp},\;\K_{212}^{\eqp},\;\K_{221}^{\eqp},\;\K_{222}^{\eqp}\Big)^{\dag},
\end{align*}
\begin{align*}
\mathbf{\Psi}\;=\;\;\Big(&\sg_{000}^\prime,\; \sg_{100}^\prime,\; \sg_{010}^\prime,\; \sg_{001}^\prime, \;\sg_{110}^\prime, \;\sg_{101}^\prime , \;\sg_{011}^\prime,\; (\sg_{200}^\prime-\sg_{020}^\prime),\;(\sg_{200}^\prime-\sg_{002}^\prime),\;(\sg_{200}^\prime+\sg_{020}^\prime+\sg_{002}^\prime),\\ & \;(\sg_{120}^\prime+\sg_{102}^\prime), \;(\sg_{210}^\prime+\sg_{012}^\prime),\;(\sg_{201}^\prime+\sg_{021}^\prime),\; (\sg_{120}^\prime-\sg_{102}^\prime),\;(\sg_{210}^\prime-\sg_{012}^\prime),\; (\sg_{201}^\prime-\sg_{021}^\prime),\;\sg_{111}^\prime,\\
&\;(\sg_{220}^\prime+\sg_{202}^\prime+\sg_{022}^\prime),\;
(\sg_{220}^\prime+\sg_{202}^\prime-\sg_{022}^\prime),\;(\sg_{220}^\prime-\sg_{202}^\prime),\;\sg_{211}^\prime,\;\sg_{121}^\prime,\;\sg_{112}^\prime,
\;\sg_{122}^\prime,\;\sg_{212}^\prime,\;\sg_{221}^\prime,\;\sg_{222}^\prime\Big)^{\dag}.
\end{align*}
Here, the distribution functions and raw moments are related via $\tensor{T}$, i.e., the basis vectors (Eq.~(\ref{eq:5})), as
\begin{equation} \label{eq:7}
\mathbf{n}= \tensor{T}\mathbf{f}, \qquad \mathbf{n^{eq}}=  \tensor{T} \mathbf{f^{eq}}, \qquad \mathbf{\Psi}= \tensor{T}\mathbf{S},
\end{equation}
\subsection{The lattice Boltzmann equation}
It should be noted that the use of a cuboid lattice would result in an anisotropic viscous tensor given in terms of the grid aspect ratios $r$ and $s$ via the second order non-equilibrium moments, and such grid-related anisotropy effects need be eliminated via carefully designed correction terms. In constructing such counteracting corrections, since the second order non-equilibrium raw components are the same as the corresponding central moments by definition, it is enough to perform an analysis on the following simpler lattice Boltzmann equation (LBE) on the raw moments, i.e., the MRT-LBE~\cite{yahia2021central}:
\begin{equation}\label{eq:8}
 \mathbf{f} (\bm{x}+\mathbf{e}\Delta t, t+\Delta t)- \mathbf{f} (\bm{x},t) =  \tensor{T^{-1}}
 \Big[  \tensor{\Lambda} \;  \left(\; \mathbf{n}^{eq}-\mathbf{n} \;\right) + \left(\tensor{I} - \frac{\tensor{\Lambda}}{2}\right)  \mathbf{\Psi}\Delta t \Big],
\end{equation}
where $\tensor{\Lambda} = \mbox{diag} \; \big(\omega_0, \omega_1,\omega_2, \dots, \omega_{26} \big)$ is the diagonal relaxation time matrix and $\tensor{I}$ is a $27\times 27$ identity matrix. Here, the right side of Eq.~(\ref{eq:8}) prescribes the relaxation of different moments $n_j$ to their equilibria $n_j^{eq}$ at a rate given by the parameter $\omega_j$, which is augmented by the effect of the source term via $\Psi_j$ ($j=0,1,2,\ldots,26$) and the results are then mapped back into the velocity space via the inverse mapping $\tensor{T^{-1}}$. This is followed by their perfect advection of the distribution functions $f_\alpha$ to the nearest node along the particle characteristics $\bm{e}_\alpha$ ($\alpha=0,1,2,\ldots,26$) as implied by the left side Eq.~(\ref{eq:8}). Then, the hydrodynamic fields are computed via appropriate moments as
\begin{eqnarray}\label{eq:12}
\rho =\sum_{\alpha=0}^{26} \mathbf{f}_{\alpha}, \quad \;\;\;\rho \bm{u} =\sum_{\alpha=0}^{26} f_{\alpha} \bm{e}_{\alpha} + \frac{1}{2} \bm{F} \Delta t.
\end{eqnarray}

\subsection{Equilibria and sources: Central moments and raw moments}
In this work, we obtain the countable discrete \emph{central moment} equilibria for the D3Q27 cuboid lattice by matching those of the continuous Maxwell distribution function~\cite{geier2006cascaded} and given by
\begin{align}\label{eq:9}
&\K_{000}^{eq}=\rho,\;\;\;\;\;\; \quad \K_{100}^{eq}=\K_{010}^{eq}=\K_{001}^{eq}=0,\;\;\;\;\;\; \K_{110}^{eq}=\K_{101}^{eq}=\K_{011}^{eq}=0,\;\;\;\;\;\; \K_{200}^{eq}=\K_{020}^{eq}=\K_{002}^{eq}=c_s^2 \rho,\nonumber\\[3pt]
&\K_{120}^{eq}=\K_{102}^{eq}=\K_{210}^{eq}=\K_{012}^{eq}=\K_{201}^{eq}=\K_{021}^{eq}=0,\;\;\;\;\;\;
\K_{111}^{eq}=0,\;\;\;\;\;\; \K_{220}^{eq}=\K_{202}^{eq}=\K_{022}^{eq}=c_s^4\rho,\nonumber\\[3pt]
&\K_{211}^{eq}=\K_{121}^{eq}=\K_{112}^{eq}=0,\;\;\;\;\;\; \K_{122}^{eq}=\K_{212}^{eq}=\K_{221}^{eq}=0,\;\;\;\;\;\;
\K_{222}^{eq}= c_s^6 \rho .
\end{align}
where $c_s$ is the speed of sound, which is an adjustable parameter of the collision model and will be related to the transport coefficients via a Chapman-Enskog analysis later. It should be noted, however, that the analysis in what follows that provides corrections for the second order moments, and the algorithm devised in the next section (Sec.~\ref{sec:3}) for the cuboid lattice are still applicable for other collision models. These include those based on the factorization property~\cite{geier2009factorized,premnath2011three} and cumulant collision model~\cite{geier2015cumulant}, which differ from the Maxwellian-based central moment model in the evolution of the higher order moments. Then, the corresponding discrete \emph{raw moment} equilibria are obtained from Eq.~(\ref{eq:9}) via binomial transforms, which read as follows:
\begin{align}\label{eq:10}
&\K_{000}^{\eqp}=\rho, \;\;\;\;\;\;\;\;\;\;\;\;\;\;\;\;\;\;\; \K_{100}^{\eqp}=\rho u_x , \;\;\;\;\;\;\;\;\;\;\; \K_{010}^{\eqp}=\rho u_y,\;\;\;\;\;\;\;\;\;\; \K_{001}^{\eqp}=\rho u_z,\nonumber\\
&\K_{110}^{\eqp}= \rho u_x u_y, \;\;\;\;\;\;\;\;\;\;\; \K_{101}^{\eqp}= \rho u_x u_z , \;\;\;\;\;\;\;\; \K_{011}^{\eqp}= \rho u_y u_z,\nonumber\\
&\K_{200}^{\eqp}= c_s^2 \rho + \rho u_x^2 , \;\;\;\;\;\;\K_{020}^{\eqp}= c_s^2 \rho + \rho u_y^2,\;\;\;\;\;\;\K_{002}^{\eqp}= c_s^2 \rho + \rho u_z^2,\nonumber\\[7pt]
&\K_{120}^{\eqp}= c_s^2 \rho u_x + \rho u_x u_y^2 , \;\;\;\;\;\;\K_{102}^{\eqp}= c_s^2 \rho u_x + \rho u_x u_z^2,\;\;\;\;\;\;\K_{210}^{\eqp}= c_s^2 \rho u_y + \rho u_x^2 u_y,\nonumber\\
&\K_{012}^{\eqp}= c_s^2 \rho u_y + \rho u_y u_z^2, \;\;\;\;\;\;\K_{201}^{\eqp}= c_s^2 \rho u_z + \rho u_x^2 u_z,\;\;\;\;\;\;\K_{021}^{\eqp}= c_s^2 \rho u_z + \rho u_y^2 u_z, \nonumber\\
&\K_{111}^{\eqp}= \rho u_x u_y u_z,\nonumber\\[7pt]
&\K_{220}^{\eqp}= c_s^4 \rho + \rho c_s^2 ( u_x^2 + u_y^2) + \rho  u_x^2 u_y^2,\;\;\;\;\;\;\K_{202}^{\eqp}= c_s^4 \rho + \rho c_s^2 ( u_x^2 + u_z^2) + \rho  u_x^2 u_z^2,\nonumber\\
&\K_{022}^{\eqp}= c_s^4 \rho + \rho c_s^2 ( u_y^2 + u_z^2) + \rho  u_y^2 u_z^2, \nonumber\\[7pt]
&\K_{211}^{\eqp}=  \rho ( c_s^2 + u_x^2) u_y u_z,  \;\;\;\;\;\; \K_{121}^{\eqp}=  \rho ( c_s^2 + u_y^2) u_x u_z,\;\;\;\;\;\; \K_{112}^{\eqp}=  \rho ( c_s^2 + u_z^2) u_x u_y, \nonumber\\[7pt]
&\K_{122}^{\eqp}= c_s^4 \rho u_x+ \rho c_s^2 u_x ( u_y^2 + u_z^2) + \rho  u_x u_y^2 u_z^2, \;\;\;\;\;\; \K_{212}^{\eqp}= c_s^4 \rho u_y+ \rho c_s^2 u_y ( u_x^2 + u_z^2) + \rho  u_x^2 u_y  u_z^2,\nonumber\\
&\K_{221}^{\eqp}= c_s^4 \rho u_z+ \rho c_s^2 u_z ( u_x^2 + u_y^2) + \rho  u_x^2 u_y^2 u_z,\nonumber\\[7pt]
&\K_{222}^{\eqp}= c_s^6 \rho + \rho c_s^4 ( u_x^2 + u_y^2 + u_z^2) + \rho c_s^2 ( u_x^2 u_y^2 + u_y^2 u_z^2+ u_x^2 u_z^2)+ \rho  u_x^2 u_y^2 u_z^2.
\end{align}

In addition, the central moments of the source terms for recovering the NS equations in 3D are given by~\cite{premnath2011three}
\begin{equation}\label{eq:11c}
\sg_{000}=0, \quad \sg_{100}=F_x, \quad \sg_{010}=F_y, \quad \sg_{001}=F_z, \quad \sg_{mnp}=0 \;\;\; \mbox{if} \;\;\; (m+n+p)\ge 2,
\end{equation}
and the corresponding raw moments follow from their binomial transform as~\cite{premnath2011three}
\begin{align}\label{eq:11}
&\sg_{000}^\prime=0,\nonumber\\
&\sg_{100}^\prime=F_x\ \;\;\;\;\;\ \sg_{110}^\prime=F_x u_y + F_y u_x,\;\;\;\;\;\  \sg_{200}^\prime= 2 F_x u_x,  \nonumber\\
&\sg_{010}^\prime=F_y,\;\;\;\;\;\ \sg_{101}^\prime= F_x u_z + F_z u_x ,\;\;\;\;\;\ \sg_{020}^\prime= 2 F_y u_y, \nonumber\\
&\sg_{001}^\prime=F_z,  \;\;\;\;\;\ \sg_{011}^\prime=F_y u_z + F_z u_y ,\;\;\;\;\;\ \sg_{002}^\prime=2 F_z u_z, \nonumber\\
&\sg_{mnp}^\prime=0 \;\;\; \mbox{if} \;\;\; (m+n+p)\ge 3;
\end{align}

\subsection{Chapman-Enskog Analysis: Identification of truncation errors due to grid anisotropy and non-Galilean invariance from aliasing effects on the D3Q27 cuboid lattice\label{subsec:CEanalysis}}
We will now perform an analysis based on the Chapman-Enskog (C-E) multiscale expansion~\cite{chapman1990mathematical}, written for the LBE in the matrix form by d'Humi\`{e}res~\cite{DHumieres92}, to identify the truncation errors due to anisotropic effects arising from using the cuboid lattice grid and non-Galilean invariant (GI) cubic velocity errors resulting from the aliasing effects associated with the D3Q27 lattice. Corrections to eliminate these errors will then be derived in what follows. The analysis for a 3D LBE formulation given above follows the approach presented in Premnath and Banjeree~\cite{premnath2009incorporating} and Hajabdollahi and Premnath~\cite{Hajabdollahi201897}, which was adopted for constructing a rectangular LB algorithm recently~\cite{yahia2021central}. In this version of the C-E analysis, we expand the moments about their equilibria by including the non-equilibrium effects as higher order perturbations and the time derivative as a multiple scale expansion as follows:
\begin{equation}\label{eq:13}
\mathbf{n}= \sum_{j=0}^{\infty} \epsilon^{j} \mathbf{n}^{(j)}, \quad \partial_t= \sum_{j=0}^{\infty} {\epsilon}^{j} {\partial_{t_j}},
\end{equation}
where $\epsilon= \Delta t$ is the perturbation parameter that serves the purpose of bookkeeping and isolating terms of the different orders. Moreover, we apply a multivariate Taylor series expansion in both space and time to the first term on the left side of Eq.~(\ref{eq:8}) and then transform all the resulting terms to the moment space via $\mathbf{f}=  \tensor{T}^{-1} \mathbf{n}$. Subsequently, substituting the C-E expansions (Eq.~(\ref{eq:13})) in the 3D LBE with the natural moment basis given in Eq.~(\ref{eq:8}) and then grouping terms of the same order of $\epsilon$, we obtain the following set of moment equations identified by $O(\epsilon^k)$ for $k=0,1$ and $2$:
\begin{subequations}
\begin{eqnarray}\label{eq:14}
 \centering
&O (\epsilon^0 ):  \mathbf{n}^{(0)} =  \mathbf{n}^{eq},  \label{eq:14a}\\
&O (\epsilon^1 ): \left({\partial_{t_0}} + \tensor{E}_i \partial_i\right)  \mathbf{n}^{(0)} =  - \tensor{\Lambda} \; \mathbf{n}^{(1)}+\mathbf{\Psi},  \label{eq:14b} \\
&O (\epsilon^2 ) : {\partial_{t_1}} \; \mathbf{n}^{(0)} +  \left({\partial_{t_0}} + \tensor{E}_i \partial_i\right) \;\left[\left( \tensor{I} - \frac{\tensor{\Lambda}} {2} \right)\mathbf{n}^{(1)}\right] =  - \tensor{\Lambda} \; \mathbf{n}^{(2)}, \label{eq:14c}
 \end{eqnarray}
 \end{subequations}
where $ \tensor{E}_i=\tensor{T} \;( \mathbf{e_i} \; \tensor{I})\tensor{T^{-1}}$ and $\mathbf{e}_i=\ket{e_{i} }$ with $i \in (x,y,z)$. We then re-express Eqs.~(\ref{eq:14b})and (\ref{eq:14c}) in the long form, which respectively read as

\underline{O($\epsilon$) moment system:}
\begin{equation}\label{eq:15}
\partial_{t_0} \mathbf{n}^{(0)}+\partial_{x} \tensor{E}_x\mathbf{n}^{(0)}+\partial_{y} \tensor{E}_y\mathbf{n}^{(0)}+\partial_{z} \tensor{E}_z\mathbf{n}^{(0)}=-\tensor{\Lambda} \mathbf{n}^{(1)}+\mathbf{\Psi}.
\end{equation}

\underline{O($\epsilon^2$) moment system:}
\begin{equation}\label{eq:16}
\partial_{t_0} \mathbf{n}^{(0)}+\partial_{t_0} \left(\tensor{I}- \frac{\tensor{\Lambda}}{2}\right)\mathbf{n}^{(1)}+\partial_{x} \tensor{E}_x\left(\tensor{I}- \frac{\tensor{\Lambda}}{2}\right)\mathbf{n}^{(1)}+\partial_{y} \tensor{E}_y\left(\tensor{I}- \frac{\tensor{\Lambda}}{2}\right)\mathbf{n^{(1)}}+ \partial_{z} \tensor{E}_z\left(\tensor{I}- \frac{\tensor{\Lambda}}{2}\right)\mathbf{n}^{(1)}= -\tensor{\Lambda}\mathbf{n}^{(2)}.
\end{equation}

Substituting the moment equilibria $\mathbf{n}^{(0)}$ listed in Eq.~(\ref{eq:10}) into the $O(\epsilon)$ Eq.~(\ref{eq:15}), we now write explicitly all the equations up to the second order moment components which are relevant to determining the fluid dynamical behavior as follows:
\begin{subequations}
 \begin{eqnarray}\label{eq:17a}
 &\partial_{t_0}\rho + \partial_x (\rho u_x) + \partial_y (\rho u_y) + \partial_z (\rho u_z) = 0,
 \end{eqnarray}
 \begin{eqnarray}\label{eq:17b}
 &\partial_{t_0}(\rho u_x) + \partial_x (\rho c_s^2  + \rho u_x^2) + \partial_y (\rho u_x u_y)+ \partial_z (\rho u_x u_z) = F_x,
 \end{eqnarray}
\begin{eqnarray}\label{eq:17c}
&\partial_{t_0}(\rho u_y) + \partial_x (\rho u_x u_y)+ \partial_y (\rho c_s^2 + \rho u_y^2)+ \partial_z (\rho u_y u_z) = F_y,
\end{eqnarray}
\begin{eqnarray}\label{eq:17d}
&\partial_{t_0}(\rho u_z) + \partial_x (\rho u_x u_z)+\partial_y (\rho u_y u_z)+ \partial_z (\rho c_s^2 + \rho u_z^2) = F_z,
\end{eqnarray}
\begin{eqnarray}\label{eq:17e}
& \partial_{t_0}(\rho u_x u_y)+  \partial_x (c_s^2 \rho u_y + \rho u_x^2 u_y)+ \partial_y (c_s^2 \rho u_x + \rho u_x u_y^2)+ \partial_z (\rho u_x u_y u_z) = \nonumber \\
&- \omega_4\;  n_4^{(1)}+ \left( F_x u_y + F_y u_x \right),
\end{eqnarray}
\begin{eqnarray}\label{eq:17f}
& \partial_{t_0}(\rho u_x u_z)+  \partial_x (c_s^2 \rho u_z + \rho u_x^2 u_z)+ \partial_y (\rho u_x u_y u_z)+ \partial_z (c_s^2 \rho u_x + \rho u_x u_z^2) = \nonumber \\
&- \omega_5\;  n_5^{(1)}+ \left( F_x u_z + F_z u_x \right),
\end{eqnarray}
\begin{eqnarray}\label{eq:17g}
& \partial_{t_0}(\rho u_y u_z)+ \partial_x (\rho u_x u_y u_z)+ \partial_y (c_s^2 \rho u_z + \rho u_y^2 u_z)+ \partial_z (c_s^2 \rho u_y + \rho u_y u_z^2) = \nonumber \\
&- \omega_6\;  n_6^{(1)}+ \left( F_y u_z + F_z u_y \right),
\end{eqnarray}
\begin{eqnarray}\label{eq:17h}
& \partial_{t_0}\left[( \rho (u_x^2- u_y^2)\right]+  \partial_x \left[(1 -c_s^2) \rho u_x - \rho u_x u_y^2\right]+  \underline{\partial_y \left[(-r^2 + c_s^2) \rho u_y + \rho u_x^2 u_y\right]} + \partial_z \left[(\rho (u_x^2- u_y^2)u_z \right] = \nonumber \\
&- \omega_7\;  n_7^{(1)}+ 2 \left( F_x u_x - F_y u_y \right),
\end{eqnarray}
\begin{eqnarray}\label{eq:17i}
& \partial_{t_0}\left[ \rho ( u_x^2- u_z^2)\right]+  \partial_x \left[(1 -c_s^2) \rho u_x - \rho u_x u_z^2\right]+ \partial_y \left[(\rho (u_x^2- u_z^2)u_y \right] + \underline{\partial_z \left[(-s^2 + c_s^2) \rho u_z + \rho u_x^2 u_z\right]}= \nonumber \\
&- \omega_8\;  n_8^{(1)}+ 2 \left( F_x u_x - F_z u_z \right),
\end{eqnarray}
\begin{eqnarray}\label{eq:17j}
& \partial_{t_0}\left[ 3 c_s^2 \rho+ \rho( u_x^2+ u_y^2+u_z^2)\right]+  \partial_x \left[(1 +2c_s^2) \rho u_x + \rho u_x (u_y^2+ u_z^2)\right] + \underline{\partial_y \left[(r^2 +2c_s^2) \rho u_y + \rho u_y (u_x^2+ u_z^2)\right]} + \nonumber \\
&\underline{\partial_z \left[(s^2 +2c_s^2) \rho u_z + \rho  u_z(u_x^2+ u_y^2)\right]} =
- \omega_9\;   n_9^{(1)}+ 2 \left( F_x u_x + F_y u_y + F_z u_z \right),
\end{eqnarray}
\end{subequations}
Clearly, at the $O(\epsilon)$ level, the evolution of the conserved moments (Eqs.~(\ref{eq:17a})-(\ref{eq:17c})) are unaffected by anisotropic lattice grid. In other words, the evolutions of the density and the components of the momentum at the $O(\epsilon)$ level, respectively, given in Eqs.~(\ref{eq:17a})-(\ref{eq:17c}) do not contain any spurious terms associated with the grid aspect ratios. However, on the other hand, it is evident from the underlined terms that the diagonal components of the non-equilibrium parts of second order moments ($n_7^{(1)}$, $n_8^{(1)}$ and $n_9^{(1)}$) are influenced by the grid aspect ratios $r$ and $s$. Hence, it follows that such underlined terms will impact the viscous stress tensor and hence the hydrodynamics. Thus, in order to obtain the complete picture, we show in the following the evolution of the conserved moments at the $O(\epsilon^2)$ level, i.e., the leading four moment components of Eq.~(\ref{eq:16}), which manifest the effect of the non-equilibrium parts on the evolution of density and momentum components at the slower time scale $t_1$:
\begin{subequations}
\begin{eqnarray}
&\partial_{t_1}\rho=0, \label{eq:18a}\\
&\partial_{t_1}\left(\rho u_x\right)+\partial_x \left[\dfrac{1}{3}\left(1-\dfrac{\omega_7}{2}\right) n_7^{(1)}+\dfrac{1}{3}\left(1-\dfrac{\omega_8}{2}\right)n_8^{(1)}
+\dfrac{1}{3}\left(1-\dfrac{\omega_9}{2}\right)n_9^{(1)}\right]
+\partial_y \left[\left(1-\dfrac{\omega_4}{2}\right)n_4^{(1)}\right] \nonumber\\
&+\partial_z \left[\left(1-\dfrac{\omega_5}{2}\right)n_5^{(1)}\right]=0, \label{eq:18b}\\
&\partial_{t_1}\left(\rho u_y\right)+\partial_x \left[\left(1-\dfrac{\omega_4}{2}\right)n_4^{(1)}\right]+\partial_y \left[-\dfrac{2}{3}\left(1-\dfrac{\omega_7}{2}\right) n_7^{(1)}+\dfrac{1}{3}\left(1-\dfrac{\omega_8}{2}\right)n_8^{(1)}
+\dfrac{1}{3}\left(1-\dfrac{\omega_9}{2}\right)n_9^{(1)}\right] \nonumber\\
&+\partial_z \left[\left(1-\dfrac{\omega_6}{2}\right)n_6^{(1)}\right]=0,\label{eq:18c}\\
&\partial_{t_1}\left(\rho u_z\right)+\partial_x \left[\left(1-\dfrac{\omega_5}{2}\right)n_5^{(1)}\right]+\partial_y \left[\left(1-\dfrac{\omega_6}{2}\right)n_6^{(1)}\right]+\partial_z \Big[-\dfrac{2}{3}\left(1-\dfrac{\omega_7}{2}\right) n_7^{(1)}
+\dfrac{1}{3}\left(1-\dfrac{\omega_8}{2}\right)n_8^{(1)}\nonumber\\
&+\dfrac{1}{3}\left(1-\dfrac{\omega_9}{2}\right)n_9^{(1)}\Big]=0,\label{eq:18d}
\end{eqnarray}
\end{subequations}
Evidently, the hydrodynamical behavior given in Eqs.~(\ref{eq:18b})-(\ref{eq:18d}) is influenced by the grid aspect ratios via the non-equilibrium moments $n_7^{(1)}$, $n_8^{(1)}$ and $n_9^{(1)}$, which needs to eliminated. Before accomplishing this, we need the complete expressions of the second order non-equilibrium moments in order to isolate the truncation errors from terms that correspond to physics. Hence, we rewrite Eqs.~(\ref{eq:17e})-(\ref{eq:17j}) after segregating the terms associated with the grid aspect ratios $r$ and $s$ (see underlined terms) from those associated with the standard cubic lattice (terms within the brackets $\{\cdots \}$) as follows:
\begin{subequations}
\begin{align}
&n_4^{(1)}= \frac{1}{\omega_4} \Big\{-{\partial_{t_0}}(\rho u_x u_y)- \partial_x (c_s^2 \rho u_y +\rho u_x^2 u_y)- \partial_y(c_s^2 \rho u_x+\rho u_x u_y^2)- \partial_z(\rho u_x u_y u_z)+ ( F_x u_y + F_y u_x) \Big\}, \label{eq:19a}\\
&n_5^{(1)}= \frac{1}{\omega_5} \Big\{-{\partial_{t_0}}(\rho u_x u_z)- \partial_x (c_s^2 \rho u_z + \rho u_x^2 u_z)- \partial_y(\rho u_x u_y u_z)- \partial_z(c_s^2 \rho u_x +\rho u_x u_z^2)+ ( F_x u_z + F_z u_x) \Big\}, \label{eq:19b}\\
&n_6^{(1)}= \frac{1}{\omega_6} \Big\{-{\partial_{t_0}}(\rho u_y u_z)- \partial_x(\rho u_x u_y u_z)-\partial_y (c_s^2 \rho u_z+\rho u_y^2 u_z)- \partial_z(c_s^2 \rho u_y+\rho u_y u_z^2)+ ( F_y u_z + F_z u_y) \Big\}, \label{eq:19c}\\
&n_7^{(1)}= \frac{1}{\omega_7} \Big\{-{\partial_{t_0}}(\rho u_x^2 - \rho u_y^2)- \partial_x \left[(1 -c_s^2) \rho u_x -\rho u_x u_y^2 \right]- \partial_y \left[(-1 +c_s^2) \rho u_y +\rho u_x^2 u_y \right]- \partial_z \left[\rho (u_x^2 - u_y^2)u_z\right]\nonumber\\
& \quad\quad\quad\quad\quad+2( F_x u_x - F_y u_y) \Big\} - \frac{1}{\omega_7}\underline{ (1-r^2) \partial_y(\rho u_y)} , \label{eq:19d}\\
&n_8^{(1)}= \frac{1}{\omega_8} \Big\{-{\partial_{t_0}}(\rho u_x^2 - \rho u_z^2)- \partial_x \left[(1 -c_s^2) \rho u_x -\rho u_x u_z^2 \right]-\partial_y \left[\rho (u_x^2 - u_z^2)u_y\right] - \partial_z \left[(-1 +c_s^2) \rho u_z +\rho u_x^2 u_z \right] \nonumber\\
&  \quad\quad\quad\quad\quad +2( F_x u_x - F_z u_z) \Big\} - \frac{1}{\omega_8}\underline {(1-s^2) \partial_z(\rho u_z)} , \label{eq:19e}\\
&n_9^{(1)}= \frac{1}{\omega_9} \Big\{-3 c_s^2 {\partial_{t_0}} \rho -{\partial_{t_0}}( \rho u_x^2 + \rho u_z^2 +\rho u_z^2)- \partial_x \left[(1 +2c_s^2) \rho u_x+\rho u_x (u_y^2+u_z^2 )\right]-\partial_y [(1 +2c_s^2) \rho u_y\nonumber\\
& \;\;\;\quad\quad\quad\quad+\rho u_y (u_x^2+u_z^2 )] - \partial_z [(1 +2c_s^2) \rho u_z +\rho u_z (u_x^2+u_y^2) ]+2( F_x u_x +F_y u_y+ F_z u_z) \Big\} \nonumber\\
& \;\;\;\quad\quad\quad\quad- \frac{1}{\omega_9}\underline{ (r^2-1)\partial_y(\rho u_y)}- \frac{1}{\omega_9} \underline{(s^2 -1) \partial_z(\rho u_z)}. \label{eq:19f}
\end{align}
\end{subequations}
It should be noted here that diagonal parts of the non-equilibrium moments $n_7^{(1)}$, $n_8^{(1)}$ and $n_9^{(1)}$ contain contributions from non-GI cubic velocity terms due to aliasing effects associated with the D3Q27 lattice (e.g., $\sum_\alpha f_\alpha e_{\alpha i}^3=\sum_\alpha f_\alpha e_{\alpha i}$, where $i \in \{x,y,z\}$) which appear together with the physical terms within those enclosed within the brackets $\{\cdots \}$) of Eqs.~(\ref{eq:19d})-(\ref{eq:19f}) in addition to the grid-dependent anisotropic errors. The grid-anisotropy related error terms will be denoted by $E_{js}$, while the non-GI cubic velocity truncation errors will be referred to as $E_{jg}$ for $j=7,8$ and $9$. The structures of the non-equilibrium moments can then be obtained from Eqs.~(\ref{eq:19a})-(\ref{eq:19f}) after substituting for the time derivatives of the conserved moments by means of terms related to the spatial derivatives using Eqs.~(\ref{eq:17a})-(\ref{eq:17d}) and simplifying the resulting equations by retaining terms of $O(u_i^3)$ (following Hajabdollahi and Premnath~\cite{Hajabdollahi201897}). The off-diagonal second order non-equilibrium moments $n_4^{(1)}=k_{110}^{\prime (1)}$, $n_5^{(1)}=k_{101}^{\prime (1)}$ and $n_6^{(1)}=k_{011}^{\prime (1)}$, which do not contain any truncation errors, then read as
\begin{subequations}
\begin{eqnarray}
n_4^{(1)}&=& -\frac{ c_s^2 \rho }{\omega_4} \left(\partial_x u_y + \partial_y u_x \right),\label{eq:20a}\\
n_5^{(1)}&=& -\frac{ c_s^2 \rho }{\omega_5} \left(\partial_x u_z + \partial_z u_x \right), \label{eq:20b}\\
n_6^{(1)}&=& -\frac{ c_s^2 \rho }{\omega_6} \left(\partial_y u_z + \partial_z u_y \right).\label{eq:20c}
\end{eqnarray}
\end{subequations}
Finally, the diagonal second order non-equilibrium moments $n_7^{(1)}=k_{200}^{\prime (1)}-k_{020}^{\prime (1)}$, $n_8^{(1)}=k_{200}^{\prime (1)}-k_{002}^{\prime (1)}$ and $n_9^{(1)}=k_{200}^{\prime (1)}+k_{020}^{\prime (1)}+k_{002}^{\prime (1)}$ are given as follows:
\begin{subequations}
\begin{eqnarray}
n_7^{(1)}&=& -\frac{ 2 c_s^2\rho }{\omega_7} \left(\partial_x u_x - \partial_y u_y \right)+E_{7s}+E_{7g},\label{eq:21a}\\
n_8^{(1)}&=& -\dfrac{2 c_s^2\rho }{\omega_8} \left(\partial_x u_x - \partial_z u_z \right)+E_{8s}+E_{8g},\label{eq:21b}\\
n_9^{(1)}&=& - \frac{2 c_s^2\rho }{\omega_9} \left(\partial_x u_x + \partial_y u_y +\partial_z u_z \right)+E_{9s}+E_{9g},\label{eq:21c}
\end{eqnarray}
\end{subequations}
where the truncation errors due to grid-anisotropy $E_{7s}$, $E_{8s}$ and $E_{9s}$ and non-GI aliasing effects $E_{7g}$, $E_{8g}$ and $E_{9g}$ can be written as
\begin{subequations}
\begin{eqnarray}
E_{7s}&=& \frac{1}{\omega_7}\left[(3c_s^2-1)\partial_x (\rho u_x)-(3c_s^2-r^2)\partial_y (\rho u_y)\right],\label{eq:22a}\\
E_{7g}&=& \frac{3\rho}{\omega_7}\left[u_x^2\partial_x u_x - u_y^2\partial_y u_y\right],\label{eq:22b}\\
E_{8s}&=& \frac{1}{\omega_8}\left[(3c_s^2-1)\partial_x (\rho u_x)-(3c_s^2-s^2)\partial_z (\rho u_z)\right],\label{eq:22c}\\
E_{8g}&=& \frac{3\rho}{\omega_8}\left[u_x^2\partial_x u_x - u_z^2\partial_z u_z\right],\label{eq:22d}\\
E_{9s}&=& \frac{1}{\omega_9}\left[(3c_s^2-1)\partial_x (\rho u_x)+(3c_s^2-r^2)\partial_y (\rho u_y)+(3c_s^2-s^2)\partial_z (\rho u_z)\right],\label{eq:22e}\\
E_{9g}&=& \frac{3\rho}{\omega_9}\left[u_x^2\partial_x u_x + u_y^2\partial_y u_y + u_z^2\partial_z u_z\right].\label{eq:22f}
\end{eqnarray}
\end{subequations}

\subsection{Elimination of errors due to lattice stretching and non-GI cubic velocity terms via extended moment equilibria}\label{subsec:correctionsviaCEanalysis}
In order to eliminate the truncation errors identified above in Eqs.~(\ref{eq:22a})-(\ref{eq:22f}) associated with Eqs.~(\ref{eq:21a})-(\ref{eq:21c}), we now extend the moment equilibria $\mathbf{n}^{eq}$ to an effective moment equilibria $\mathbf{n}^\mathit{eq,eff}$ as
\begin{equation} \label{eq:29}
\mathbf{n}^\mathit{eq,eff}= \mathbf{n}^{eq}+ \Delta t \mathbf{n}^{eq(1)},
\end{equation}
where $\mathbf{n}^{eq(1)}$ are the required correction terms. Since the truncation errors appear in the diagonal parts of the second order non-equilibrium moments only, it suffices to consider non-zero correction terms only for those corresponding components identified by the indices 7, 8 and 9 in the effective equilibria. Hence, we write
\begin{eqnarray}\label{eq:30}
\mathbf{n}^{eq(1)} =
\begin{cases}
\theta_{7x} \partial_x u_x  -  \theta_{7y} \partial_y u_y + \lambda_{7x} \partial_x \rho + \lambda_{7y} \partial_y \rho,  & \quad j =7\\
\theta_{8x} \partial_x u_x  -  \theta_{8z} \partial_z u_z + \lambda_{8x} \partial_x \rho + \lambda_{8z} \partial_z \rho,  & \quad j =8\\
\theta_{9x} \partial_x u_x  + \theta_{9y} \partial_y u_y  +\theta_{9z} \partial_z u_z + \lambda_{9x} \partial_x \rho + \lambda_{9y} \partial_y \rho +\lambda_{9z} \partial_z \rho, & \quad j =9\\
0, & \quad \mbox{otherwise}\\
\end{cases}
\end{eqnarray}
The expressions in Eq.~(\ref{eq:30}) are inspired by an inspection of Eqs.~(\ref{eq:22a})-(\ref{eq:22f}) and involve the diagonal components of the velocity gradient tensor $\partial_x u_x$ $\partial_y u_y$ and $\partial_z u_z$ and the density gradients $\partial_x \rho$, $\partial_y \rho$ and $\partial_z \rho$ and their unknown coefficients $\theta_{7x}$, $\theta_{7y}$, $\theta_{8x}$, $\theta_{8z}$, $\theta_{9x}$, $\theta_{9y}$, and $\theta_{9z}$ and $\lambda_{7x}$, $\lambda_{7y}$, $\lambda_{8x}$, $\lambda_{8z}$, $\lambda_{9x}$, $\lambda_{9y}$, and $\lambda_{9z}$. The formulas for these coefficients will now be established by performing a modified C-E analysis that accounts for the effective moment equilibria introduced in Eq.~(\ref{eq:29}) into the expansion. That is,
\begin{eqnarray*}\label{eq:31}
\mathbf{n}&=&  \mathbf{n}^{\mathit{eq,eff}}+ \epsilon \mathbf{n}^{(1)}+ \epsilon^2 \mathbf{n}^{(2)}+ \cdots = \mathbf{n}^{eq}+\epsilon \underline{\mathbf{n}^{eq(1)}}+ \epsilon \mathbf{n}^{(1)}+ \epsilon^2 \mathbf{n}^{(2)}+ \cdots, \nonumber \\
\partial_t &=& {\partial_{t_0}}+ \epsilon {\partial_{t_1}} + \epsilon^2 {\partial_{t_2}} + \cdots.
\end{eqnarray*}
Then, repeating the steps performed in Sec.~\ref{subsec:CEanalysis}, the respective moment equations of $O(\epsilon^k)$ for $k=0,1$ and $2$, respectively, in Eqs.~\eqref{eq:14a}-\eqref{eq:14c} are replaced with the following:
\begin{subequations}
\begin{eqnarray}\label{eq:32}
 \centering
&O (\epsilon^0 ):  \mathbf{n}^{(0)} =  \mathbf{n}^\mathit{eq},  \label{eq:32a}\\
&O (\epsilon^1 ): \left({\partial_{t_0}} + \tensor{E}_i \partial_i\right)  \mathbf{n}^{(0)} =  - \tensor{\Lambda} \left[ \mathbf{n}^{(1)}- \underline{\mathbf{n}^{eq(1)}}\right]  +\mathbf{\Psi},  \label{eq:32b} \\
&O (\epsilon^2 ): {\partial_{t_1}} \mathbf{n}^{(0)} +  \left({\partial_{t_0}} + \tensor{E}_i \partial_i\right) \left[\left( \tensor{I} - \frac{\tensor{\Lambda}} {2} \right)\mathbf{n}^{(1)}\right] +  \left({\partial_{t_0}} + \tensor{E}_i \partial_i\right) \underline{\left[ \frac{\tensor{\Lambda}} {2} \mathbf{n}^{eq(1)}\right]}=  - \tensor{\Lambda}\mathbf{n}^{(2)}. \label{eq:32c}
\end{eqnarray}
\end{subequations}
Notice that the moment equilibria corrections appear in both $O(\epsilon)$ and $O(\epsilon^2)$ equations in Eqs.~(\ref{eq:32b}) and (\ref{eq:32c}), respectively, via the underlined terms. Following the steps in Sec.~\ref{subsec:CEanalysis}, we simplify Eq.~\eqref{eq:32b} for the leading set of moments up to the second order that are related to the derivation of the hydrodynamical equations. These lead to the following results for the effective non-equilibrium parts of the second order moments:
\begin{subequations}\label{eq:33}
 \begin{eqnarray}
&n_4^{(1)}= -\cfrac{c_s^2 \rho}{\omega_4} \left(\partial_x u_y + \partial_y u_x \right),   \label{eq:33a}\\
&n_5^{(1)}= -\cfrac{c_s^2 \rho}{\omega_5} \left(\partial_x u_z + \partial_z u_x \right),
\label{eq:33b}\\
&n_6^{(1)}= -\cfrac{c_s^2\rho}{\omega_6} \left(\partial_y u_z + \partial_z u_y \right),
\label{eq:33c}\\
&n_7^{(1)}= -\cfrac{2c_s^2\rho }{\omega_7} \left(\partial_x u_x - \partial_y u_y \right) +(E_{7s}+E_{7g})+ \underline{n_7^{eq(1)}},
\label{eq:33d}\\
&n_8^{(1)}= -\cfrac{2c_s^2\rho}{\omega_8} \left(\partial_x u_x - \partial_z u_z \right) +(E_{8s}+E_{8g})+\underline{n_8^{eq(1)}},
\label{eq:33e}\\
&n_9^{(1)}= -\cfrac{2c_s^2\rho}{\omega_9} \left(\partial_x u_x +\partial_y u_y+\partial_z u_z \right)+ (E_{9s}+E_{9g})+ \underline{n_9^{eq(1)}},
\label{eq:33f}
\end{eqnarray}
\end{subequations}
where $E_{jg}$ and $E_{js}$  $(j=7,8$ and $9)$ are truncation error terms associated with grid anisotropy and non-GI due to aliasing, which are given in Eqs.~\eqref{eq:22a}-\eqref{eq:22f} and $n_7^{eq(1)}$, $n_8^{eq(1)}$ and $n_9^{eq(1)}$ are the corresponding yet to be determined corrections.

It now remains to obtain the desired expressions for the correction terms. In this regard, we combine the moment equations (Eq.~(\ref{eq:32b})), in particular, those for the hydrodynamic moments (mass and momentum components) evolving at time scale $t_0$ with $\epsilon$ times the corresponding equations (Eq.~(\ref{eq:32c})) based on the time scale $t_1$ variations. Then, using $\partial_t={\partial_t}_0 + \epsilon {\partial_t}_1 $, we obtain the effective evolution equations for all the moments, including the macroscopic fields related to the mass and momentum components. In particular, the momentum equations will then contain the truncation error terms and the counteracting correction terms along with those associated with the physics. We can then isolate the terms related to the truncation errors and corrections from those related to the desired NS equations and set the combined effect of the former to be zero. This would then lead to constraint relations between the errors and the required corrections as discussed in detail in Hajabdollahi and Premnath~\cite{Hajabdollahi201897}. Thus, if we define the following vector containing the truncation errors as
\begin{equation}\label{eq:34}
\mathbf{\Xi}=\left(\varphi_{0},\varphi_{1},\varphi_{2},\dots,\varphi_{26}\right)^{\dag},
\end{equation}
where
\begin{eqnarray}\label{eq:35}
\varphi_j =
\begin{cases}
 E_{7s}+ E_{7g}& \quad j =7\\
 E_{8s}+ E_{8g}& \quad j =8\\
 E_{9s}+ E_{9g}& \quad j =9\\
 0 & \quad \mbox{otherwise},
\end{cases}
\end{eqnarray}
Then, it follows from Eqs.~\eqref{eq:32b},\eqref{eq:32c} and  Eqs.~\eqref{eq:33a}-\eqref{eq:33f}, the required constraint equation between the moment equilibria corrections vector $\mathbf{n}^{eq(1)}$ identified in Eq.~(\ref{eq:29}) and the vector of truncation errors $\mathbf{\Xi}$ is given by
\begin{equation}\label{eq:36}
\mathbf{n}^{eq(1)}+ \left( \tensor{I} - \frac{\tensor{\Lambda}} {2} \right)\mathbf{\Xi}=0,
\end{equation}
Specifically, it reduces to
\begin{equation}\label{eq:37}
n_j^{eq(1)}+ \left( 1 - \frac{\omega_j} {2} \right)(E_{js}+ E_{jg})=0, \quad j=7,8,9.
\end{equation}
In Eq.~(\eqref{eq:37}), we do not assume the summation convention of repeated indices.

Now, evaluating Eq.~\eqref{eq:37} by applying Eq.~(\ref{eq:30}) for $j=7$ and using Eqs.~(\ref{eq:22a}) and (\ref{eq:22b}), we get
\begin{equation*}
\theta_{7x} \partial_x u_x  -  \theta_{7y} \partial_y u_y + \lambda_{7x} \partial_x \rho + \lambda_{7y} \partial_y \rho  = -\left( 1 - \dfrac{\omega_7} {2} \right)E_{7s} -\left( 1 - \dfrac{\omega_7} {2} \right) E_{7g}.
\end{equation*}
Comparing, we finally obtain the following required expressions for the coefficients $\theta_{7x}$, $\theta_{7y}$, $\lambda_{7x}$ and $\lambda_{7y}$:
\begin{subequations}\label{eq:38}
\begin{eqnarray}
\theta_{7x}&=& -\left( \frac{1}{\omega_7}- \frac{1}{2}\right)\rho\left[(3 c_s^2 - 1) + 3 u_x^2 \right],\\
\theta_{7y}&=& -\left(\frac{1}{\omega_7}- \frac{1}{2}\right)\rho \left[(3 c_s^2- r^2) + 3 u_y^2 \right],\\
\lambda_{7x}&=& -\left(\frac{1}{\omega_7}- \frac{1}{2}\right)\left(3 c_s^2 - 1 \right)u_x,\\
\lambda_{7y}&=& +\left(\frac{1}{\omega_7}- \frac{1}{2}\right)\left(3 c_s^2 - r^2\right)u_y,
\end{eqnarray}
\end{subequations}
Repeating these steps by applying Eqs.~\eqref{eq:37} and (\ref{eq:30}) together for $j=8$ and $j=9$ and invoking Eqs.~(\ref{eq:22c}) and (\ref{eq:22d}), and Eqs.~(\ref{eq:22e}) and (\ref{eq:22f}), respectively, i.e.,
\begin{equation*}
\theta_{8x} \partial_x u_x  -  \theta_{8z} \partial_z u_z + \lambda_{8x} \partial_x \rho + \lambda_{8z} \partial_z \rho  = -\left( 1 - \dfrac{\omega_8}{2} \right)E_{8s} -\left( 1 - \dfrac{\omega_8}{2} \right) E_{8g},
\end{equation*}
and
\begin{eqnarray*}
&\theta_{9x} \partial_x u_x  +  \theta_{9y} \partial_y u_y +  \theta_{9z} \partial_z u_z + \lambda_{9x} \partial_x \rho + \lambda_{9y} \partial_y \rho+ \lambda_{9z} \partial_z \rho  = -\left( 1 - \dfrac{\omega_9}{2} \right)E_{9s} -\left( 1 - \dfrac{\omega_9}{2} \right) E_{9g},
\end{eqnarray*}
we arrive at the required formulas for the coefficients $\theta_{8x}$, $\theta_{8z}$, $\lambda_{8x}$, $\lambda_{8z}$, and $\theta_{9x}$, $\theta_{9y}$, $\theta_{9z}$, $\lambda_{9x}$, $\lambda_{9y}$, and $\lambda_{9z }$ as follows:
\begin{subequations}\label{eq:39}
\begin{eqnarray}
\theta_{8x}&=& -\left(\dfrac{1}{\omega_8}- \dfrac{1}{2}\right)\rho\left[(3 c_s^2 - 1) + 3 u_x^2\right],\\
\theta_{8z}&=& -\left(\dfrac{1}{\omega_8}- \dfrac{1}{2}\right)\rho \left[(3 c_s^2 - s^2) + 3 u_z^2 \right] ,\\
\lambda_{8x}&=& -\left(\dfrac{1}{\omega_8}- \dfrac{1}{2}\right)\left(3 c_s^2 - 1 \right)u_x,\\
\lambda_{8z}&=& +\left(\dfrac{1}{\omega_8}- \dfrac{1}{2}\right)\left(3 c_s^2 - s^2\right)u_z,
\end{eqnarray}
\end{subequations}
and
\begin{subequations}\label{eq:40}
\begin{eqnarray}
\theta_{9x}&=& -\left( \frac{1}{\omega_9}- \frac{1}{2}\right)\rho\left[(3 c_s^2 - 1) + 3 u_x^2 \right],\\
\theta_{9y}&=& -\left(\frac{1}{\omega_9}- \frac{1}{2}\right)\rho\left[(3 c_s^2 - r^2) + 3 u_y^2 \right],\\
\theta_{9z}&=& -\left(\frac{1}{\omega_9}- \frac{1}{2}\right)\rho\left[(3 c_s^2 - s^2) + 3 u_z^2 \right],\\
\lambda_{9x}&=& - \left(\frac{1}{\omega_9}- \frac{1}{2}\right)\left(3 c_s^2 - 1 \right)u_x,\\
\lambda_{9y}&=& - \left(\frac{1}{\omega_9}- \frac{1}{2}\right)\left(3 c_s^2 - r^2\right)u_y,\\
\lambda_{9z}&=& - \left(\frac{1}{\omega_9}- \frac{1}{2}\right)\left(3 c_s^2 - s^2\right)u_z,
\end{eqnarray}
\end{subequations}
When the extended moment equilibria (Eq.~(\ref{eq:29})) using the corrections (Eq.~\ref{eq:30})) with these above coefficients are used in the LBE formulation based on the D3Q27 cuboid lattice, it recovers the 3D NS equations given by
\begin{eqnarray}\label{eq:51}
\partial_t \rho + {\bm{\nabla}}. (\rho \bm{u})   &=& 0, \\
\partial_t (\rho \bm{u})+ {\bm{\nabla}} \cdot (\rho \bm{u}\bm{u}) &=& -{\bm{\nabla}} p+{\bm{\nabla}} \cdot \left\{\rho\left[\nu\left(\bm{\nabla}\bm{u}+(\bm{\nabla}\bm{u})^\dag\right)+\left(\xi-\frac{2}{3}\nu\right)\bm\nabla\cdot\bm{u}\tensor{I}\right]\right\}+\bm{F},
\end{eqnarray}
where $p = c_s^2\rho$ represents the pressure field, and the bulk viscosity $\xi$ and shear viscosity $\nu$ are related to the relaxation parameters of the second order moments as
\begin{equation}\label{eq:53}
\xi= \frac{2c_s^2}{3} \left( \frac{1}{\omega_9}- \frac{1}{2} \right)\Delta t, \quad \quad \;\;\; \nu =c_s^2 \left( \frac{1}{\omega_j}- \frac{1}{2} \right)\Delta t,\;\;\; j=4,5,\dots, 8. \quad
\end{equation}

Before concluding this section, the following important remarks related to our above derivation are in order:
\begin{itemize}
\item The expressions for the moment equilibria corrections and the transport coefficients involve just the minimally required set of free parameters, viz., the grid aspect ratios $r$ and $s$ and the speed of sound $c_s$, and are considerably simpler than those used in a recent work~\cite{wang2019lattice} that used a set of orthogonalized moment basis on the D3Q19 cuboid lattice and many additional parameters in the specification of the equilibria. Moreover, unlike the previous study~\cite{wang2019lattice}, the equilibria used in our approach given in Eq.~(\ref{eq:10}) involve higher order velocity terms obtained from matching with the moments of the continuous Maxwell distribution thereby naturally eliminating the non-GI truncation errors. This aspect along with the use of non-orthogonal moment basis in the present derivation that avoids a spurious coupling of the evolution of the higher order moments with those of the lower moments existing in approaches that use orthogonal moment basis~\cite{dubois2015stability} result in a robust LBE formulation on a cuboid lattice.
\item The results derived here for D3Q27 cuboid lattice for the counteracting corrections and the transport coefficients are also equally valid for the D3Q15 and D3Q19 cuboid lattice sets when they use subsets of the same non-orthogonal moment basis.
\item The formulas derived in this section for the equilibria corrections necessary to eliminate the grid anisotropy and non-GI error are general and applicable for a wide variety of collision models. For example, if $k_{mnp}^\prime$, $k_{mnp}$ and $c_{mnp}$ correspond to raw moments, central moments and cumulants, respectively, of order $(m+n+p)$, owing to the definitions of these quantities, it follows that their second order non-equilibrium components are identical, i.e., $k_{mnp}^{(1)\prime}=\;k_{mnp}^{(1)}= c_{mnp}^{(1)}$ for $(m+n+p)=2$. From this it follows that the formulas obtained above can be directly incorporated into either raw moment-, central moment- or cumulant-based 3D LB algorithms extended to using a cuboid lattice. Moreover, we point out that our derivation can be readily used to construct even an SRT-based LB scheme using a cuboid lattice by setting all the relaxation parameters equal to one another and then constructing the equilibrium distribution functions including the necessary corrections in the velocity space via using $\mathbf{f}^\mathit{eq,eff}=\tensor{T}^{-1}\mathbf{n}^\mathit{eq,eff}= \tensor{T}^{-1}(\mathbf{n}^{eq} + \Delta t \mathbf{n}^{eq(1)})$. However, as demonstrated in our recent work on the rectangular lattice~\cite{yahia2021central}, the SRT- and raw moment-based LB schemes are less stable compared to those based on central moments; hence the present investigation will focus on further developing the derivation presented here into an efficient and robust 3D LB algorithm on the D3Q27 cuboid lattice and their numerical study in the following sections. Such an implementation for a cumulant collision kernel on a cuboid lattice is a subject for a future study.
\item When $r=s=1$ and $c_s^2=1/3$, all the correction terms shown above become equal to zero, and the previous formulations applicable for the cubic lattice sets are recovered.
\item Finally, when the interest in simulating relatively very low Mach number flows, the non-GI cubic velocity errors can become relatively small and can thus be neglected. Under such conditions, the formulas derived in the above for the corrections lead to further simplifications. In particular, the coefficients determined above reduce to the following:
    \begin{eqnarray*}
        \theta_{7x}&=& -\left( \frac{1}{\omega_7}- \frac{1}{2}\right)\rho(3 c_s^2 - 1),\\
        \theta_{7y}&=& -\left(\frac{1}{\omega_7}- \frac{1}{2}\right)\rho (3 c_s^2- r^2),\\
        \lambda_{7x}&=& \lambda_{7y} \; \; = \; \; 0,
    \end{eqnarray*}
    \begin{eqnarray*}
        \theta_{8x}&=& -\left( \frac{1}{\omega_8}- \frac{1}{2}\right)\rho(3 c_s^2 - 1),\\
        \theta_{8z}&=& -\left(\frac{1}{\omega_8}- \frac{1}{2}\right)\rho (3 c_s^2- s^2),\\
        \lambda_{8x}&=& \lambda_{8z} \; \; = \; \; 0,
    \end{eqnarray*}
    \begin{eqnarray*}
        \theta_{9x}&=& -\left( \frac{1}{\omega_9}- \frac{1}{2}\right)\rho(3 c_s^2 - 1),\\
        \theta_{9y}&=& -\left(\frac{1}{\omega_9}- \frac{1}{2}\right)\rho (3 c_s^2- r^2),\\
        \theta_{9z}&=& -\left(\frac{1}{\omega_9}- \frac{1}{2}\right)\rho (3 c_s^2- s^2),\\
        \lambda_{9x}&=& \lambda_{9y} \; \; = \; \; \lambda_{9z} \; \; = \; \; 0,
    \end{eqnarray*}
    and, thus, we do not require the computation of the density gradients; moreover, the local expressions for the velocity gradients derived in the next section will become more compact and can be readily re-derived when required.
\end{itemize}

\subsection{Strain rate tensor components based on non-equilibrium moments for the D3Q27 cuboid lattice}
The moment equilibria corrections in Eq.~(\ref{eq:30}) depend on the diagonal parts of the strain rate tensor $\partial_x u_x$, $\partial_y u_y$ and $\partial_z u_z$. These components, along with the off-diagonal components, i.e., $\frac{1}{2}(\partial_x u_y+\partial_y u_x)$, $\frac{1}{2}(\partial_x u_z+\partial_z u_x)$ and $\frac{1}{2}(\partial_y u_z+\partial_z u_y)$ can be obtained locally from the second-order non-equilibrium moments as follows. From Eq.~\eqref{eq:33d} and Eq.~\eqref{eq:37} for $j=7$ and using Eqs.~(\ref{eq:22a}) and (\ref{eq:22b}), and after rearranging, we get
\begin{eqnarray}\label{eq:41}
&\left[-\dfrac{2 c_s^2}{\omega_7} + \dfrac{1}{2}(3 c_s^2-1)+\dfrac{3u_x^2}{2} \right] \rho \partial_x u_x + \left[\dfrac{2 c_s^2}{\omega_7} - \dfrac{1}{2}(3 c_s^2-r^2)- \dfrac{3u_y^2}{2}\right] \rho \partial_y u_y \nonumber \\
&=n_7^{(1)}-\dfrac{1}{2}(3 c_s^2-1) u_x \partial_x \rho + \dfrac{1}{2}(3 c_s^2-r^2 ) u_y \partial_y\rho,
\end{eqnarray}
Also, combining Eq.~\eqref{eq:33e} and Eq.~\eqref{eq:37} for $j=8$ and applying  Eqs.~(\ref{eq:22c}) and (\ref{eq:22d}), and after rearranging it follows that
\begin{eqnarray}\label{eq:42}
&\left[-\dfrac{2 c_s^2}{\omega_8} + \dfrac{1}{2}(3 c_s^2-1)+ \dfrac{3}{2}u_x^2 \right] \rho \partial_x u_x + \left[\dfrac{2 c_s^2}{\omega_8} -\dfrac{1}{2}(3 c_s^2-s^2)- \dfrac{3}{2}u_z^2\right] \rho \partial_z u_z \nonumber \\
&=n_8^{(1)}-\dfrac{1}{2} (3 c_s^2-1)u_x \partial_x \rho + \dfrac{1}{2}(3 c_s^2-1) u_z \partial_z \rho,
\end{eqnarray}
Finally, using Eq.~\eqref{eq:33f} and Eq.~\eqref{eq:37} together for $j=9$ along with Eqs.~(\ref{eq:22e}) and (\ref{eq:22f}), we obtain the following:
\begin{eqnarray}\label{eq:43}
&\left[-\dfrac{2 c_s^2}{\omega_9} + \dfrac{1}{2}(3 c_s^2-1)+ \dfrac{3}{2}u_x^2 \right] \rho \partial_x u_x + \left[-\dfrac{2 c_s^2}{\omega_9}+\dfrac{1}{2}(3 c_s^2-r^2)+\dfrac{3}{2}u_y^2 \right] \rho \partial_y u_y\nonumber\\
&+\left[-\dfrac{2 c_s^2}{\omega_9} + \dfrac{1}{2}(3 c_s^2-s^2) + \dfrac{3}{2}u_z^2\right] \rho \partial_z u_z   = n_9^{(1)}-\dfrac{1}{2}(3 c_s^2-1)u_x \partial_x \rho\nonumber\\
&-\dfrac{1}{2}(3 c_s^2-r^2) u_y \partial_y \rho- -\dfrac{1}{2}(3 c_s^2-s^2) u_z \partial_z \rho.
\end{eqnarray}
We can use the above three equations to solve for the diagonal parts of the strain rate tensor as follows. First, we define the following variables:
\begin{subequations}\label{eq:44}
\begin{eqnarray}
 A&=& \dfrac{1}{2}\left( 3 c_s^2 - 1 \right) u_x,\quad B =  \dfrac{1}{2}\left(3 c_s^2 - r^2 \right) u_y, \quad C = \dfrac{1}{2}\left(3 c_s^2 - s^2 \right) u_z ,\\
 e_{7\rho}&=& -A \partial_x \rho + B \partial_y \rho,\\
 e_{8\rho}&=& -A \partial_x \rho + C \partial_z \rho,\\
 e_{9\rho}&=& -A \partial_x \rho - B \partial_y \rho - C \partial_z \rho.
\end{eqnarray}
\end{subequations}
Here, the density gradients $\partial_x \rho$, $\partial_y \rho$ and $\partial_z \rho$ are based on a (isotropic) second order finite-difference scheme (see the work by Kumar~\cite{kumar2004isotropic} and Leclaire~\cite{leclaire2015forward} for details). The Eqs.~(\ref{eq:41})-(\ref{eq:43}) require the non-equilibrium moments $n_7^{(1)}$, $n_8^{(1)}$ and $n_9^{(1)}$, respectively, which can be obtained from either using raw moments or central moments as follows:
\begin{eqnarray*}
&n_7^{(1)}= \left(k_{200}^\prime - k_{020}^\prime \right)- \left(k_{200}^{\eqp} - k_{020}^{\eqp} \right)= \left(k_{200} - k_{020} \right)- \left(k_{200}^{eq} - k_{020}^{eq} \right),\\
&n_8^{(1)}= \left(k_{200}^\prime - k_{002}^\prime \right)- \left(k_{200}^{\eqp} - k_{002}^{\eqp} \right)= \left(k_{200} - k_{002} \right)- \left(k_{200}^{eq} - k_{002}^{eq} \right),\\
&n_9^{(1)}= \left(k_{200}^\prime + k_{020}^\prime + k_{002}^\prime\right)- \left(k_{200}^{\eqp} + k_{020}^{\eqp} + k_{002}^{\eqp} \right)= \left(k_{200} + k_{020}+ k_{002} \right)- \left(k_{200}^{eq} + k_{020}^{eq} + k_{002}^{eq}\right),
\end{eqnarray*}
where the equilibrium central moments $k_{200}^{eq}$, $k_{020}^{eq}$ and $k_{002}^{eq}$ are given in Eq.~(\ref{eq:9}), while the corresponding raw moments are listed in Eq.~(\ref{eq:10}). For example,
\begin{subequations}\label{eq:45}
\begin{eqnarray}
n_7^{(1)}&=& \left(k_{200} - k_{020} \right),\\
n_8^{(1)}&=& \left(k_{200} - k_{002} \right),\\
n_9^{(1)}&=& \left(k_{200} + k_{02 0}+ k_{002} \right) -3 \rho c_s^2.
\end{eqnarray}
\end{subequations}
Based on these considerations, we can rewrite the right and left sides Eqs.~\eqref{eq:41}-\eqref{eq:43}, respectively, by means of the following variables:
\begin{subequations}\label{eq:46}
\begin{eqnarray}
R_7 &=& n_7^{(1)}+ e_{7\rho} = \left(k_{200} - k_{020} \right)+ e_{7\rho},\\
R_8 &=& n_8^{(1)}+ e_{8\rho} = \left(k_{200} - k_{002} \right)+ e_{8\rho},\\
R_9 &=& n_9^{(1)}+ e_{9\rho} = \left[\left(k_{200} + k_{020}+ k_{002} \right)-3 \rho c_s^2\right]+ e_{9\rho},
\end{eqnarray}
\end{subequations}
and,
\begin{align}\label{eq:47}
&C_{7x}=\left[-\dfrac{2 c_s^2}{\omega_7} +\dfrac{1}{2}\left(3 c_s^2 - 1 \right)+ \dfrac{3}{2} u_x^2 \right] \rho, &C_{7y}= \left[+\dfrac{2 c_s^2}{\omega_7} -\dfrac{1}{2}\left(3 c_s^2 - r^2 \right)- \dfrac{3}{2}u_y^2\right] \rho,\nonumber\\
&C_{8x}=\left[-\dfrac{2 c_s^2}{\omega_8} +\dfrac{1}{2}\left(3 c_s^2 - 1 \right)+  \dfrac{3}{2} u_x^2 \right] \rho, &C_{8z}= \left[+\dfrac{2 c_s^2}{\omega_8} -\dfrac{1}{2}\left(3 c_s^2 - s^2 \right)- \dfrac{3}{2}u_z^2 \right] \rho,\nonumber\\
&C_{9x}= \left[-\dfrac{2 c_s^2}{\omega_9}+\dfrac{1}{2}\left(3 c_s^2 - 1 \right)+ \dfrac{3}{2} u_x^2 \right] \rho, & C_{9y}= \left[-\dfrac{2 c_s^2}{\omega_9}+\dfrac{1}{2}\left(3 c_s^2 - r^2 \right)+ \dfrac{3}{2}u_y^2\right] \rho,\nonumber\\
&  &C_{9z}= \left[-\dfrac{2 c_s^2}{\omega_9}+\dfrac{1}{2}\left(3 c_s^2 - s^2 \right)+ \dfrac{3}{2}u_z^2\right] \rho.
\end{align}
Then, Eqs.~\eqref{eq:41}-\eqref{eq:43} reduce to the following system of equations
\begin{subequations}\label{eq:48}
\begin{eqnarray}
C_{7x}\partial_x u_x  + C_{7y}\partial_y u_y &=& R_7,\\
C_{8x}\partial_x u_x  + C_{8z}\partial_z u_z &=& R_8,\\
C_{9x}\partial_x u_x  + C_{9y}\partial_y u_y+ C_{9z}\partial_z u_z &=& R_9,
\end{eqnarray}
\end{subequations}
which can be readily solved as follows, thus providing the local expressions for the diagonal components of the strain rate tensor on a cuboid lattice:
\begin{subequations}\label{eq:49}
\begin{eqnarray}
&\partial_x u_x = {\cfrac{\left[- C_{8z}C_{9y}R_7 - C_{7y} \left( C_{9z} R_8  - C_{8z} R_9 \right)\right]}{\left[- C_{7x}C_{8z}C_{9y} - C_{7y} \left( C_{8x} C_{9z}  - C_{8z}C_{9x} \right)\right]}},\\
&\partial_y u_y = \left[ R_7 - C_{7x} \partial_x u_x \right]/C_{7y},\quad \quad \;\;\;\ \partial_z u_z = \left[ R_8 - C_{8x} \partial_x u_x\right]/C_{8z}.
\end{eqnarray}
\end{subequations}
For completeness, we note that the off-diagonal components of the strain rate tensor can be obtained from Eqs.~\eqref{eq:33a}-\eqref{eq:33c} as
\begin{eqnarray}\label{eq:50}
\frac{1}{2}(\partial_x u_y+ \partial_y u_x) &=& - \dfrac{\omega_4}{2\rho c_s^2} n^{(1)}_4, \quad  n^{(1)}_4=k_{110}^\prime- \rho u_x u_y = k_{110},\\
\frac{1}{2}(\partial_x u_z+ \partial_z u_x) &=& - \dfrac{\omega_5}{2\rho c_s^2} n^{(1)}_5, \quad  n^{(1)}_5=k_{101}^\prime- \rho u_x u_z = k_{101}\\
\frac{1}{2}(\partial_y u_z+ \partial_z u_y) &=& - \dfrac{\omega_6}{2\rho c_s^2} n^{(1)}_6, \quad  n^{(1)}_6=k_{011}^\prime- \rho u_y u_z = k_{011}.
\end{eqnarray}

Before concluding this section, we provide a guideline in choosing the speed of sound $c_s$ for the cuboid lattice-based LB formulations. The optimal value for the cubic lattice $c_s^2=1/3$. However, in general cases for the cuboid lattice, the particle speeds in the $y$ and $z$ directions are, respectively, $r$ and $s$ times the particle speed in the $x$ direction. With different particle speeds in different coordinate directions, based on physical considerations, the speed of sound is then chosen as $c_s^2=(1/3)\mbox{min}(r^2,s^2)$. This choice maintains the physically consistent isotropy requirements, reduces the number of spurious terms to be eliminated via the corrections derived earlier, and recovers the optimal value for the cubic lattice.

\section{3D cuboid central moment LBM (3DCCM-LBM) using a non-orthogonal moment basis on the D3Q27 lattice}\label{sec:3}\par
Before constructing a 3D central moment LBM on the cuboid lattice grid using the derivation presented in the previous section, we first note a challenge in directly using the moment basis $\tensor{T}$ given in Eqs.~(\ref{eq:4}) and (\ref{eq:5}) and the resulting definition of the moments in Eq.~(\ref{eq:7}), and then present its simple resolution that is more suitable for devising an efficient algorithm. In particular, the moment basis $\tensor{T}$ (Eqs.~(\ref{eq:4}) and (\ref{eq:5})) involves linear combinations of moments of their second order diagonal components that are written to separate the trace of the diagonal elements from the others in order to allow an independent specification of the bulk and shear viscosities and are also accordingly parameterized by the grid aspect ratios $r$ and $s$. This was necessary in showing consistency of our approach to the 3D NS equations and in the derivation of the required corrections for using the cuboid lattice. Now, since the collision step to be performed in terms of the relaxations of either the raw moments or the central moments and whose effect need to be transformed back into the velocity space in terms of the distribution functions via $\tensor{T}^{-1}$ (see e.g., Eq.~(\ref{eq:8}) for the raw moment representation of the LBE), a direct inversion of $\tensor{T}$ would lead to cumbersome expressions involving the grid aspect ratios $r$ and $s$. However, we will now show that by using a re-defined moment basis, the segregation of the evolution of the moments for independent adjustments of bulk and shear viscosities can be still be achieved by confining it only within the collision step for the relaxation process and not for the mappings, which would have the same overall effect as the original moment basis. Such equivalent but considerably simpler re-formulations of the moment basis and the associated LBE given in Eq.~(\ref{eq:8}) will then form the foundation for the constructions of the 3D LB scheme on the cuboid lattice based on raw moments as well as its generalization in terms of central moments.

\subsection{Efficient formulation of LBE on the cuboid lattice and its interpretation based on various matrices}
First, we define the following moment basis $\tensor{Q}$ involving just the linearly independent bare moments for the D3Q27 \emph{cuboid} lattice (i.e., without any linear combinations of them as in $\tensor{T}$ -- see Eqs.~(\ref{eq:4}) and (\ref{eq:5})):
\begin{align}\label{eq:57}
&\tensor{Q}=\big[\ket{1}, \ket{e_x},\ket{e_y},\ket{e_z},\ket{e_x e_y},\ket{e_x e_z},\ket{e_y e_z},
\ket{e_x^2 },\ket{e_y^2},\ket{e_z^2},\ket{e_x e_y^2},
\ket{e_x e_z^2 },\ket{e_x^2 e_y},\nonumber\\
&\ket{e_y e_z^2}, \ket{e_x^2 e_z},\ket{e_y^2 e_z}, \ket{e_x e_y e_z},\ket{e_x^2 e_y^2},
\ket{e_x^2 e_z^2},\ket{e_y^2 e_z^2}, \ket{e_x^2e_y e_z},\ket{e_x e_y^2 e_z},\ket{e_x e_y e_z^2},\nonumber\\
&\ket{e_x e_y^2 e_z^2},\ket{e_x^2 e_y e_z^2},
\ket{e_x^2 e_y^2e_z},\ket{e_x^2 e_y^2 e_z^2}\big].
\end{align}
where the particle velocities $\ket{e_x}$, $\ket{e_x}$ and $\ket{e_z}$ appearing in Eq.~(\ref{eq:57}) are given in Eqs.~\eqref{eq:1a}-\eqref{eq:1c} and thus $\tensor{Q}$ depends on the grid aspect ratios $r$ and $s$. We can then express a correspondence between this re-defined moment basis $\tensor{Q}$ and the original one, i.e, $\tensor{T}$ given Eqs.~(\ref{eq:4}) and (\ref{eq:5}) using
\begin{equation}\label{eq:relationbetweenTandQ}
\tensor{T}= \tensor{B} \tensor{Q},
\end{equation}
where $\tensor{B}$ represents those operations that combine the various moments according to Eq.~(\ref{eq:5}) and has the form of a block diagonal matrix. While an explicit specification of $\tensor{B}$ is not necessary for the following discussion, Eq.~(\ref{eq:relationbetweenTandQ}) would be helpful in recasting the LBE in an equivalent form that leads to an efficient implementation. Moreover, let us see how this moment basis $\tensor{Q}$ for the cuboid lattice can be related to that for the \emph{cubic} lattice that involves the following set of particle velocities so as to construct a modular LB scheme:
\begin{subequations}
\begin{eqnarray}
\ket{\bar{e}_{x}} &=& (0,1,-1,0,0,0,0,1,-1,1,-1,1,-1,1,-1,0,0,0,0,1,-1,1,-1,1,-1,1,-1)^\dag, \label{eq:cubiclatticevelocitya}\\
\ket{\bar{e}_{y}} &=& (0,0,0,1,-1,0,0,1,1,-1,-1,0,0,0,0,1,-1,1,-1,1,1,-1,-1,1,1,-1,-1)^\dag,\label{eq:cubiclatticevelocityb}\\
\ket{\bar{e}_{z}} &=& (0,0,0,0,0,1,-1,0,0,0,0,1,1,-1,-1,1,1,-1,-1,1,1,1,1,-1,-1,-1,-1)^\dag.\label{eq:cubiclatticevelocityc}
\end{eqnarray}
\end{subequations}
Such a moment basis for the cubic lattice, referred to as $\tensor{P}$ in the following obviously does not depend on the grid aspect ratios, and can be written as
\begin{align}\label{eq:momentbasiscubiclattice}
&\tensor{P}=\big[\ket{1}, \ket{\bar{e}_x},\ket{\bar{e}_y},\ket{\bar{e}_z},\ket{\bar{e}_x \bar{e}_y},\ket{\bar{e}_x \bar{e}_z},\ket{\bar{e}_y \bar{e}_z},
\ket{\bar{e}_x^2 },\ket{\bar{e}_y^2},\ket{\bar{e}_z^2},\ket{\bar{e}_x \bar{e}_y^2},
\ket{\bar{e}_x \bar{e}_z^2 },\ket{\bar{e}_x^2 \bar{e}_y},\nonumber\\
&\ket{\bar{e}_y \bar{e}_z^2}, \ket{\bar{e}_x^2 \bar{e}_z},\ket{\bar{e}_y^2 \bar{e}_z}, \ket{\bar{e}_x \bar{e}_y \bar{e}_z},\ket{\bar{e}_x^2 \bar{e}_y^2},
\ket{\bar{e}_x^2 \bar{e}_z^2},\ket{\bar{e}_y^2 \bar{e}_z^2}, \ket{\bar{e}_x^2\bar{e}_y \bar{e}_z},\ket{\bar{e}_x \bar{e}_y^2 \bar{e}_z},\ket{\bar{e}_x \bar{e}_y \bar{e}_z^2},\nonumber\\
&\ket{\bar{e}_x \bar{e}_y^2 \bar{e}_z^2},\ket{\bar{e}_x^2 \bar{e}_y \bar{e}_z^2},
\ket{\bar{e}_x^2 \bar{e}_y^2\bar{e}_z},\ket{\bar{e}_x^2 \bar{e}_y^2 \bar{e}_z^2}\big].
\end{align}
It follows from the above definitions that $\tensor{Q}$ and $\tensor{P}$ matrices are related via
\begin{equation}\label{eq:58}
\tensor{Q}= \tensor{S} \tensor{P},
\end{equation}
where $\tensor{S}$ is a simple diagonal scaling matrix, and for the D3Q27 lattice it can be expressed as
\begin{equation}\label{eq:59}
\tensor{S} = \mbox{diag}\big(\;1 ,  1  ,  r  ,   s  , r  ,   s  , r s ,   1  , r^2  , s^2  ,   r^2  , s^2  ,   r  , r s^2  , s  , r^2 s , r s   ,   r^2 ,   s^2  ,  r^2 s^2  ,  rs , r^2 s  , r s^2  ,  r^2s^2  , r s^2  ,   r^2 s , r^2 s^2   \;  \big).
\end{equation}
Moreover, the inverse mapping for the cuboid lattice $\tensor{Q}^{-1}$ can be related to that for the cubic lattice $\tensor{P}^{-1}$ using
\begin{equation}\label{eq:60}
\tensor{Q}^{-1}= \tensor{S}^{-1} \tensor{P}^{-1},
\end{equation}
where $\tensor{S}^{-1}$ is another diagonal matrix whose elements are the reciprocals of the corresponding elements of $\tensor{S}$. Thus,
\begin{eqnarray}\label{eq:61}
&\tensor{S}^{-1} = \mbox{diag}\big(\;1 ,  1  ,  r^{-1}  ,   s^{-1}  , r^{-1}  ,   s^{-1}  , r^{-1} s^{-1} ,   1  , r^{-2}  , s^{-2}  ,   r^{-2} , s^{-2}  ,   r^{-1}  , r^{-1} s^{-2}  , s^{-1}  , r^{-2} s^{-1} , r^{-1} s^{-1}   ,   r^{-2} , \nonumber\\
&  s^{-2}  ,  r^{-2} s^{-2}  ,  r^{-1}s^{-1} , r^{-2} s^{-1}  , r^{-1} s^{-2}  ,  r^{-2}s^{-2}  , r^{-1} s^{-2} ,   r^{-2} s^{-1} , r^{-2} s^{-2}   \; \big).
\end{eqnarray}
As we will see below, these considerations would facilitate the construction of the 3D cuboid LB schemes involving both forward and inverse transformations around collision with effort similar to that for the common cubic lattice that is augmented with some simple scalings of moments based on $r$ and $s$ rather than using lengthy formulas. Next, using the simpler moment basis $\tensor{Q}$ we can relate the vectors of the bare moments (i.e., without involving any linear combinations) and the distribution functions via
\begin{eqnarray}\label{eq:56}
&\mathbf{m}= \tensor{Q}\mathbf{f}\label{eq:56a}, \quad \mathbf{f}={\tensor{Q}}^{-1}\mathbf{m},\label{eq:56b}
\end{eqnarray}
based on which we now list the countable raw moments for the D3Q27 lattice as follows:
\begin{eqnarray}
&\mathbf{m}=\big(k_{000}^{\prime},k_{100}^{\prime},k_{010}^{\prime},k_{001}^{\prime},k_{110}^{\prime},k_{101}^{\prime},
k_{011}^{\prime},k_{200}^{\prime},k_{020}^{\prime},k_{002}^{\prime},k_{120}^{\prime},k_{102}^{\prime},k_{210}^{\prime}, k_{012}^{\prime}, k_{201}^{\prime},k_{021}^{\prime},k_{111}^{\prime},\nonumber \\ &k_{220}^{\prime},k_{202}^{\prime},k_{022}^{\prime},k_{211}^{\prime},k_{121}^{\prime},k_{112}^{\prime},k_{122}^{\prime},k_{212}^{\prime},k_{221}^{\prime},k_{222}^{\prime}\big)^{\dag}.\label{eq:rawmomentlist}
\end{eqnarray}
and similarly using $\mathbf{m}^{eq}= \tensor{Q}\mathbf{f}^{eq}$ and $\mathbf{\Phi}= \tensor{Q}\mathbf{S}$, we can write the respective equilibria and the source terms in the moment space as
\begin{eqnarray}
&\mathbf{m}^{eq}=\big(k_{000}^{eq\prime},k_{100}^{eq\prime},k_{010}^{eq\prime},k_{001}^{eq\prime},k_{110}^{eq\prime},k_{101}^{eq\prime},
k_{011}^{eq\prime},k_{200}^{eq\prime},k_{020}^{eq\prime},k_{002}^{eq\prime},k_{120}^{eq\prime},k_{102}^{eq\prime},k_{210}^{eq\prime}, k_{012}^{eq\prime}, k_{201}^{eq\prime},k_{021}^{eq\prime},k_{111}^{eq\prime},\nonumber \\ &k_{220}^{eq\prime},k_{202}^{eq\prime},k_{022}^{eq\prime},k_{211}^{eq\prime},k_{121}^{eq\prime},k_{112}^{eq\prime},k_{122}^{eq\prime},k_{212}^{eq\prime},k_{221}^{eq\prime},k_{222}^{eq\prime}\big)^{\dag},\label{eq:rawmomentequilibrialist}
\end{eqnarray}
\begin{eqnarray}
&\mathbf{\Phi}=\big(\sigma_{000}^{\prime},\sigma_{100}^{\prime},\sigma_{010}^{\prime},\sigma_{001}^{\prime},\sigma_{110}^{\prime},\sigma_{101}^{\prime},
\sigma_{011}^{\prime},\sigma_{200}^{\prime},\sigma_{020}^{\prime},\sigma_{002}^{\prime},\sigma_{120}^{\prime},\sigma_{102}^{\prime},\sigma_{210}^{\prime}, \sigma_{012}^{\prime}, \sigma_{201}^{\prime},\sigma_{021}^{\prime},\sigma_{111}^{\prime},\nonumber \\ &\sigma_{220}^{\prime},\sigma_{202}^{\prime},\sigma_{022}^{\prime},\sigma_{211}^{\prime},\sigma_{121}^{\prime},\sigma_{112}^{\prime},\sigma_{122}^{\prime},\sigma_{212}^{\prime},\sigma_{221}^{\prime},\sigma_{222}^{\prime}\big)^{\dag}.\label{eq:rawmomentsourcelist}
\end{eqnarray}
Here, the components of the raw moments, their equilibria and the source moments $k_{mnp}^\prime$ and $k_{mnp}^{eq\prime}$ and $\sigma_{mnp}^\prime$, respectively, are defined in Eq.~(\ref{eq:3b}).

With these preliminaries, we now rewrite the LBE given in Eq.~(\ref{eq:8}) and using $\mathbf{f}=\tensor{T}^{-1}\mathbf{n}$ in terms of the following collide-and-stream steps with the goal of constructing an efficient LB algorithm on the cuboid lattice:
\begin{eqnarray}
\tilde{\mathbf{f}} (\bm{x},t) &=&  \tensor{T^{-1}}
 \Big[\mathbf{n} + \tensor{\Lambda} \;  \left(\; \mathbf{n}^{eq}-\mathbf{n} \;\right) + \left(\tensor{I} - \frac{\tensor{\Lambda}}{2}\right)  \mathbf{\Psi}\Delta t \Big],\label{eq:LBEcollide}\\
 \mathbf{f} (\bm{x}+\mathbf{e}\Delta t, t+\Delta t)&=&\tilde{\mathbf{f}} (\bm{x},t).\label{eq:LBEstream}
\end{eqnarray}
Note that here and in what follows, we use `tilde' ($\tilde{\cdot}$) to refer to the post-collision state of any variable. Then, we combine Eqs.~(\ref{eq:7}) and (\ref{eq:56}), i.e., to relate the sets of moments $\mathbf{m}$ and $\mathbf{n}$ as $\mathbf{n}=\tensor{T}\mathbf{f}=\tensor{T}\tensor{Q}^{-1}\mathbf{m}$, and, similarly, $\mathbf{\Phi}$ and $\mathbf{\Psi}$ as $\mathbf{\Phi}=\tensor{T}\tensor{Q}^{-1}\mathbf{\Psi}$. Substituting these expressions for $\mathbf{n}$ and $\mathbf{\Psi}$ and eliminating the use of the problematic $\tensor{T}$ in favor of $\tensor{Q}$ via $\tensor{T}=\tensor{B}\tensor{Q}$ (see Eq.~(\ref{eq:relationbetweenTandQ})) in Eq.~(\ref{eq:LBEcollide}), the post-collision vector $\mathbf{f}$ then modifies to
\begin{equation}
\tilde{\mathbf{f}} (\bm{x},t) = (\tensor{B}\tensor{Q})^{-1}
 \Big[\tensor{T}\tensor{Q}^{-1}\mathbf{m} + \tensor{\Lambda} \tensor{T}\tensor{Q}^{-1}\;  \left(\; \mathbf{m}^{eq}-\mathbf{m} \;\right) + \left(\tensor{I} - \frac{\tensor{\Lambda}}{2}\right)  \tensor{T}\tensor{Q}^{-1}\mathbf{\Phi}\Delta t \Big].\label{eq:LBEcollidemodified1}
\end{equation}
Exploiting the fact that $\tensor{T}\tensor{Q}^{-1}=\tensor{B}$ and $(\tensor{B}\tensor{Q})^{-1}=\tensor{Q}^{-1}\tensor{B}^{-1}$ based on their definitions, and then performing the product of $\tensor{B}^{-1}$ with the resulting terms inside the brackets $[\cdots]$ in the above Eq.~(\ref{eq:LBEcollidemodified1}) and invoking $\tensor{B}^{-1}\tensor{B}=\tensor{I}$, it follows that
\begin{equation}
\tilde{\mathbf{f}} (\bm{x},t) = \tensor{Q}^{-1}
 \Big[\mathbf{m} + \tensor{B}^{-1}\tensor{\Lambda} \tensor{B}\;  \left(\; \mathbf{m}^{eq}-\mathbf{m} \;\right) + \tensor{B}^{-1}\left(\tensor{I} - \frac{\tensor{\Lambda}}{2}\right)  \tensor{B}\mathbf{\Phi}\Delta t \Big].\label{eq:LBEcollidemodified2}
\end{equation}
As a further simplification, in order to bring this above LB formulation for the cuboid lattice (Eq.~(\ref{eq:LBEcollidemodified2})) closer to that for the common cubic lattice, we replace the transformations based on $\tensor{Q}$ in terms $\tensor{P}$ via Eq.~(\ref{eq:58}), i.e., $\tensor{Q}= \tensor{S} \tensor{P}$ and $\tensor{Q}^{-1}$ in terms $\tensor{P}^{-1}$ using Eq.~(\ref{eq:60}), i.e., $\tensor{Q}^{-1}= \tensor{P}^{-1}\tensor{S}^{-1}$. Thus, using these in Eq.~(\ref{eq:LBEcollidemodified2}) and writing the change in moments under collision as $\tensor{B}^{-1}\tensor{\Lambda} \tensor{B}\;  \left(\; \mathbf{m}^{eq}-\mathbf{m} \;\right)=\tensor{B}^{-1}\tensor{\Lambda}\; \left(\; \tensor{B}\mathbf{m}^{eq}-\tensor{B}\mathbf{m} \;\right)$, the equation for the post-collision vector $\mathbf{f}$ reduces finally to
\begin{equation}
\tilde{\mathbf{f}} (\bm{x},t) = \tensor{P}^{-1}\tensor{S}^{-1}
 \Big[\mathbf{m} + \tensor{B}^{-1}\tensor{\Lambda}\;\left(\; \tensor{B}\mathbf{m}^{eq}-\tensor{B}\mathbf{m} \;\right) + \tensor{B}^{-1}\left(\tensor{I} - \frac{\tensor{\Lambda}}{2}\right)  \tensor{B}\mathbf{\Phi}\Delta t \Big].\label{eq:LBEcollidemodified3}
\end{equation}
Evidently, the linear combinations of moments $\tensor{B}$ and their inverse operation to retrieve the individual bare moments via $\tensor{B}^{-1}$ are now confined just around the steps in determining the changes due to collision in terms of moments only and not for any mappings to or from the velocity space, which are handled by operations similar to those for the cubic lattice augmented with simple scalings of moments based on the grid aspect ratios. These aspects greatly simplify the implementation of the cuboid LB algorithm. Noting that the terms within the brackets $[\cdots]$ in Eq.~(\ref{eq:LBEcollidemodified3}) correspond to the post-collision raw moments, which we write as $\tilde{\mathbf{m}}=\mathbf{m} + \tensor{B}^{-1}\left\{\tensor{\Lambda}\;\left(\; \tensor{B}\mathbf{m}^{eq}-\tensor{B}\mathbf{m} \;\right) + \left(\tensor{I} - \frac{\tensor{\Lambda}}{2}\right)  \tensor{B}\mathbf{\Phi}\Delta t\right\}$, where the pre-collision raw moments $\mathbf{m}$ can be obtained from the distribution functions $\mathbf{f}$ via $\mathbf{m}=\tensor{S}\tensor{P}\mathbf{f}$, we can finally obtain the following equivalent but significantly more efficient cuboid LB formulation of that given in Eqs.~(\ref{eq:LBEcollide})-(\ref{eq:LBEstream}) based on raw moments:
\begin{eqnarray}
\mathbf{m}&=&\tensor{S}\tensor{P}\mathbf{f},\nonumber\\
\tilde{\mathbf{m}}&=&\mathbf{m} + \tensor{B}^{-1}\left\{\tensor{\Lambda}\;\left(\; \tensor{B}\mathbf{m}^{eq}-\tensor{B}\mathbf{m} \;\right) + \left(\tensor{I} - \frac{\tensor{\Lambda}}{2}\right)  \tensor{B}\mathbf{\Phi}\Delta t\right\},\nonumber\\
\tilde{\mathbf{f}} (\bm{x},t) &=&  \tensor{P}^{-1}\tensor{S}^{-1}\tilde{\mathbf{m}},\nonumber\\
 \mathbf{f} (\bm{x}+\mathbf{e}\Delta t, t+\Delta t)&=&\tilde{\mathbf{f}} (\bm{x},t).\label{eq:LBErawmomentcuboidlattice}
\end{eqnarray}
Note that when we discuss the implementation details of each of the steps involved here in what follows, we will write down the explicit details of performing $\tensor{P}$ and $\tensor{P}^{-1}$ related to mappings between the raw moments (which can be readily constructed based on the usual cubic lattice), $\tensor{S}$ and $\tensor{S}^{-1}$ related to scalings of raw moments by the grid aspect ratios, and $\tensor{B}$ and $\tensor{B}^{-1}$ related to combining and segregating the moments around the collision steps involving the relaxation process and augmented by the source terms.

As discussed in the introduction earlier, schemes constructed using the central moments have been shown to be more robust than those based on raw moments. Hence, a natural generalization of the above is to consider next in using the central moments in lieu of the raw moments in performing the collision step. In this regard, we first list the countable central moments $\mathbf{m}^c$ and their equilibria $\mathbf{m}^{c,eq}$ as well as the source central moments $\mathbf{\Phi}^c$ for the D3Q27 lattice as follows:
\begin{eqnarray}
&\mathbf{m}^c=\big(k_{000},k_{100},k_{010},k_{001},k_{110},k_{101},
k_{011},k_{200},k_{020},k_{002},k_{120},k_{102},k_{210}, k_{012}, k_{201},k_{021},k_{111},\nonumber \\ &k_{220},k_{202},k_{022},k_{211},k_{121},k_{112},k_{122},k_{212},k_{221},k_{222}\big)^{\dag},\label{eq:centralmomentlist}
\end{eqnarray}
\begin{eqnarray}
&\mathbf{m}^{c,eq}=\big(k_{000}^{eq},k_{100}^{eq},k_{010}^{eq},k_{001}^{eq},k_{110}^{eq},k_{101}^{eq},
k_{011}^{eq},k_{200}^{eq},k_{020}^{eq},k_{002}^{eq},k_{120}^{eq},k_{102}^{eq},k_{210}^{eq}, k_{012}^{eq}, k_{201}^{eq},k_{021}^{eq},k_{111}^{eq},\nonumber \\ &k_{220}^{eq},k_{202}^{eq},k_{022}^{eq},k_{211}^{eq},k_{121}^{eq},k_{112}^{eq},k_{122}^{eq},k_{212}^{eq},k_{221}^{eq},k_{222}^{eq}\big)^{\dag},\label{eq:ce tralmomentequilibrialist}
\end{eqnarray}
\begin{eqnarray}
&\mathbf{\Phi}^c=\big(\sigma_{000},\sigma_{100},\sigma_{010},\sigma_{001},\sigma_{110},\sigma_{101},
\sigma_{011},\sigma_{200},\sigma_{020},\sigma_{002},\sigma_{120},\sigma_{102},\sigma_{210}, \sigma_{012}, \sigma_{201},\sigma_{021},\sigma_{111},\nonumber \\ &\sigma_{220},\sigma_{202},\sigma_{022},\sigma_{211},\sigma_{121},\sigma_{112},\sigma_{122},\sigma_{212},\sigma_{221},\sigma_{222}\big)^{\dag},\label{eq:centralmomentsourcelist}
\end{eqnarray}
where the components of the central moment components $k_{mnp}$ and $k_{mnp}^{eq}$ and $\sigma_{mnp}$, respectively, are defined in Eq.~(\ref{eq:3a}). We note that the mappings between the central moments and raw moments can be accomplished via the frame transformation matrix $\tensor{\mathcal{F}}$ and its inverse $\tensor{\mathcal{F}}^{-1}$ as
\begin{equation}
\mathbf{m}^c= \tensor{\mathcal{F}}\; \mathbf{m}, \quad \mathbf{m}=\tensor{\mathcal{F}}^{-1}\; \mathbf{m}^c. \label{eq:centralmomentrawmomentmappings}
\end{equation}
Here, $\tensor{\mathcal{F}}$ and $\tensor{\mathcal{F}}^{-1}$ are both lower triangular matrices dependent on the fluid velocity components $(u_x, u_y, u_z)$ and are the same for the cuboid lattice as well as the cubic lattice. The elements of $\tensor{\mathcal{F}}$ can be readily obtained by listing the binomial transforms of all the linearly independent central moments for the D3Q27 lattice in terms of the raw moments and collecting them in a matrix-vector representation. As noted in our recent work on the rectangular central moment LB scheme~\cite{yahia2021central},once $\tensor{\mathcal{F}}=\tensor{\mathcal{F}}(u_x, u_y, u_z)$ is known, its inverse follows from it by exploiting an interesting property of the generating function representation of the binomial expansion, which involves simply reversing the signs of the velocity components appearing in $\tensor{\mathcal{F}}$, i.e., $\tensor{\mathcal{F}}^{-1}=\tensor{\mathcal{F}}(-u_x, -u_y, -u_z)$.

Then, as an extension of Eq.~(\ref{eq:LBEcollidemodified3}), the changes under collision as well as the source term in the LBE can be prescribed in terms of central moments, and the resulting post-collision central moments are then mapped back to raw moments via $\tensor{\mathcal{F}}^{-1}$, which leads to the following equation for the post-collision vector $\mathbf{f}$:
\begin{equation}
\tilde{\mathbf{f}} (\bm{x},t) = \tensor{P}^{-1}\tensor{S}^{-1}\tensor{\mathcal{F}}^{-1}
 \Big[\mathbf{m}^c + \tensor{B}^{-1}\tensor{\Lambda}\;\left(\; \tensor{B}\mathbf{m}^{eq,c}-\tensor{B}\mathbf{m}^c \;\right) + \tensor{B}^{-1}\left(\tensor{I} - \frac{\tensor{\Lambda}}{2}\right)  \tensor{B}\mathbf{\Phi}^c\Delta t \Big].\label{eq:LBEcollidemodified4}
\end{equation}
In view of Eq.~(\ref{eq:LBEcollidemodified4}), we can write the 3D cuboid central moment LB method (3DCCM-LBM) as a generalization of the previous raw moment formulation given in Eq.~(\ref{eq:LBErawmomentcuboidlattice}) as follows:
\begin{eqnarray}
\mathbf{m}^c&=&\tensor{\mathcal{F}}\tensor{S}\tensor{P}\mathbf{f},\nonumber\\
\tilde{\mathbf{m}}^c&=&\mathbf{m}^c + \tensor{B}^{-1}\left\{\tensor{\Lambda}\;\left(\; \tensor{B}\mathbf{m}^{c,eq}-\tensor{B}\mathbf{m}^c \;\right) + \left(\tensor{I} - \frac{\tensor{\Lambda}}{2}\right)  \tensor{B}\mathbf{\Phi}^c\Delta t\right\},\nonumber\\
\tilde{\mathbf{f}} (\bm{x},t) &=&  \tensor{P}^{-1}\tensor{S}^{-1}\tensor{\mathcal{F}}^{-1}\tilde{\mathbf{m}}^c,\nonumber\\
 \mathbf{f} (\bm{x}+\mathbf{e}\Delta t, t+\Delta t)&=&\tilde{\mathbf{f}} (\bm{x},t).\label{eq:LBEcentralmomentcuboidlattice}
\end{eqnarray}
The expressions associated with $\tensor{\mathcal{F}}$ and $\tensor{\mathcal{F}}^{-1}$ in addition to the other matrices will be written explicitly when we discuss the implementation details of the central moment LB scheme for the cuboid lattice next.

\subsection{Algorithmic details of the 3DCCM-LBM}\label{subsec:algorithmicdetails3DCCM-LBM}
We will now discuss the implementation details of the 3DCCM-LBM based on Eq.~(\ref{eq:LBEcentralmomentcuboidlattice}). While the main steps will be identified systematically in the following, the formulas associated with each of steps will be presented in respective appendices.

\begin{itemize}

\item Compute pre-collision raw moments
\begin{equation*}
\mathbf{m}=\tensor{P}\mathbf{f},
\end{equation*}
where $\tensor{P}$ is given in Eq.~(\ref{eq:momentbasiscubiclattice}), and the components of $\mathbf{f}$, i.e., $f_\alpha$ are at time level $t$, i.e., $f_\alpha=f_\alpha(\bm{x},t)$. This provides intermediate values for all the pre-collision raw moments $k_{mnp}^\prime$ for the D3Q27 cubic lattice. The implementation of this step is given in~\ref{sec:appendix1}.

\item Perform scaling of pre-collision raw moments for cuboid lattice
\begin{equation*}
\mathbf{m}\leftarrow\tensor{S}\mathbf{m},
\end{equation*}
where $\tensor{S}$ is given in Eq.~(\ref{eq:59}) and this step yields the pre-collision raw moments for the cuboid lattice $k_{mnp}^\prime$  from those for the cubic lattice computed in the previous step. The equations representing this operation is presented in~\ref{sec:appendix2}.

\item Compute pre-collision central moments
\begin{equation*}
\mathbf{m}^c=\tensor{\mathcal{F}}\mathbf{m},
\end{equation*}
where $\tensor{\mathcal{F}}$ transforms the pre-collision raw moments $k_{mnp}^\prime$ into central moments $k_{mnp}$, and the associated details are listed in~\ref{sec:appendix3}.

\item Compute post-collision central moments: Relaxation under collision using extended equilibria for corrections and sources for body force
\begin{equation*}
\tilde{\mathbf{m}}^c=\mathbf{m}^c + \tensor{B}^{-1}\left\{\tensor{\Lambda}\;\left(\; \tensor{B}\mathbf{m}^{c,eq}-\tensor{B}\mathbf{m}^c \;\right) + \left(\tensor{I} - \frac{\tensor{\Lambda}}{2}\right)  \tensor{B}\mathbf{\Phi}^c\Delta t\right\}.
\end{equation*}
Here, we obtain the post-collision central moments $\tilde{k}_{mnp}$ as a result of the relaxation of the central moments to their effective equilibria and augmented with source terms due to the body force. The details on the effective central moment equilibria accounting for the corrections due to grid anisotropy and non-GI velocity errors using the results from Sec.~\ref{subsec:correctionsviaCEanalysis}, the selection of the relaxation parameters appearing in $\tensor{\Lambda}$ and how certain moments are combined and segregated prior to and after the relaxation step under collision as required via $\tensor{B}$ and $\tensor{B}^{-1}$ are given in~\ref{sec:appendix4}.

\item Compute post-collision raw moments
\begin{equation*}
\tilde{\mathbf{m}}=\tensor{\mathcal{F}}^{-1}\tilde{\mathbf{m}}^c.
\end{equation*}
The expressions for the mappings of the post-collision central moments $\tilde{k}_{mnp}$ to raw moments $\tilde{k}_{mnp}^\prime$ performed via the inverse frame transformation matrix $\tensor{\mathcal{F}}^{-1}$ are provided in~\ref{sec:appendix5}.

\item Perform inverse scaling of post-collision raw moments for cuboid lattice
\begin{equation*}
\tilde{\mathbf{m}}\leftarrow\tensor{S}^{-1}\tilde{\mathbf{m}}.
\end{equation*}
This scales down the post-collision raw moments $\tilde{k}_{mnp}^\prime$ and facilitates their more efficient transformation to distribution functions via the moment basis for the cubic lattice $\tensor{P}^{-1}$ in the next step. Here, the inverse of the scaling matrix, i.e., $\tensor{S}^{-1}$ is presented in Eq.~(\ref{eq:61}) and the operations involved in this step are listed in~\ref{sec:appendix6}.

\item Compute post-collision distribution functions
\begin{equation*}
\tilde{\mathbf{f}}=\tensor{P}^{-1}\tilde{\mathbf{m}},
\end{equation*}
where the computations involved in the inverse of the simpler moment basis $\tensor{P}$ is given in Eq.~(\ref{eq:momentbasiscubiclattice}) are shown in~\ref{sec:appendix7}.

\item Perform streaming of distribution functions
\begin{equation*}
f_\alpha(\bm{x},t+\Delta t)=\widetilde{f}_\alpha(\bm{x}-\bm{e}_\alpha\Delta t,t).
\end{equation*}

\item Update hydrodynamic fields
\newline
Using $f_\alpha(\bm{x},t+\Delta t)$ at the new time level $t+\Delta t$ from the previous step, fluid density and velocities can be obtained as
\begin{equation*}
\rho =\sum_{\alpha=0}^{26} f_\alpha, \quad     \rho \bm{u} =\sum_{\alpha=0}^{26} f_\alpha \bm{e}_\alpha + \frac{1}{2} \mathbf{F}\Delta t.
\end{equation*}

\end{itemize}
The following comments regarding the algorithm given above is now in order. (i) When compared to the central moment LB scheme for the common D3Q27 cubic lattice, we emphasize that the only extra computations incurred by the above discussed method for the D3Q27 cuboid lattice are those involving the corrections in the diagonal components of the second order moments in the collision step and the applications of scalings on the raw moments before and after the collision step (i.e., $\tensor{S}$ and $\tensor{S}^{-1}$); these result in a minor additional overhead, but as will be demonstrated in the simulations of an inhomogeneous and anisotropic shear flow case study later, it will be offset by far due to the overall significant computational advantages of using the cuboid lattice in lieu of the cubic lattice that accommodates the nature of the flow much more efficiently. (ii) The algorithm given above based on central moments is modular in nature. It can be readily extended to other collision models. For example, a raw moment-based cuboid scheme as a modification of the above would simply involve the elimination of the mappings between the raw moments and central moments (i.e., those involving $\tensor{\mathcal{F}}$ and $\tensor{\mathcal{F}}^{-1}$) around collision and performing the collision step in terms of raw moments rather than central moments, but using the same corrections via the extended moment equilibria. However, as will be numerically shown later that the central moment formulation is more robust than that based on raw moments for the cuboid lattice. Schemes based on other collision models such as those based on cumulants~\cite{geier2015cumulant} can be readily constructed from the above algorithm by introducing additional mappings between central moments and cumulants and performing the collision step in terms of relaxations of cumulants by using the same required corrections for the cuboid lattice as in the above algorithm. Such an extension can be considered in a future study. (iii) Finally, the present approach devised for the D3Q27 cuboid lattice readily extends for other lattice sets, such as D3Q15 and D3Q27 cuboid lattices, by using only a subset of the moment basis as necessary while incorporating the corrections to the equilibria derived in this work.

\section{Boundary conditions on moving walls: Momentum-augmented bounce back scheme on cuboid lattice grids}\label{sec:4}
Before discussing the numerical results based on the 3DCCM-LBM constructed above, we note that the use of cuboid lattice requires some modifications to the implementation of the standard link-based half-way bounce back boundary condition for moving boundaries. Since we will be investigating shear driven flows due to the motion of the walls in this work, we will now derive the necessary changes to such a boundary scheme. If $\bm{x}_f$ and $\bm{x}_w$ represent the fluid node nearest to the boundary and the wall node, respectively, and $\alpha$ and $\bar{\alpha}$ denote, respectively, the outgoing and incoming particle directions, where $\mathbf{e}_{\bar{\alpha}}=-\mathbf{e}_{\alpha}$, then the general form of the momentum-augmented half-way bounce back scheme can be written as
\begin{equation}\label{eq:momentumaugmentedBBscheme}
f_{\bar{ \alpha}}(\bm{x}_f, t+\Delta t)= \tilde{f}_{\alpha}(\bm{x}_f,t) - \left[ f_{\alpha}^{eq}(\bm{x}_w) -f_{\bar{\alpha}}^{eq}(\bm{x}_w) \right].
\end{equation}
Here, the equilibrium distribution functions $f_{\alpha}^{eq}$ and $f_{\bar{\alpha}}^{eq}$ associated with $\mathbf{f}^{eq}$ at $\bm{x}_w$ depend on the wall density and velocities. It can be obtained via an inverse mapping from the equilibrium moments $\mathbf{m}^{eq}$ using $\mathbf{f}^{eq}=\tensor{Q}^{-1}\mathbf{m}^{eq}$, where the components of the equilibrium moments are evaluated using Eq.~(\ref{eq:10}) from those based on the imposed wall conditions. Note that since $\tensor{Q}^{-1}$ is dependent on the grid aspect ratios, the resulting formulas will be parameterized by $r$ and $s$. Specifically, consider a moving boundary located in the $x-z$ plane, whose outward normal is along the $y$ direction and pointing into the plane of the paper (see Fig.~\ref{fig:1c}). Let this boundary plane be moving along the $x$ direction with an imposed velocity $U$; thus based on the no-slip boundary conditions for the fluid, we can then write $u_x=U$, $u_y=0$, $u_z=0$, and, as usual for the bounce back scheme, we approximate the wall node density to that based on the known fluid, i.e., $\rho=\rho_f$. For the case under consideration, referring to Fig.~\ref{fig:1c}, the unknown or incoming distribution functions are $f_{\bar{\alpha}}$, where $\bar{\alpha}=4,9,10,16,18,21,22,25,26$, and can be written in terms of the known directions using Eq.~(\ref{eq:momentumaugmentedBBscheme}) as
\begin{align*}
  f_4(\bm{x}_f, t+\Delta t)= \tilde{f}_3(\bm{x}_f,t)- \left[ f_3^{eq}(\bm{x}_w) -f_4^{eq}(\bm{x}_w) \right] ,&\\
  f_9(\bm{x}_f, t+\Delta t)= \tilde{f}_8(\bm{x}_f,t)- \left[ f_8^{eq}(\bm{x}_w) -f_9^{eq}(\bm{x}_w) \right] , & \quad\quad f_{21}(\bm{x}_f, t+\Delta t)= \tilde{f}_{24}(\bm{x}_f,t)- \left[ f_{24}^{eq}(\bm{x}_w) -f_{21}^{eq}(\bm{x}_w) \right], \\
  f_{10}(\bm{x}_f, t+\Delta t)= \tilde{f}_7(\bm{x}_f,t)- \left[ f_7^{eq}(\bm{x}_w) -f_{10}^{eq}(\bm{x}_w) \right],& \quad\quad f_{22}(\bm{x}_f, t+\Delta t)= \tilde{f}_{23}(\bm{x}_f,t)- \left[ f_{23}^{eq}(\bm{x}_w) -f_{22}^{eq}(\bm{x}_w) \right], \\
  f_{16}(\bm{x}_f, t+\Delta t)= \tilde{f}_{17}(\bm{x}_f,t)- \left[ f_{17}^{eq}(\bm{x}_w) -f_{16}^{eq}(\bm{x}_w) \right] ,& \quad\quad f_{25}(\bm{x}_f, t+\Delta t)= \tilde{f}_{20}(\bm{x}_f,t)- \left[ f_{20}^{eq}(\bm{x}_w) -f_{25}^{eq}(\bm{x}_w) \right] ,\\
  f_{18}(\bm{x}_f, t+\Delta t)= \tilde{f}_{15}(\bm{x}_f,t)- \left[ f_{15}^{eq}(\bm{x}_w) -f_{18}^{eq}(\bm{x}_w) \right], & \quad\quad f_{26}(\bm{x}_f, t+\Delta t)= \tilde{f}_{19}(\bm{x}_f,t)- \left[ f_{19}^{eq}(\bm{x}_w) -f_{26}^{eq}(\bm{x}_w) \right].
\end{align*}
Then, we determine the expressions of the known equilibrium distribution functions using $\mathbf{f}^{eq}=\tensor{Q}^{-1}\mathbf{m}^{eq}$, where we evaluate the components of the moment equilibria on the wall via Eq.~(\ref{eq:10}) based on the imposed wall conditions mentioned above. Using these, we then finally obtain the momentum-augmented bounce-back scheme for the cuboid lattice as
\begin{align}\label{eq:69}
\begin{matrix}
    f_4(\bm{x}_f, t+\Delta t)= \tilde{f}_3(\bm{x}_f,t),& \\
    f_{16}(\bm{x}_f, t+\Delta t)=  \tilde{f}_{17}(\bm{x}_f,t),& f_{21}(\bm{x}_f,t+\Delta t)= \tilde{f}_{24}(\bm{x}_f,t)+ \left[\dfrac{c_s^4}{4 r^2 s^2}\right] \rho U, \\
    f_{18}(\bm{x}_f, t+\Delta t)= \tilde{f}_{15}(\bm{x}_f,t),& f_{22}(\bm{x}_f, t+\Delta t)= \tilde{f}_{23}(\bm{x}_f,t)- \left[\dfrac{c_s^4}{4 r^2 s^2}\right] \rho U,\\
    f_9(\bm{x}_f, t+\Delta t)= \tilde{f}_8(\bm{x}_f,t)- \left[\dfrac{(c_s^2 - s^2)c_s^2}{2 r^2 s^2}\right]\; \rho U,& f_{25}(\bm{x}_f, t+\Delta t)= \tilde{f}_{24}(\bm{x}_f,t)+ \left[\dfrac{c_s^4}{4 r^2 s^2}\right] \rho U,\\
    f_{10}(\bm{x}_f, t+\Delta t)= \tilde{f}_7(\bm{x}_f,t)+\left[\dfrac{(c_s^2 - s^2)c_s^2}{2 r^2 s^2}\right]\; \rho U,& f_{26}(\bm{x}_f, t+\Delta t)= \tilde{f}_{19}(\bm{x}_f,t)- \left[\dfrac{c_s^4}{4 r^2 s^2}\right] \rho U.
\end{matrix}
\end{align}
Clearly, the formulas given in Eq.~(\ref{eq:69}), which will be utilized in the next section, depend on the grid aspect ratios $r$ and $s$ in addition to the specified wall velocity $U$.

\section{Results and discussion: Numerical validation} \label{sec:5}
In this section, we will perform numerical validations of our new 3D cuboid central moment LBM via simulations of an assortment of canonical fluid flow problems including flows in square duct, pulsatile flow in a square duct driven by a periodic body force, and lid-driven flow within a cubic cavity at various characteristic parameters and grid aspect ratios.

\subsection{Flow through a square duct driven by a constant body force}
\begin{figure}[H]
\centering
 \includegraphics[width=0.5\textwidth] {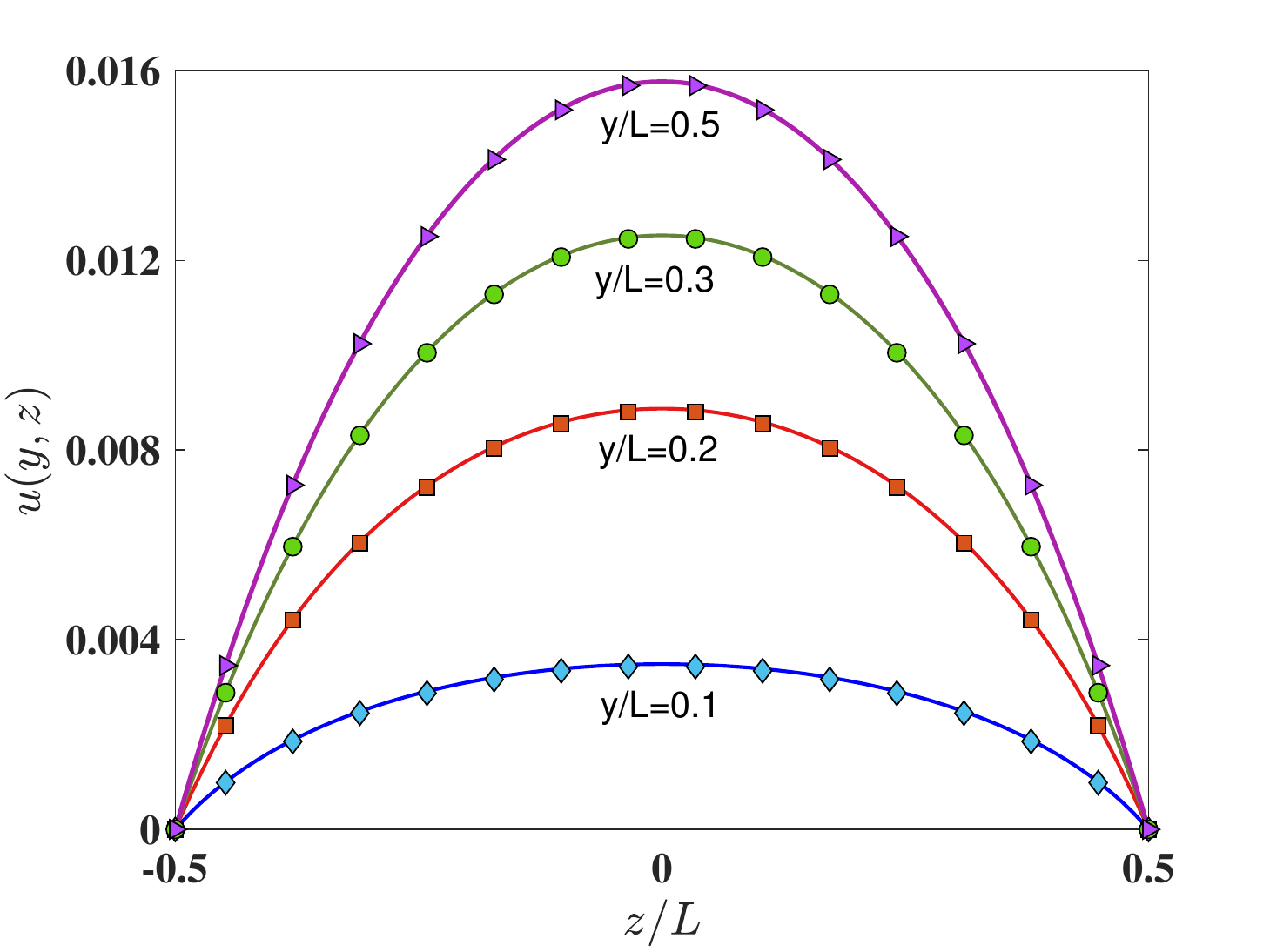}
  \label{fig:1}  
 \caption{Comparison of the velocity profiles in the $y-z$ plane along the reference lines at $y=0.1L$, $y=0.2L$, $y=0.3L$ and $y=0.5L$ of a square duct of side $L$ computed using cuboid 3DCCM-LBM (lines) with aspect ratios of $(r,s)=(0.5,0.33)$ with the analytical solution shown in Eq.~(\ref{eq:70}) (symbols).} \label{fig:duct1}
\end{figure}

First, we perform simulations of a fully developed flow through a square duct driven by a constant body force using the 3DCCM-LBM. The analytical solution of this flow problem for the velocity field $u(y,z)$ can be written as~\cite{white2006viscous}
\begin{equation}\label{eq:70}
   u(y,z)= \frac{4 L^2 F_x}{\pi^3\rho \nu}\sum_{n=1,3,5,...}^{\infty}(-1)^{(n-1/2)}\left[1- \frac{\cosh(n\pi z/L)}{\cosh(n\pi/2)}\right] \frac{\cos(n\pi y/L)}{n^3},
\end{equation}
where $L$ represents the side of the square duct, $\nu$ is the kinematic viscosity and $F_x$ is the body force imposed in the direction of the fluid flow, the x-direction. We resolve the computational domain using $N_x\times N_y \times N_z$ grid nodes and the grid aspect ratios in the $y$ and $z$ directions, i.e., $r$ and $s$, respectively, are then calculated using $r= L/N_y$ and $s= L/N_z$. Periodic boundary condition is applied along the flow direction ($x$-direction) and no slip boundary conditions at the four side walls are imposed using the half-way bounce back boundary conditions (see e.g., Eqs.~\eqref{eq:69}). We define the Reynolds number $\mbox{Re}$ for this flow based on the maximum flow velocity occurring at the midpoint within the duct cross section and the duct side length as the characteristic velocity and length scales, respectively. Table~\ref{table:1} presents a list of the model parameters used for different combinations of the aspect ratios $r$ and $s$ for flow simulations at $L=30$ and $\mbox{Re}=50$, including the choices for the body force, speed of sound $c_s$, the relaxation time $\tau=1/\omega_\nu$ that specifies the viscosity $\nu$.
\begin{table}[h]
\centering
\captionsetup{justification=centering}
\caption{Parameters used in the simulations of 3D square duct flow using 3DCCM-LBM at different lattice grid aspect ratios with a constant Reynolds number $\mbox{Re}=50$.}
\begin{tabular}{c | c c | c c c c}
\hline
 $ N_x \times N_y\times N_z$ &$r$ & $s$ & $F_x$  & $ c_s^2 $ & $\tau$ & $\nu$ \\[0.5ex]
\hline\hline
 $3 \times 30\times 60$ & 1.0 & 0.5 & $3.82 \times 10^{-6}$ & 0.10   & 0.6    & 0.01\\[0.5ex]
 $3 \times 30\times 90$ & 1.0 & 0.33 & $5.50 \times 10^{-6}$ & 0.04   & 0.8    & 0.012\\[0.5ex]
 $3 \times 60\times 30$ & 0.5 & 1.0 & $5.09\times 10^{-9}$ & 0.04   & 1.0    & 0.02  \\[0.5ex]
 $3 \times 60\times 90$ & 0.5 & 0.33 & $5.09 \times 10^{-6}$  & 0.04  & 1.0    & 0.02 \\[0.5ex]
\hline
\end{tabular} \label{table:1}
\end{table}
As an example of the results obtained, Fig.~\ref{fig:duct1} shows the comparisons of the velocity profiles computed using 3DCCM-LBM with $r=0.5$ and $s=0.33$ along the $z$ direction for four different selected locations in the $y$ direction. Excellent agreement can be seen between the computed results using the cuboid lattice and the analytical solution. Moreover, though not shown in this figure, the 3DCCM-LBM results were also found to be in similar agreements with the exact solution for all the other cases of the grid aspect ratios tested according to Table~\ref{table:1}.
We note here that due to the similarity of the governing equations for this Stokes flow problem with the anisotropic advection diffusion equation (ADE), it can be alternatively solved by the LBMs developed in this regard by Ginzburg.~\cite{ginzburg2005equilibrium,ginzburg2013multiple}  via a coordinate grid transform, as in e.g., ~\cite{ginzburg2005generic, ginzburg2006variably}, for which the role of the associated anisotropy on the numerical stability was studied in~\cite{ginzburg2012truncation}. More generally, the idea of the coordinate grid transform used for the ADE LBM may also be adopted for flow LBM.

\subsection{Pulsatile flow in a square duct driven by a periodic body force}
\begin{figure}[ht]
\centering
\advance\leftskip-1.7cm
    \subfloat[] {
        \includegraphics[width=0.5\textwidth] {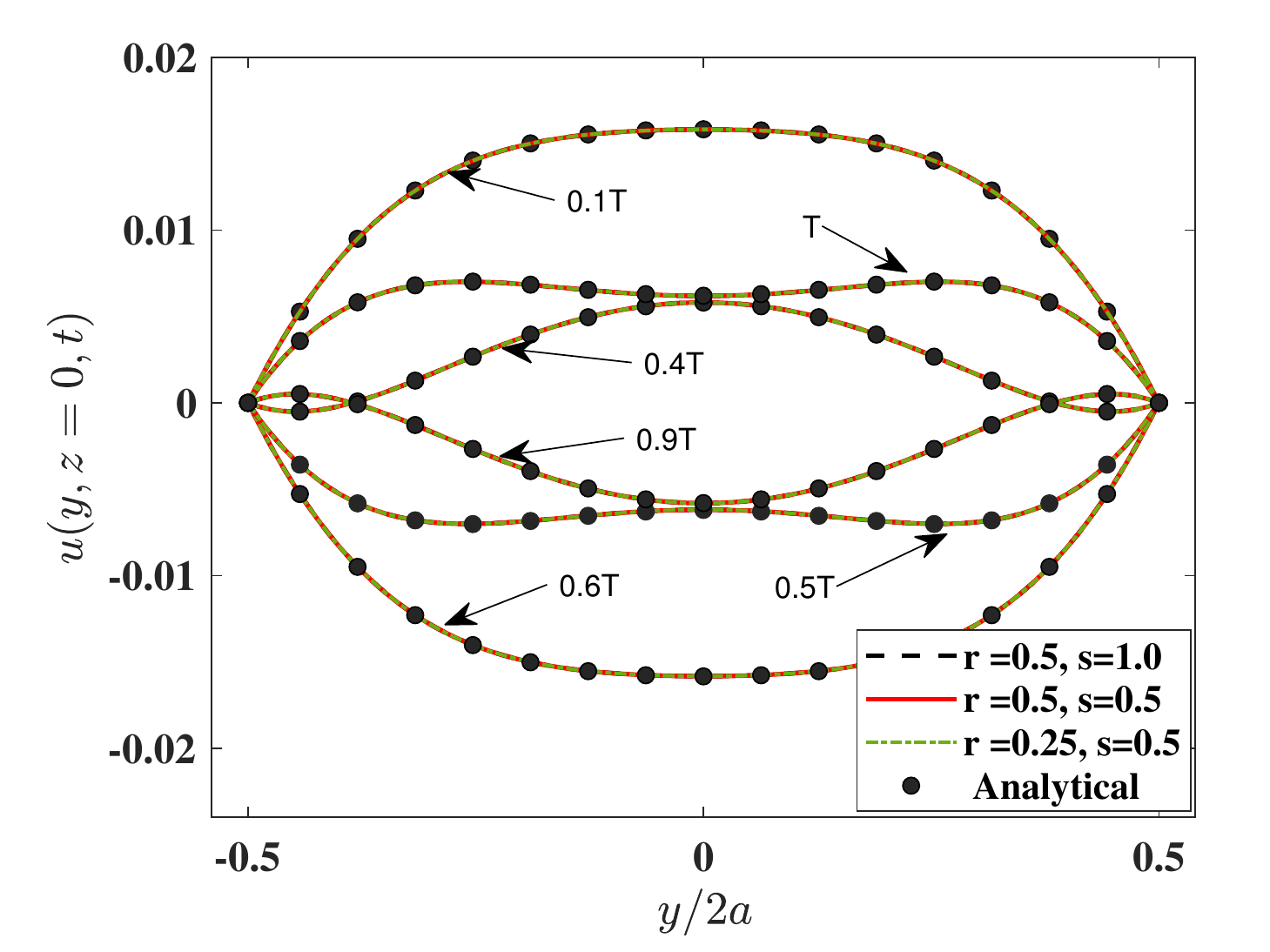}
        \label{fig:3a} } 
    \subfloat[] {
        \includegraphics[width=0.5\textwidth] {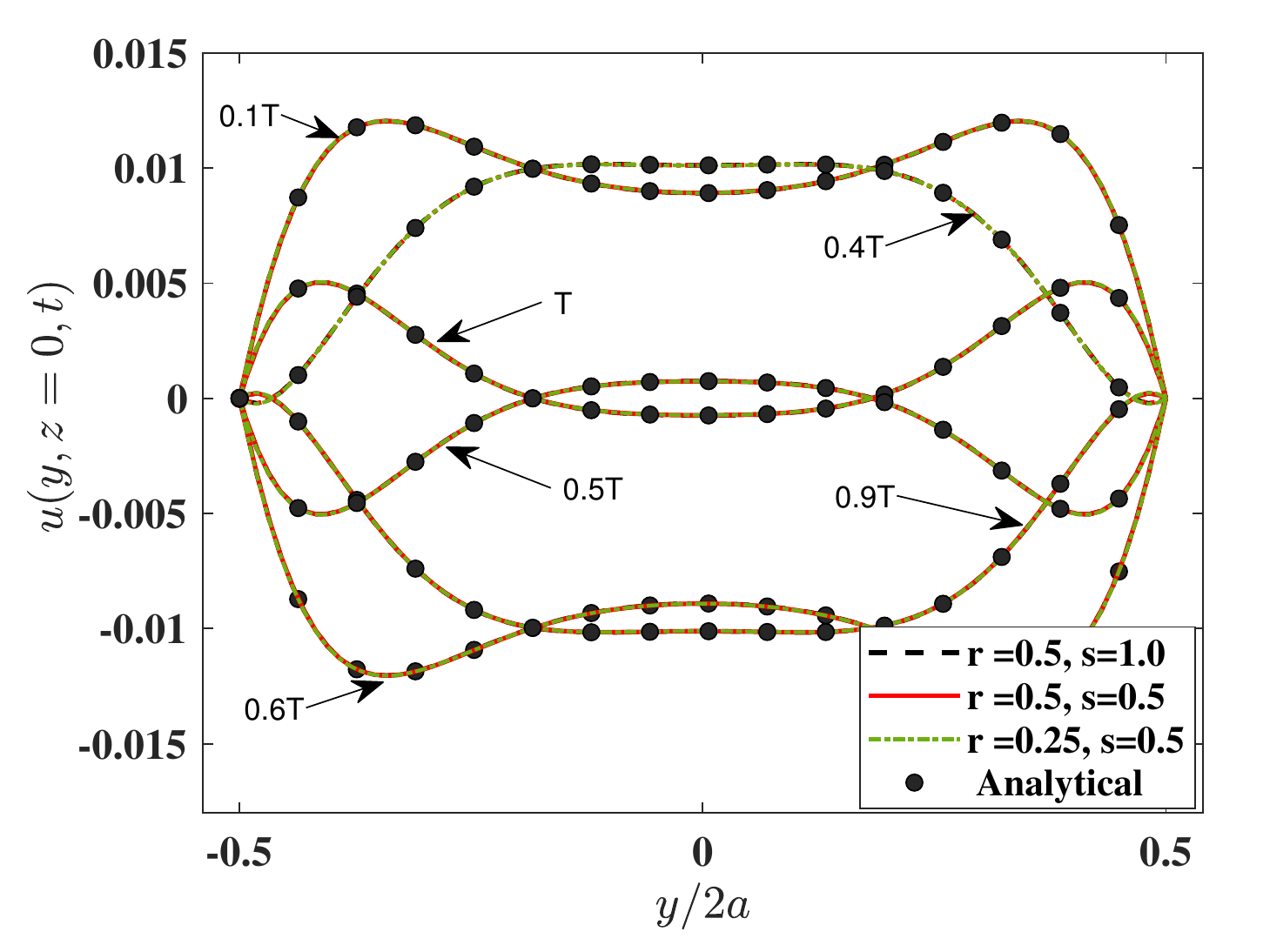}
        \label{fig:3b} } \\
                \advance\leftskip0cm
    \caption{The velocity profiles for the pulsatile flow through a square duct at different instants within the time period $T$ with Womersley numbers of (a) $\text{Wo}=3.09$ and  (b) $\text{Wo}=6.25$ computed using 3DCCM-LBM (lines) with different aspect ratios ($(r,s)=(0.5,1.0)$, $(r,s)=(0.5,0.5)$, and $(r,s)=(0.25,0.5)$) compared with the analytical solution given in Eq.~(\ref{eq:pulsatileflowanalyticalsolution}) (symbols).}
    \label{fig:Wo3}
\end{figure}

\begin{figure}[H]
\centering
\advance\leftskip-1.7cm
    \subfloat[] {
        \includegraphics[width=0.5\textwidth] {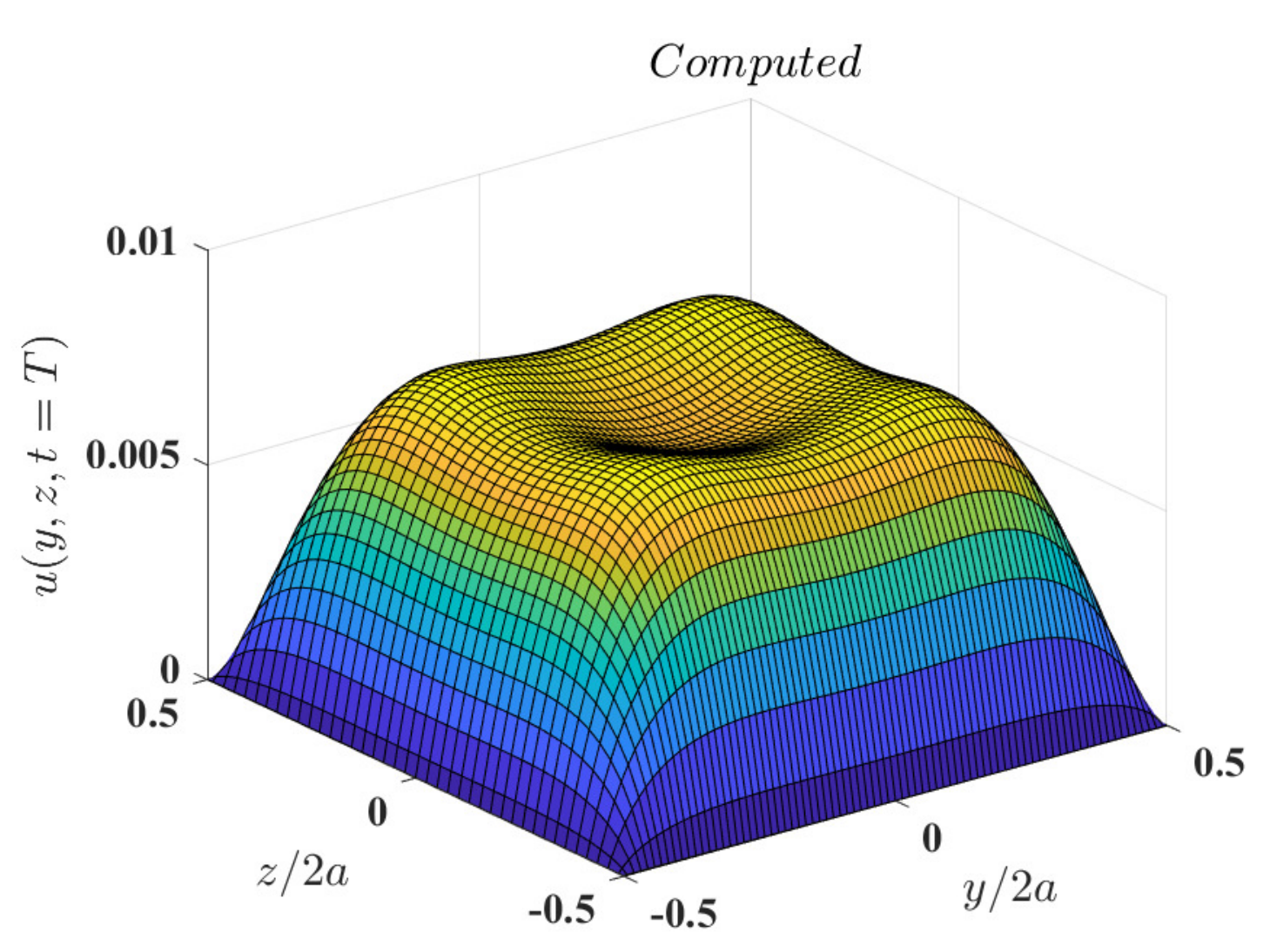}
        \label{fig:4a} } 
    \subfloat[] {
        \includegraphics[width=0.5\textwidth] {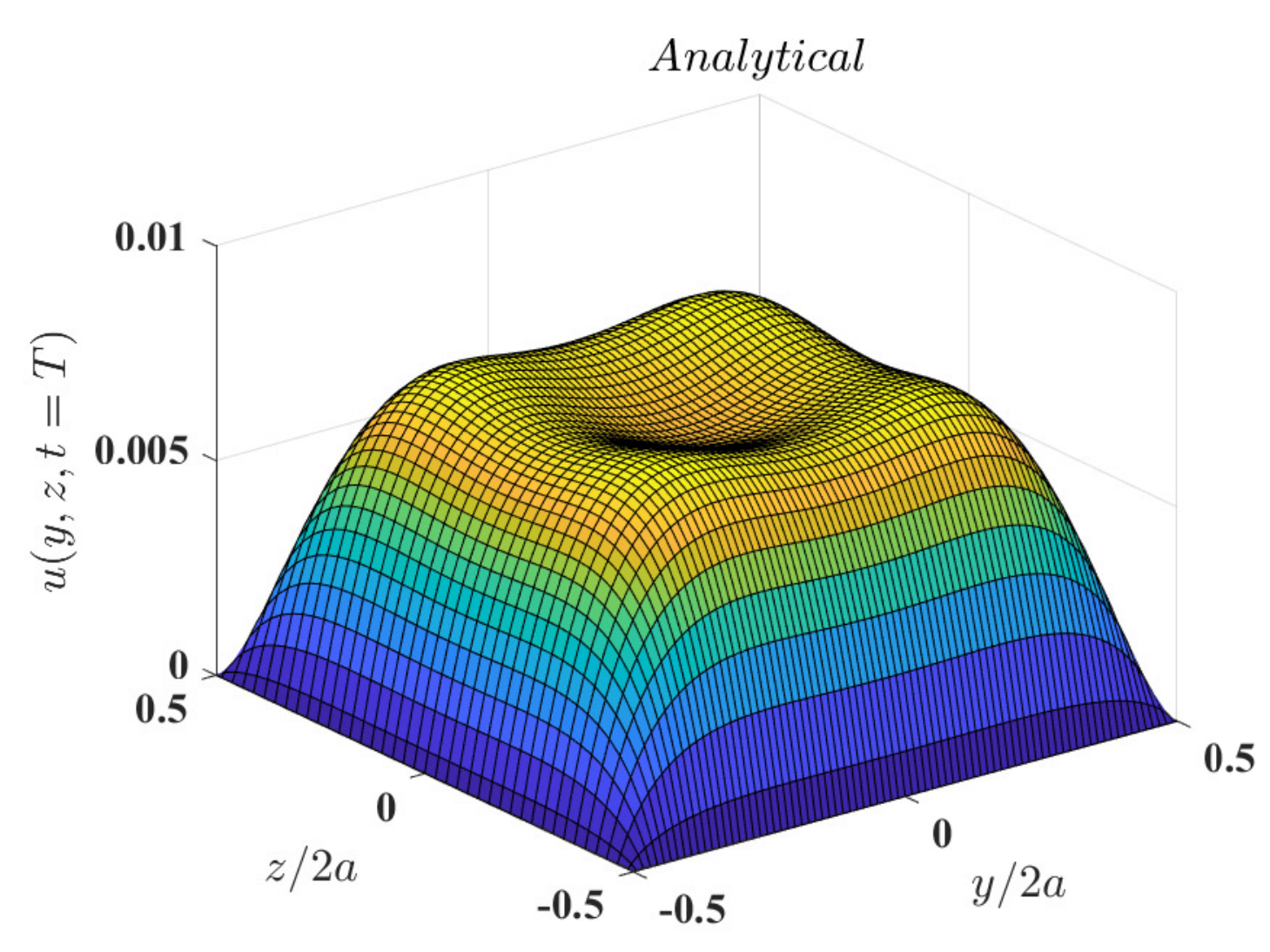}
        \label{fig:4b} } \\
                \advance\leftskip0cm
    \caption{Distribution of the velocity field across the $y-z$ plane of the pulsatile flow through a square duct with the Womersley number $\text{Wo}=3.09$ at the instant $t=T$ (a) computed using 3DCCM-LBM with aspect ratios of $(r,s)=(0.5,1.0)$ and (b) obtained from the exact solution given in Eq.~(\ref{eq:pulsatileflowanalyticalsolution}) (symbols).}
    \label{fig:Wo4}
\end{figure}

\begin{figure}[H]
\centering
\advance\leftskip-1.7cm
    \subfloat[] {
        \includegraphics[width=0.5\textwidth] {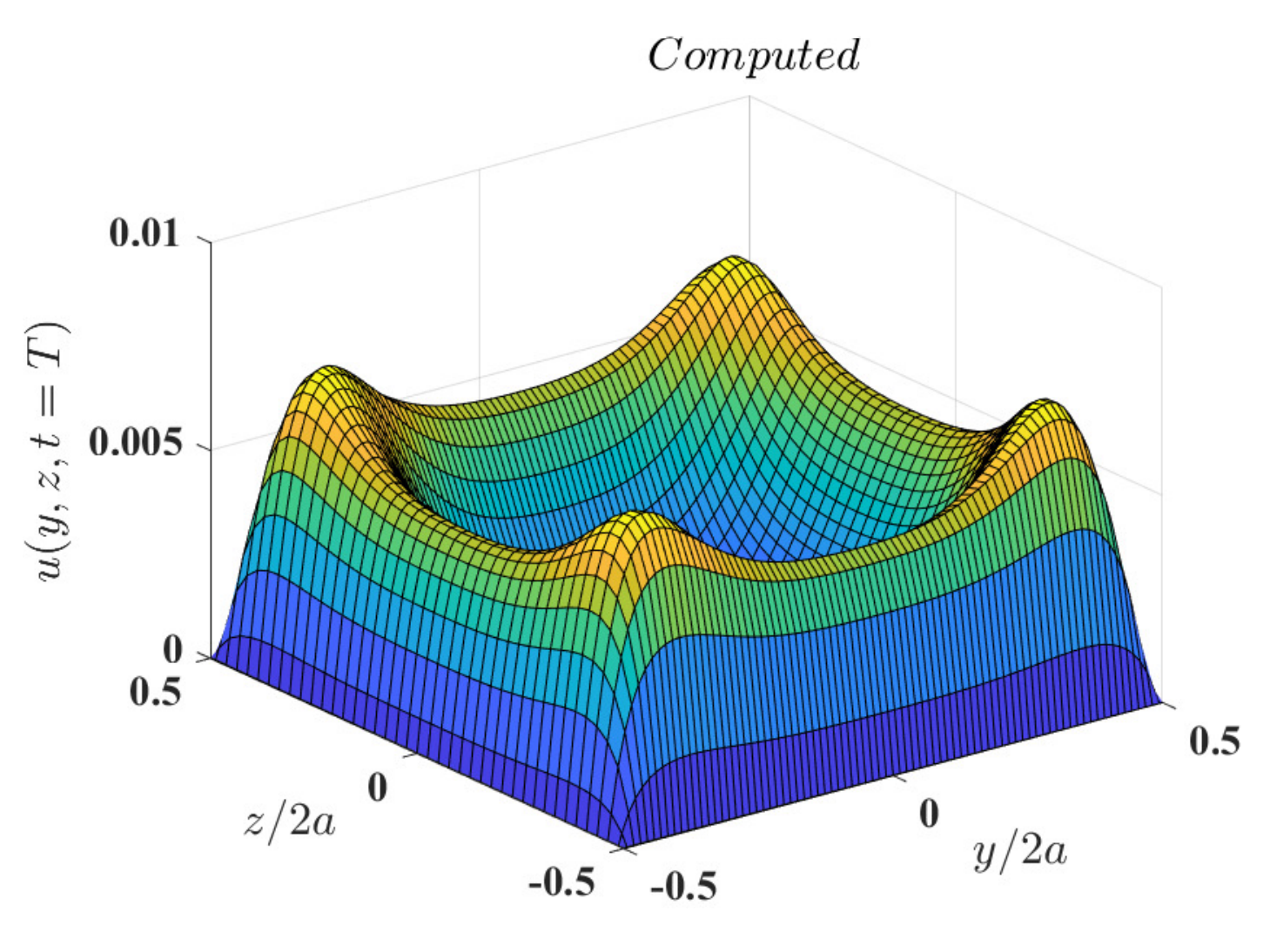}
        \label{fig:5a} } 
    \subfloat[] {
        \includegraphics[width=0.5\textwidth] {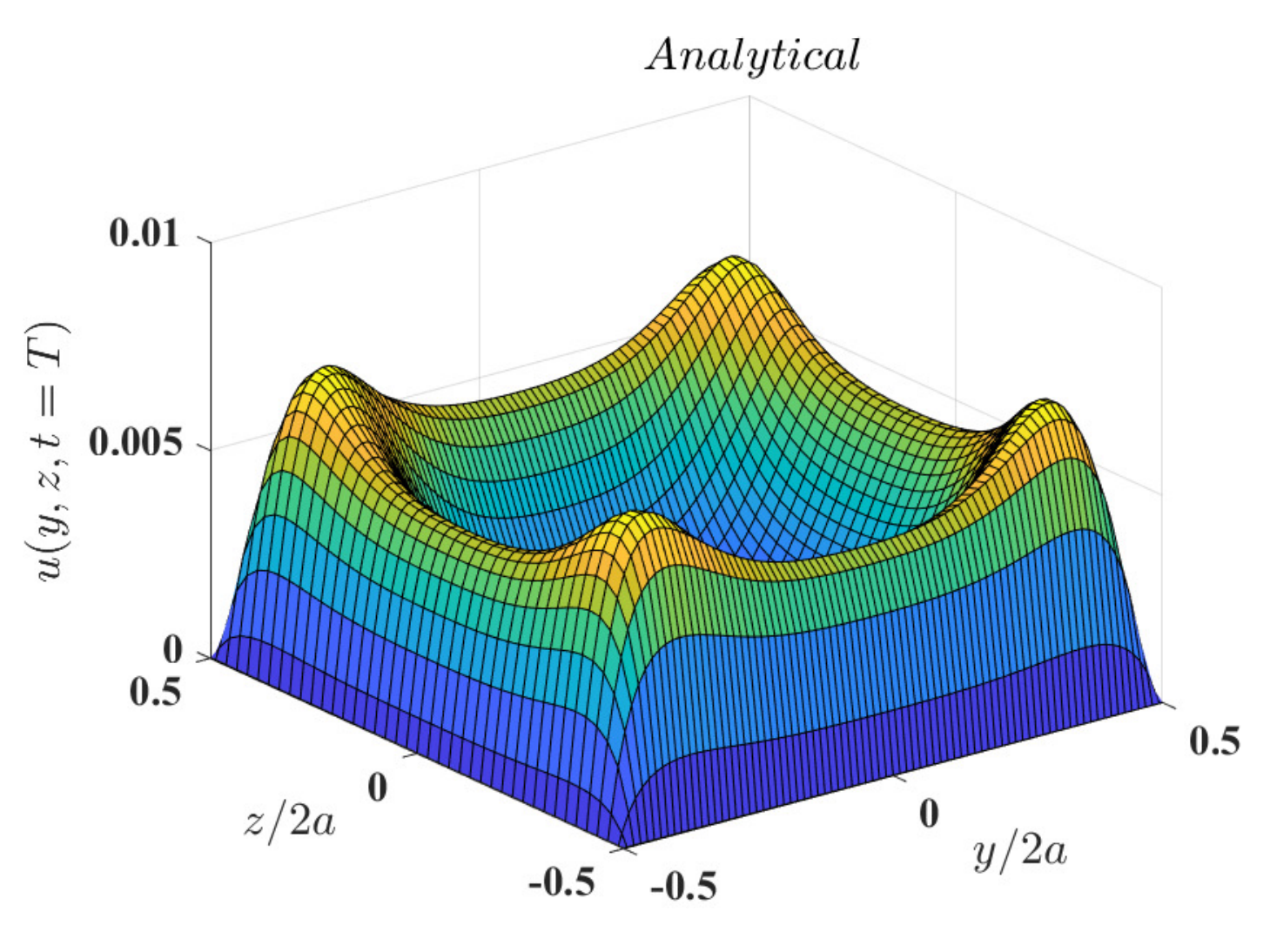}
        \label{fig:5b} } \\
                \advance\leftskip0cm
    \caption{Distribution of the velocity field across the $y-z$ plane of the pulsatile flow through a square duct with the Womersley number $\text{Wo}=6.25$ at the instant $t=T$ (a) computed using 3DCCM-LBM with aspect ratios of $(r,s)=(0.5,1.0)$ and (b) obtained from the exact solution given in Eq.~(\ref{eq:pulsatileflowanalyticalsolution}) (symbols).}
    \label{fig:Wo5}
\end{figure}

Next, we will assess the ability of the 3DCCM-LBM in accurately representing a time-dependent flow problem for which an analytical solution is available. In this regard, we simulate time-periodic flow through a square duct of side length $2a$ driven by a sinusoidally varying body force $ F_x=F_m \cos \omega t$, where $F_m$ is its peak amplitude and $\omega=2\pi/T$ is the angular frequency of the applied pulsations and with $T$ being its time period. This case study admits an analytical solution for the spatially and temporally varying velocity field based on Fourier series and given by~\cite{o1975pulsatile}
\begin{eqnarray}\label{eq:pulsatileflowanalyticalsolution}
   &u(y,z,t)= \mathcal{R} \Biggl[\cfrac{F_m}{\omega}\Bigg( 1 - 2 \sum_{n=0}^{\infty}\dfrac{(-1)^{n}}{p^n}\Bigg\{ \dfrac{\cosh(\gamma_n y/a)\cos(p_n z/a)}{\cosh(\gamma_n)} + \dfrac{\cosh(\gamma_n z/a)\cos(p_n y/a)}{\cosh(\gamma_n)} \Bigg\}\; \Bigg) e^{i \omega t}\Biggr],\\
   &\gamma_n = \sqrt{p_n^2+ i \mbox{Wo}^2}, \;\; \; \; \; \; \quad p_n=(2n+1)\pi/2. \nonumber
\end{eqnarray}
where $-a \le y,z \le a$ and $\mathcal{R}[\cdots]$ represents the real part of the terms within the brackets. Here, $\mbox{Wo}$ is referred to as the Womersley number and defined by $\mbox{Wo} = a \sqrt{\omega/ \nu}$, which is an indication of the ratio of the flow time scale due to viscous diffusion and the time scale of the variations of the imposed force. Considering $2a=40$, $F_m=1 \times 10^{-5}$ and $T=10,000$, we simulated this flow at two different values of the Womersley number, $\text{Wo}=3.09$ and $\text{Wo}=6.25$. For a grid resolution of $N_x\times N_y \times N_z$ the grid aspect ratios $r$ and $s$ are chosen via $r= 2a/N_y$ and $s= 2a/N_z$. We considered three different choices for the grid resolution, viz., $40\times80\times40$, $40\times80\times80$ and $40\times160\times80$ using the cuboid lattice, which correspond to the cases with $(r,s)=(0.5,1.0)$, $(r,s)=(0.5,0.5)$, and $(r,s)=(0.25,0.5)$, respectively. Figure~\ref{fig:Wo3} presents the velocity profiles at different instants within the time period $T$ computed using 3DCCM-LBM and compared with the analytical solution (Eq.~(\ref{eq:pulsatileflowanalyticalsolution})). The computed results obtained using the cuboid lattice for all choices and combinations of the grid aspect ratios are in very good agreement with the analytical solution at any given instant in time.
In addition, Figs.~\ref{fig:Wo4} and \ref{fig:Wo5} show the computed velocity distributions for the entire cross section of the duct obtained using $(r,s)=(0.5,1.0)$, $(r,s)=(0.5,0.5)$, and $(r,s)=(0.25,0.5)$ for $\text{Wo}=3.09$ and $\text{Wo}=6.25$, respectively, at the time instant $t=T$, which compare well with those given by the exact solution. These results provide evidence that the 3DCCM-LBM can perform relatively accurate numerical computations of temporally varying flows.

\subsection{3D lid-driven shear flow in a cubic cavity}
\begin{figure}[ht]
\centering
\advance\leftskip-1.7cm
    \subfloat[] {
        \includegraphics[width=0.5\textwidth] {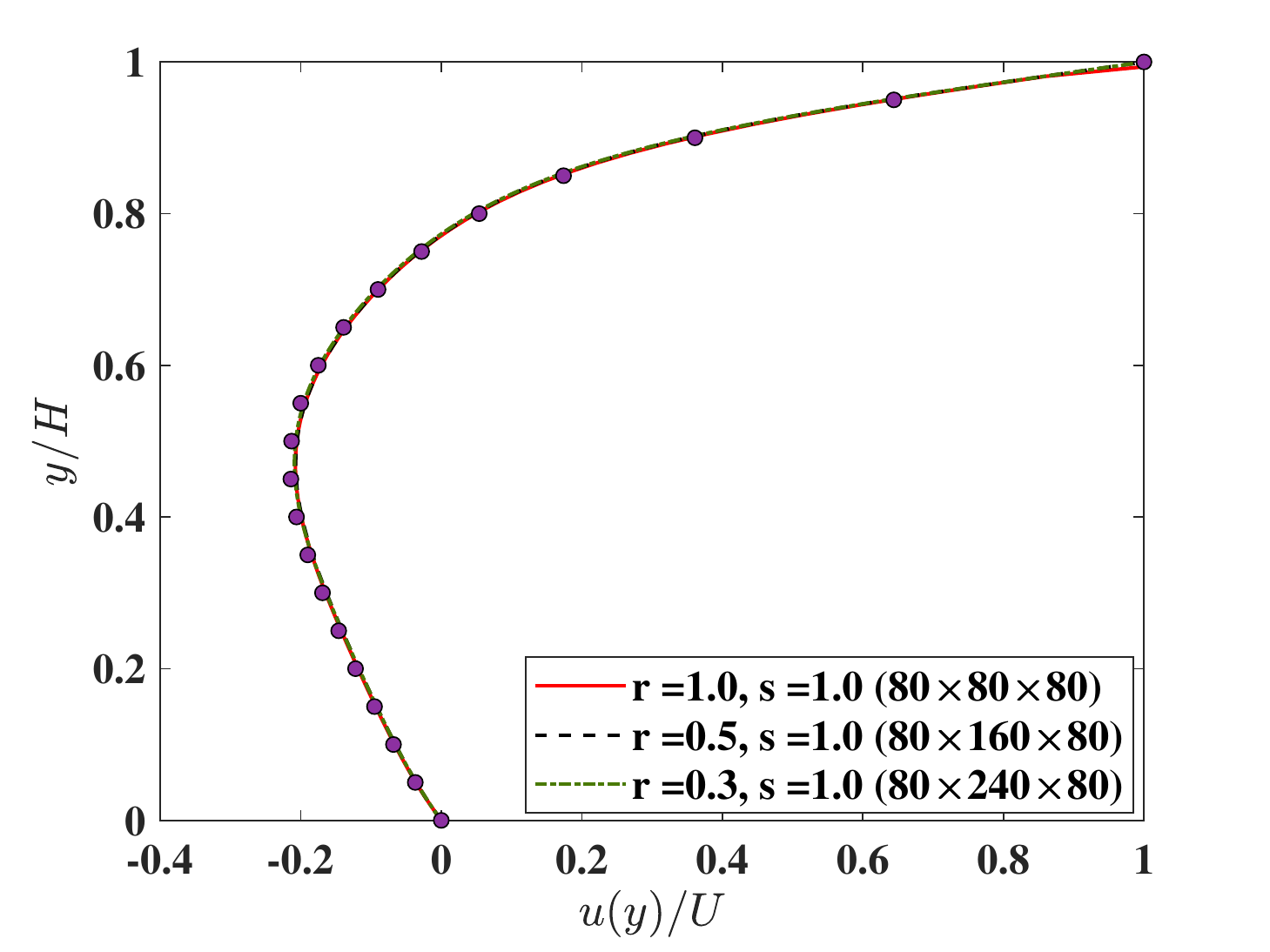}
        \label{fig:3a} } 
    \subfloat[] {
        \includegraphics[width=0.5\textwidth] {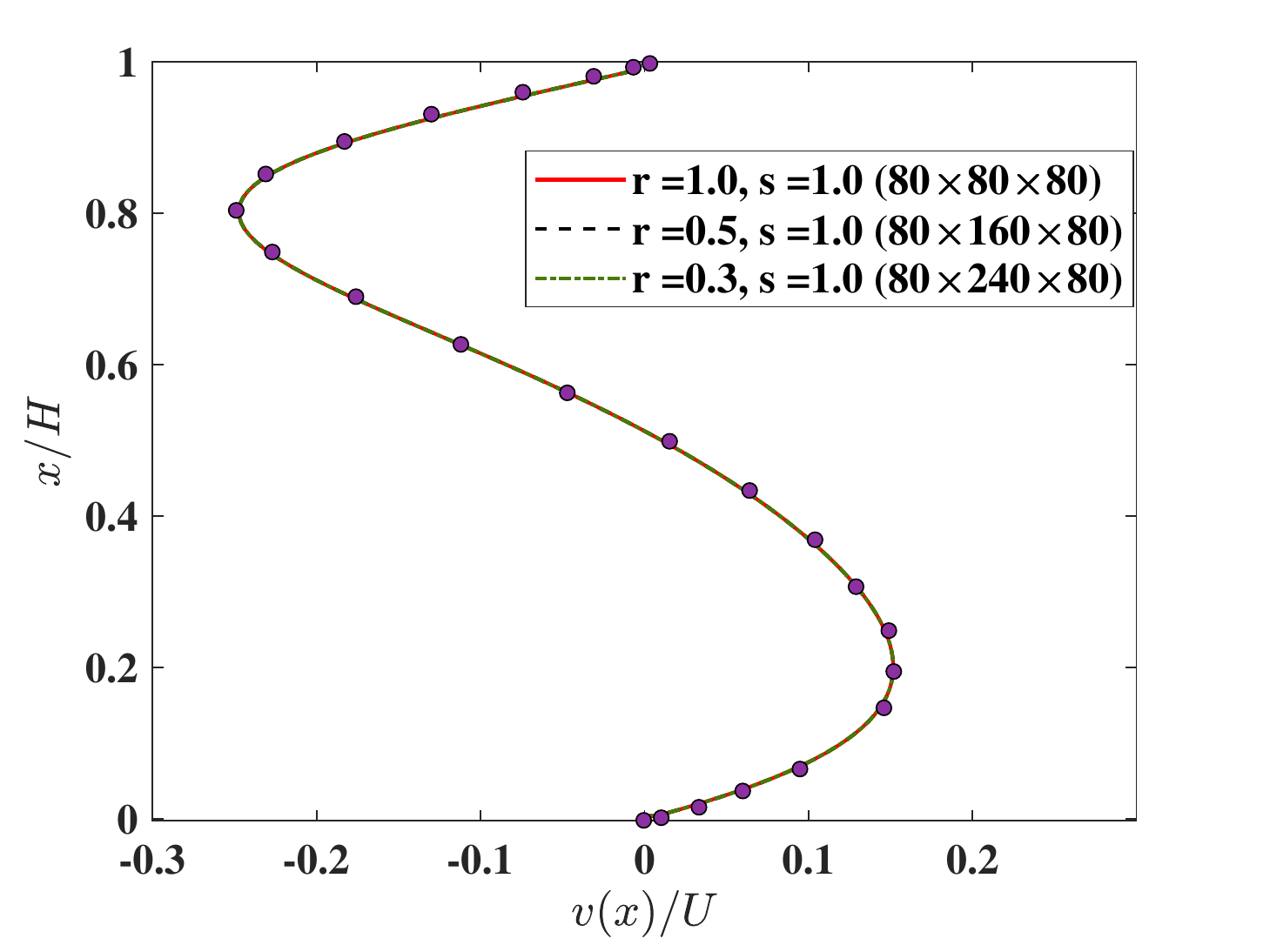}
        \label{fig:3b} } \\
                \advance\leftskip0cm
    \caption{The velocity profiles along the centerlines of the 3D lid driven cavity flow for (a) $u$ component  along the $y$ direction at $x=0.5H$ and $z=0.5H$, and (b) $v$ component along the $x$ direction at $y=0.5H$ and $z=0.5H$ computed using the 3DCCM-LBM using three different aspect ratios and compared with the reference numerical solution from Shu \emph{et al}~\cite{shu2003numerical} (symbols).}
    \label{fig:vprofile}
\end{figure}

\begin{figure}[H]
\centering
\advance\leftskip-1.7cm
    \subfloat[] {
        \includegraphics[width=0.5\textwidth] {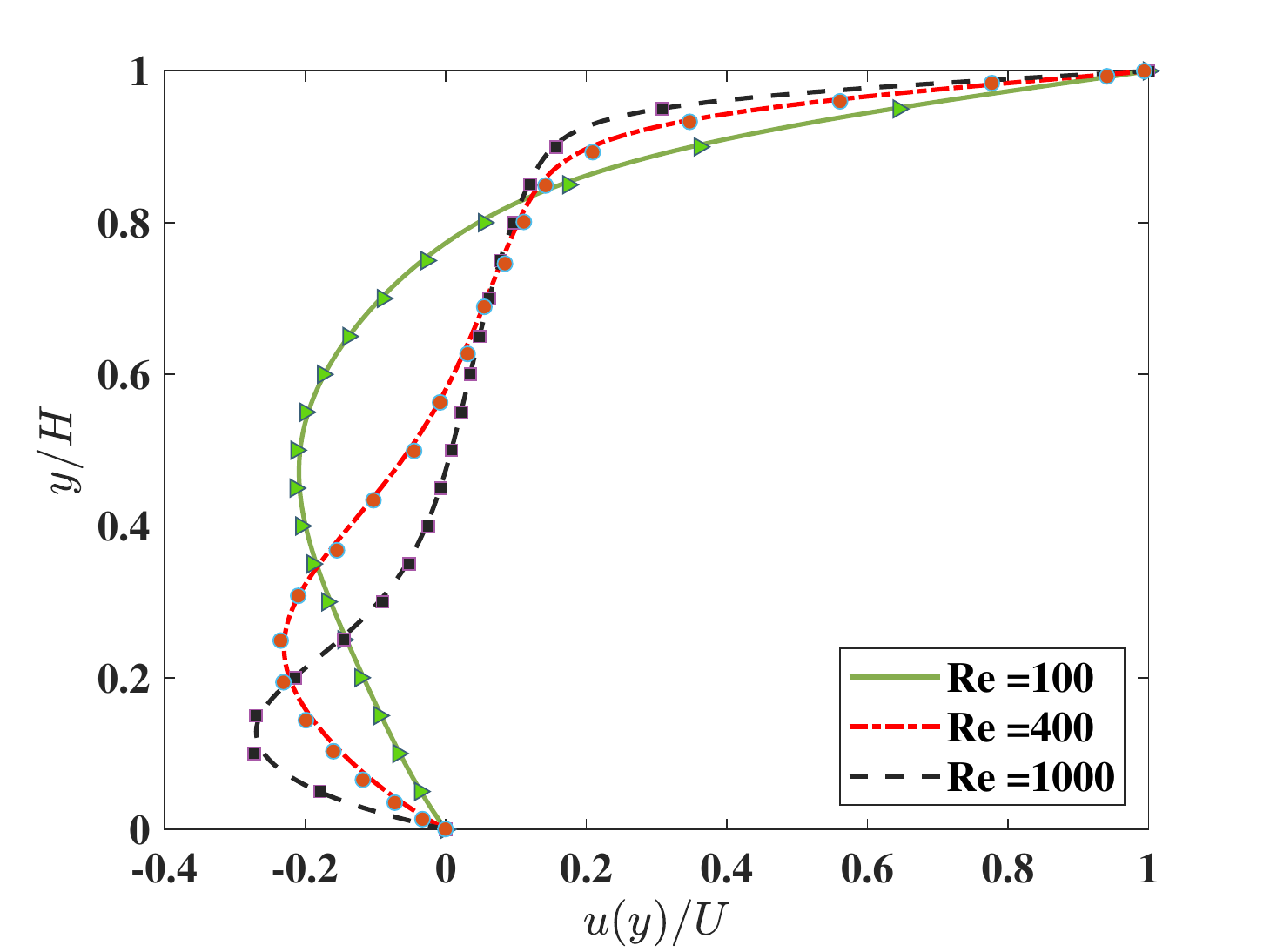}
        \label{fig:4a} } 
    \subfloat[] {
        \includegraphics[width=0.5\textwidth] {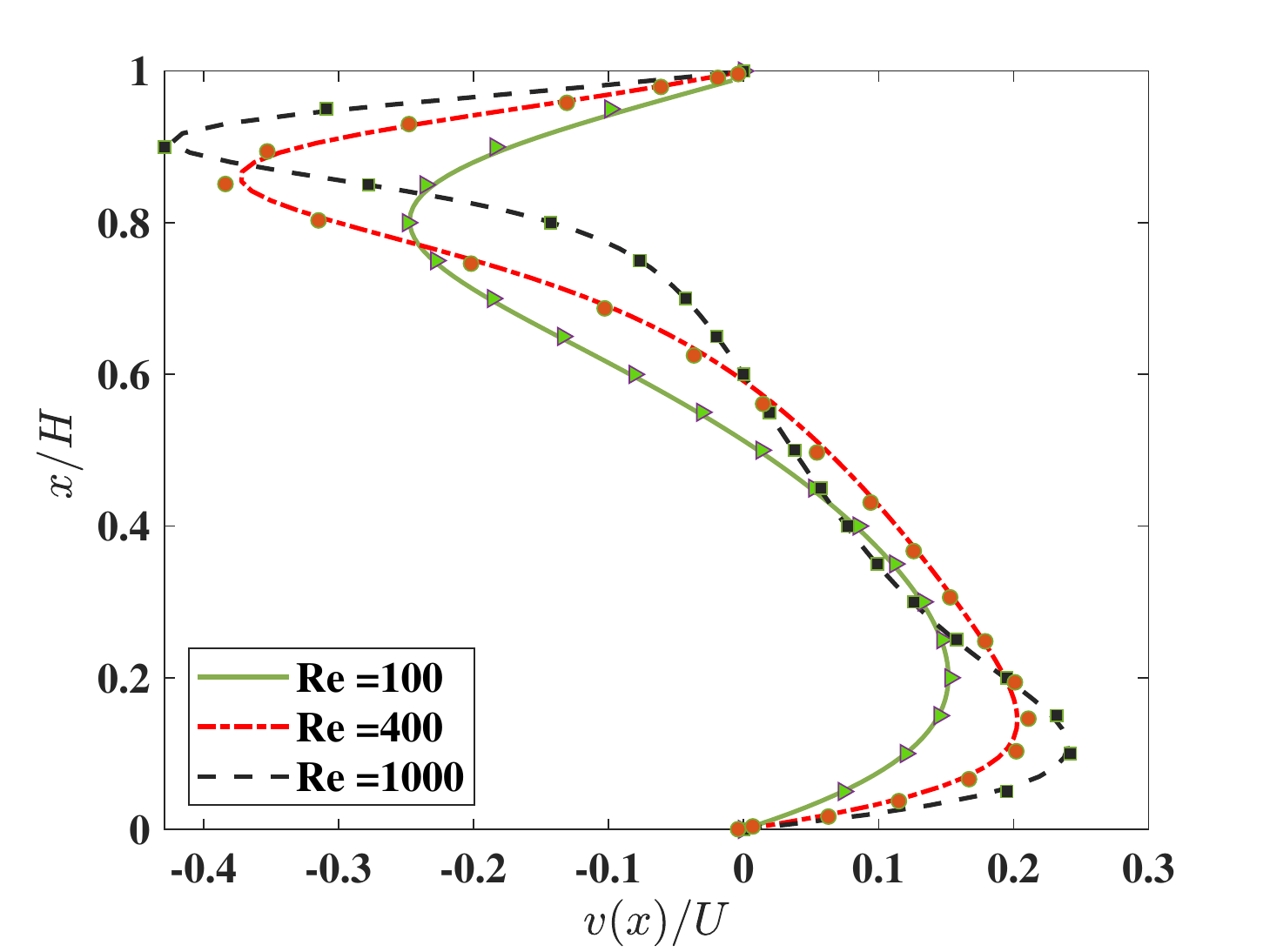}
        \label{fig:4b} } \\
                \advance\leftskip0cm
    \caption{The velocity profiles along the centerlines of the 3D lid driven cavity flow for (a) $u$ component along the $y$ direction at $x=0.5H$ and $z=0.5H$, and (b) $v$ component along the $x$ direction at $y=0.5H$ and $z=0.5H$ computed using the 3DCCM-LBM on a cuboid lattice with the grid aspect ratios of $r=0.5$ and $s=1.0$ at three different Reynolds numbers $\mbox{Re}=100$, $400$ and $1000$ and compared with the reference numerical solution from Shu \emph{et al}~\cite{shu2003numerical} (symbols).}
    \label{fig:Revelocity}
\end{figure}

\begin{figure}[htb]
    \centering 
\begin{subfigure}{0.3\textwidth}
  \includegraphics[width=\linewidth]{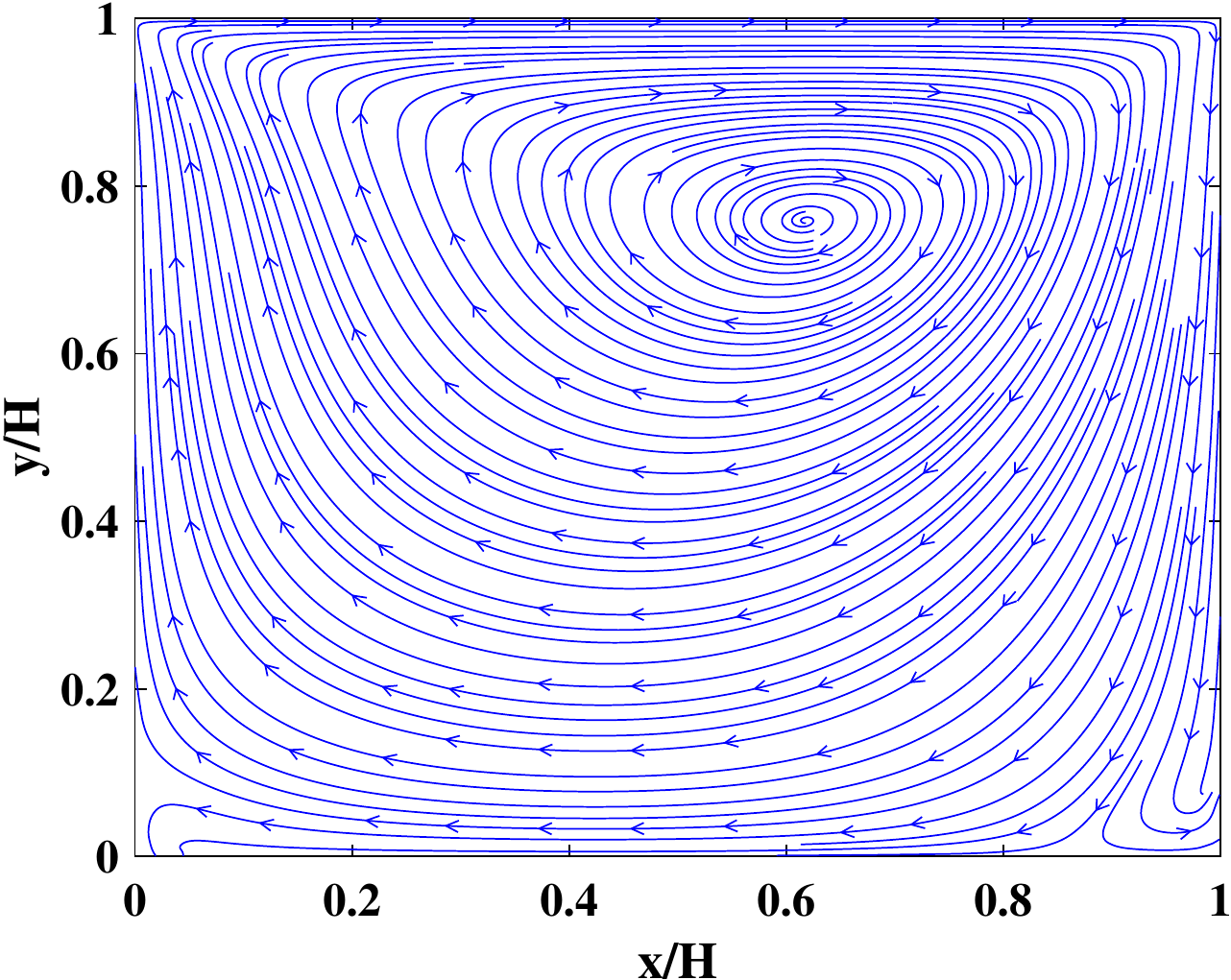}
  \caption{$\mbox{Re}=100, \mbox{along}\; z=0.5H$}
  \end{subfigure}\hfil 
\begin{subfigure}{0.3\textwidth}
  \includegraphics[width=\linewidth]{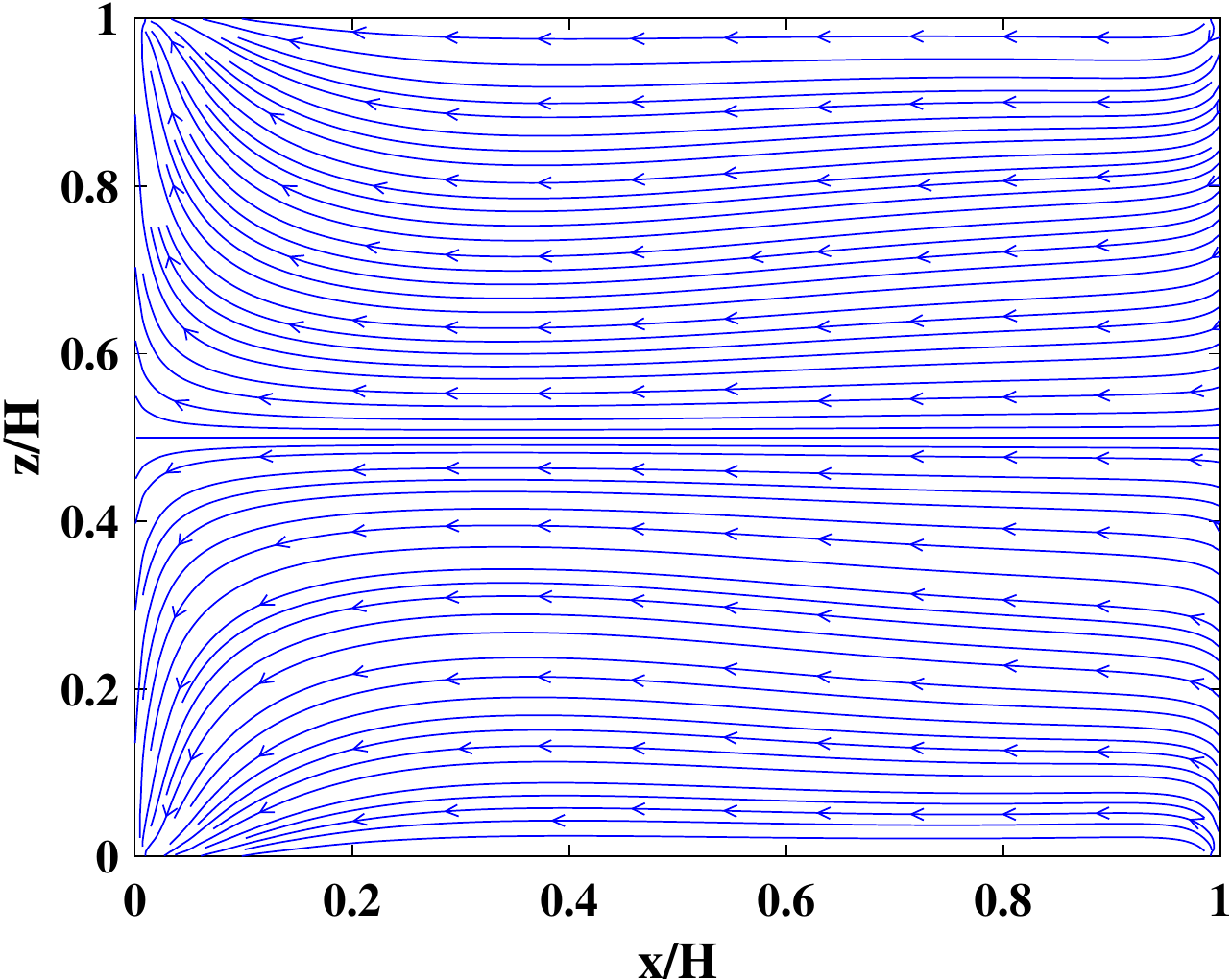}
  \caption{$\mbox{Re}=100, \mbox{along}\; y=0.5H$}
  \end{subfigure}\hfil 
\begin{subfigure}{0.3\textwidth}
  \includegraphics[width=\linewidth]{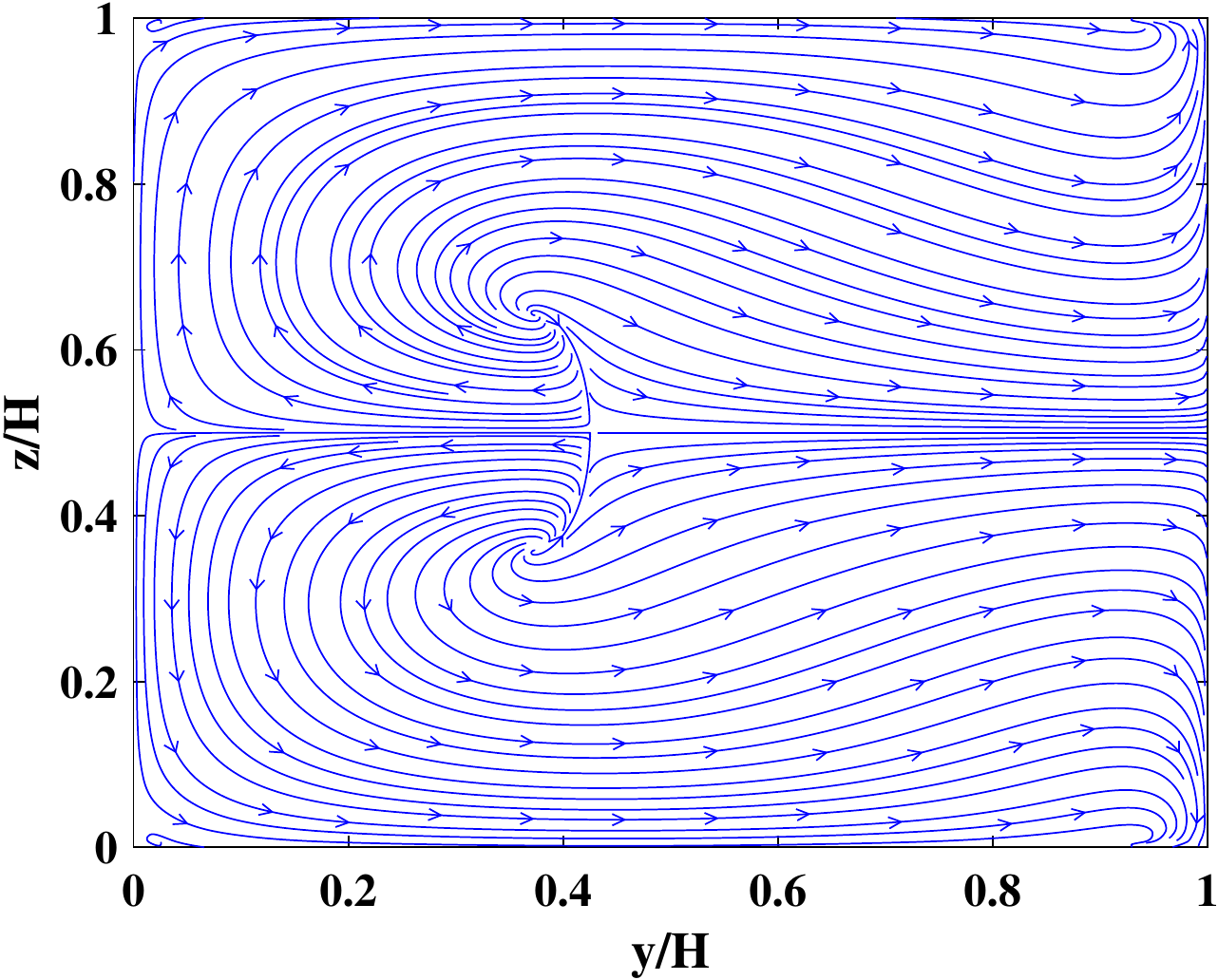}
  \caption{$\mbox{Re}=100, \mbox{along}\; x=0.5H$}
 \end{subfigure}

\medskip
\begin{subfigure}{0.3\textwidth}
  \includegraphics[width=\linewidth]{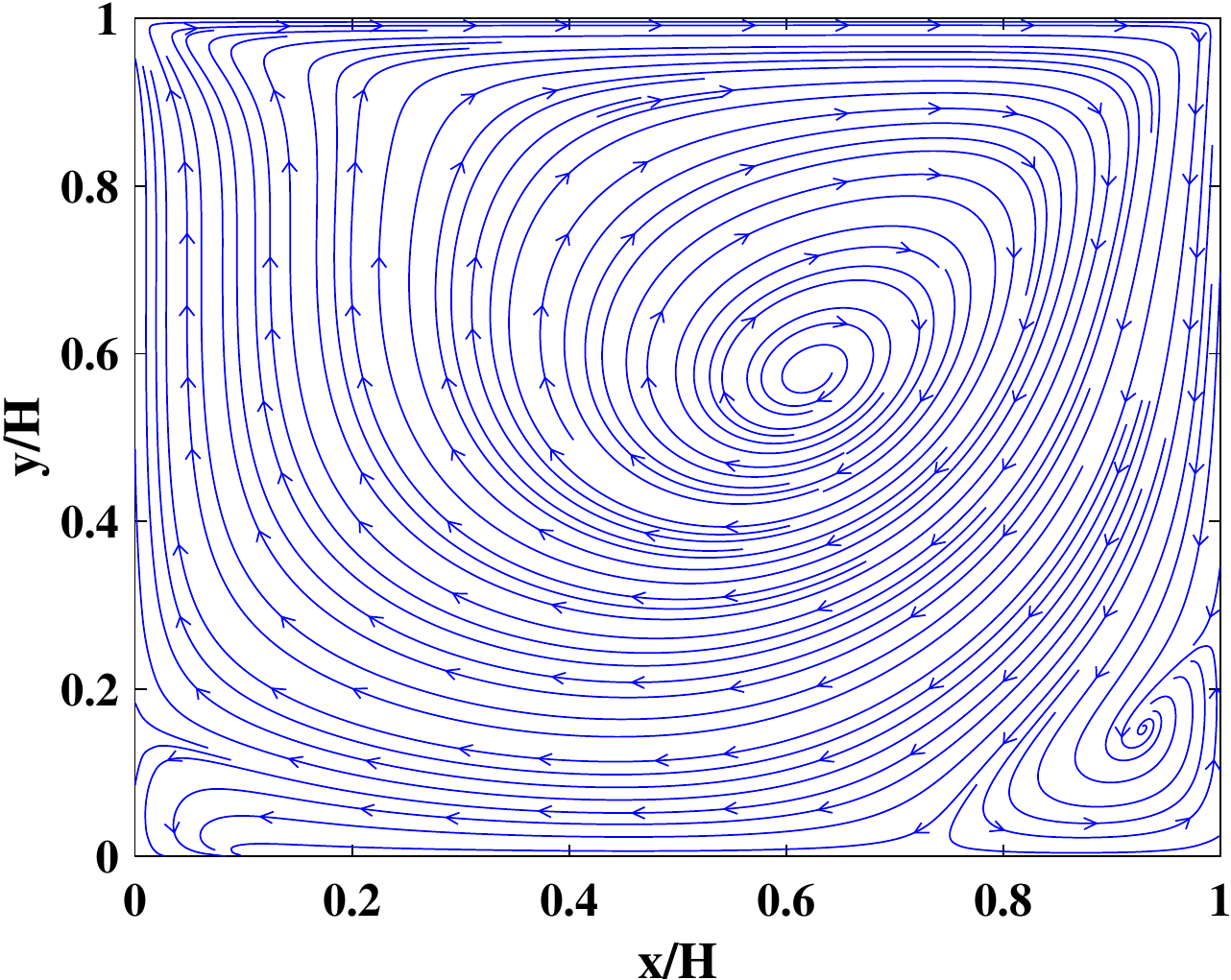}
  \caption{$\mbox{Re}=400, \mbox{along}\; z=0.5H$}
  \end{subfigure}\hfil 
\begin{subfigure}{0.3\textwidth}
  \includegraphics[width=\linewidth]{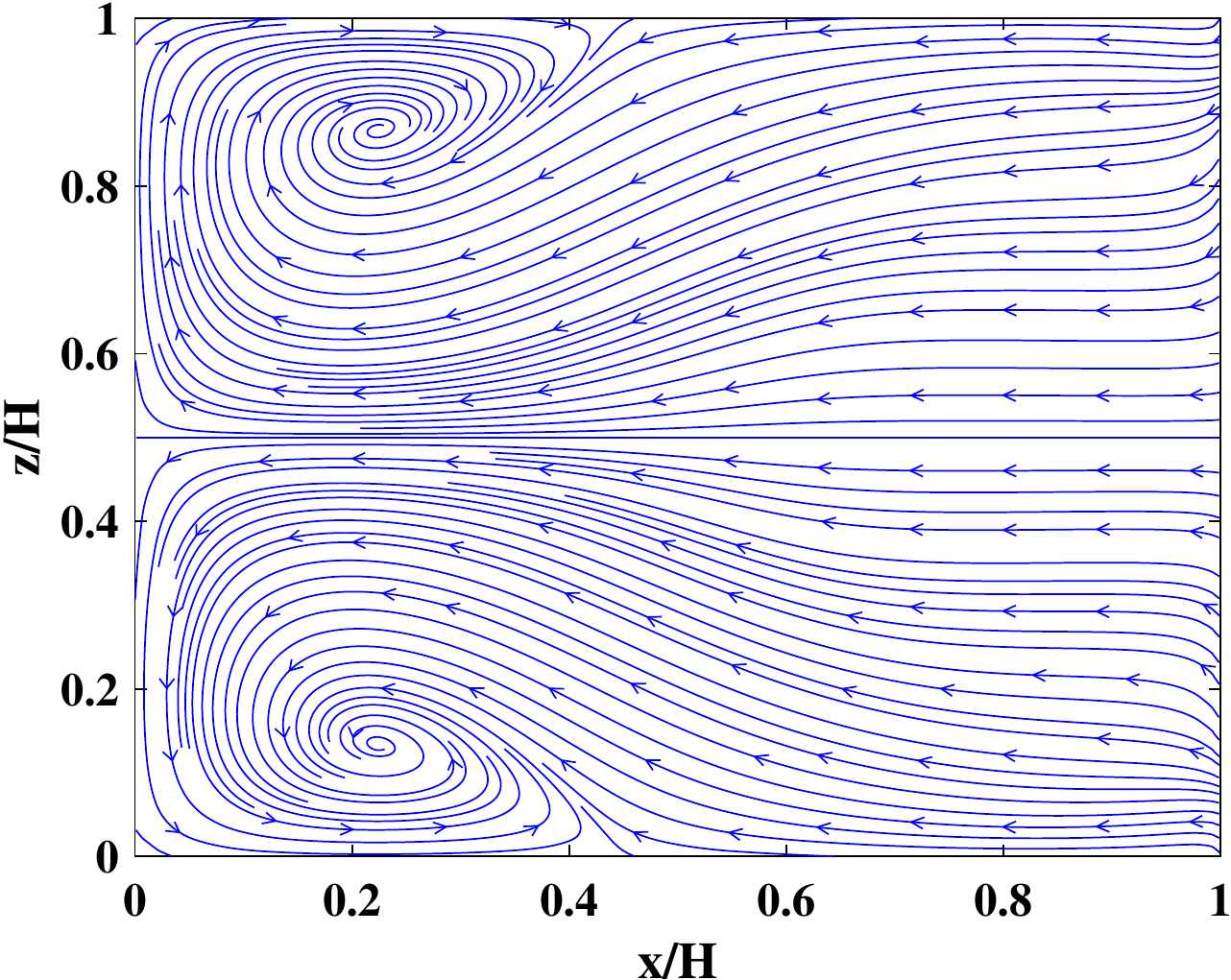}
  \caption{$\mbox{Re}=400, \mbox{along}\; y=0.5H$}
 \end{subfigure}\hfil 
\begin{subfigure}{0.3\textwidth}
  \includegraphics[width=\linewidth]{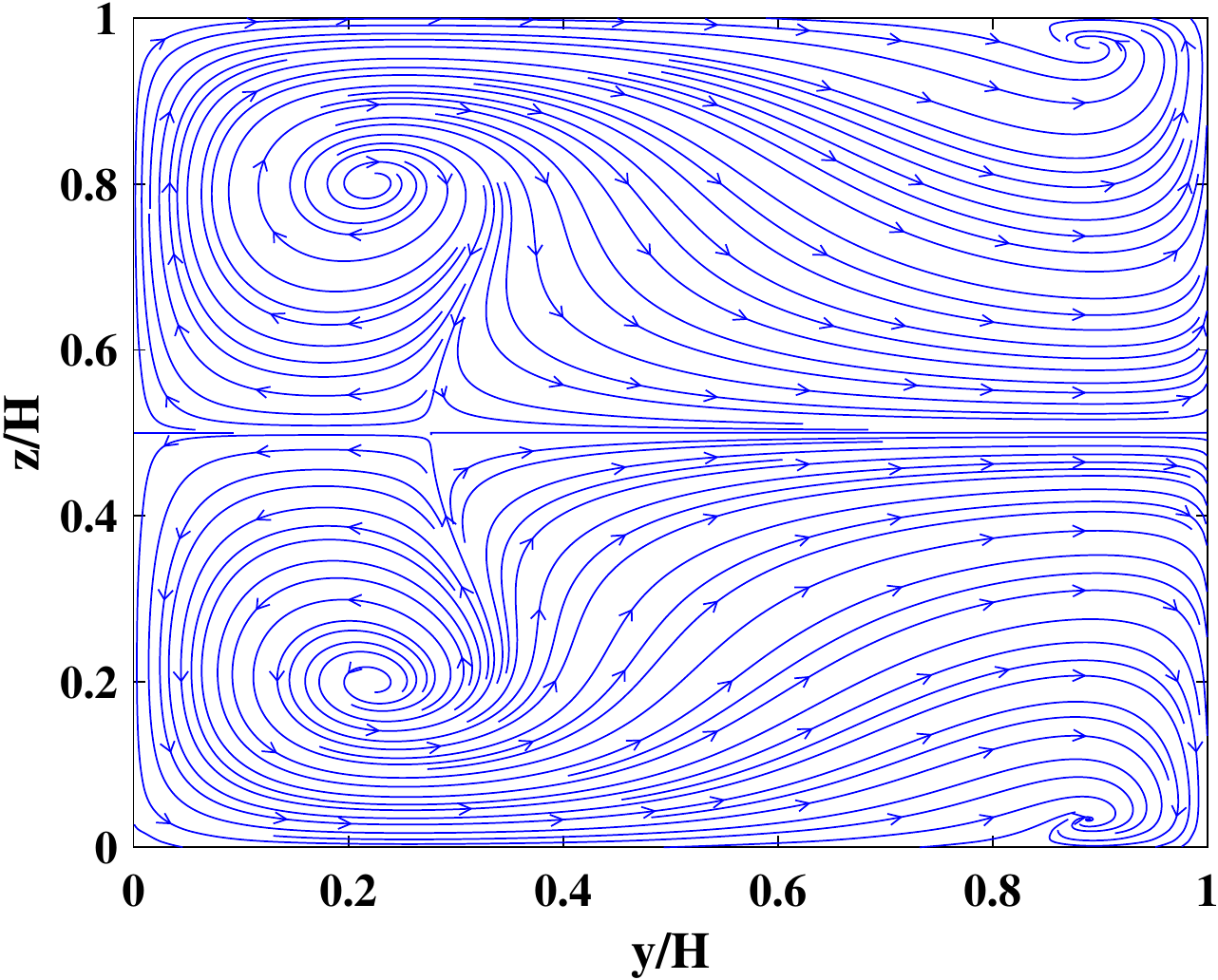}
  \caption{$\mbox{Re}=400, \mbox{along}\; x=0.5H$}
  \end{subfigure}\hfil 

  \medskip
\begin{subfigure}{0.3\textwidth}
  \includegraphics[width=\linewidth]{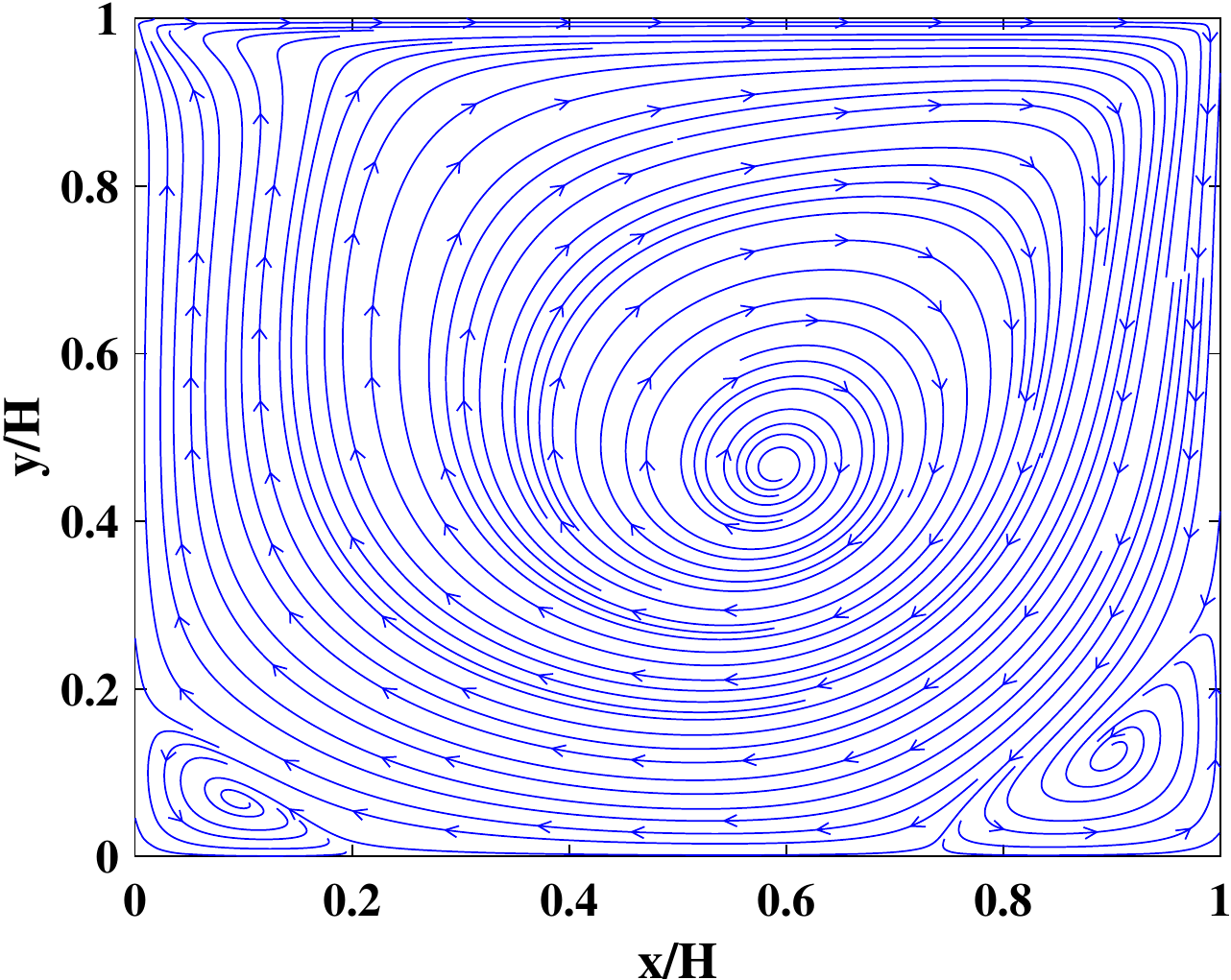}
  \caption{$\mbox{Re}=1000, \mbox{along}\; z=0.5H$}
  \end{subfigure}\hfil 
\begin{subfigure}{0.3\textwidth}
  \includegraphics[width=\linewidth]{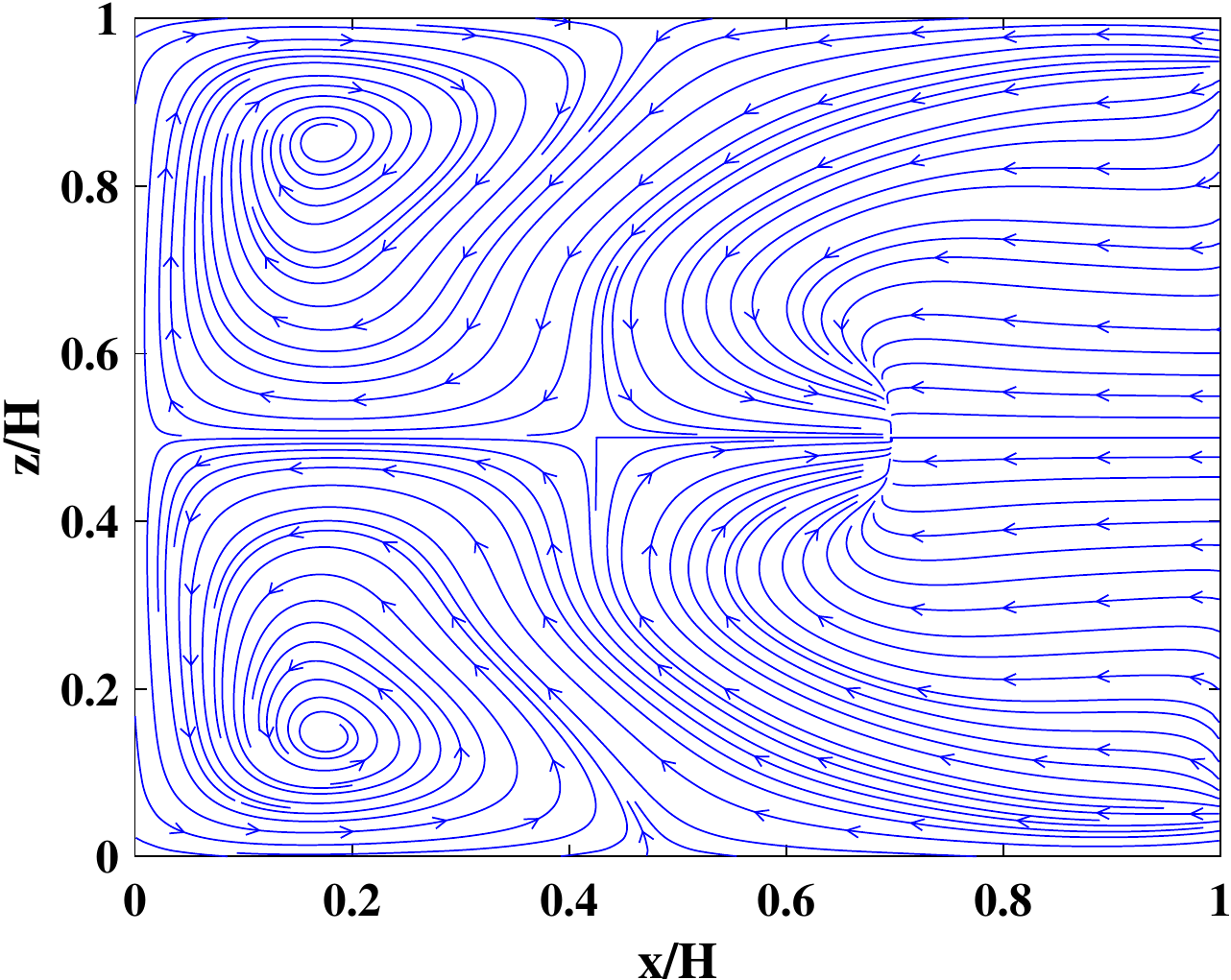}
  \caption{$\mbox{Re}=1000, \mbox{along}\; y=0.5H$}
  \end{subfigure}\hfil 
\begin{subfigure}{0.3\textwidth}
  \includegraphics[width=\linewidth]{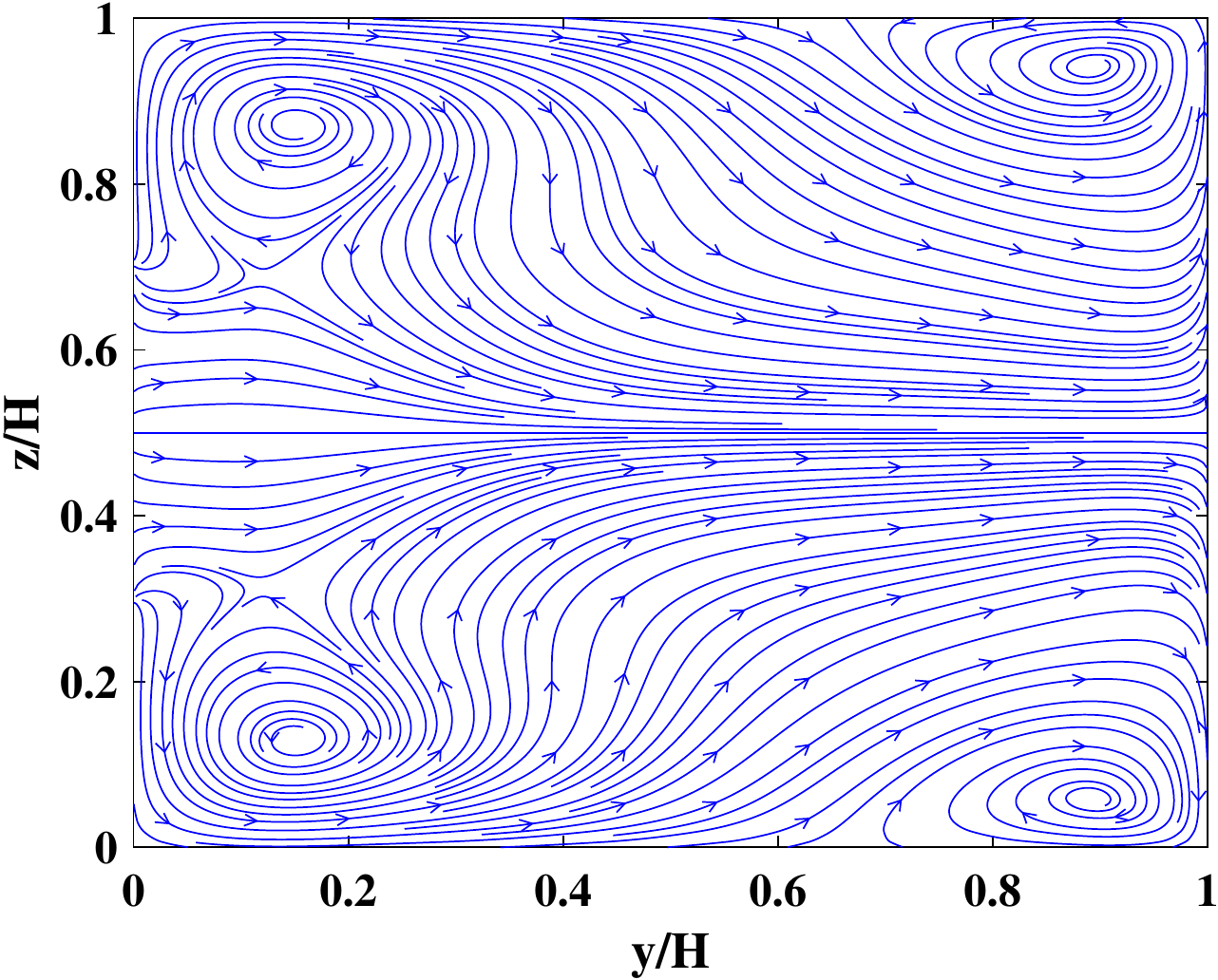}
  \caption{$\mbox{Re}=1000, \mbox{along}\; x=0.5H$}
  \end{subfigure}\hfil 
  \caption{Projections of the streamlines in the 3D lid-driven cubic cavity flow along three different mid-planes computed using 3DCCM-LBM with a grid aspect ratio of $r=0.5$ and $s=1.0$ at \mbox{Re}=100 (a,b,c), \mbox{Re}=400 (d,e,f) and \mbox{Re}=1000 (g,h,i) at $z=0.5H$, $y=0.5H$, and $x=0.5H$, respectively.}
\label{fig:Restream}
\end{figure}

Finally, we investigate our 3DCCM-LBM for simulating a fully 3D shear flow case with complex flow features for which only some prior benchmark numerical solutions are available. In this regard, we consider the motion of the fluid enclosed within a cubic cavity of side length $H$ driven by the motion of the top lid located parallel to the $x-z$ plane at $y=H$ at a velocity $U$ along the $x$ direction. Such a shear-driven flow involves circulation patterns with 3D vortical structures whose details depend on the Reynolds number $\mbox{Re}$ defined as $\mbox{Re}=UH/\nu$~\cite{shankar2000fluid}. In this paper, we will perform simulation of this flow at $\mbox{Re}=100, 400$ and $1000$ for which the reference numerical solutions for benchmarking can be found from various prior studies (see e.g.~\cite{ku1987pseudospectral,jiang1994large,shu2003numerical}). For applying the 3DCCM-LBM in this regard, we used the following three different grid aspect ratios $r=\left\{1.0, 0.5, 0.33\right\}$ and $s=1$, so that the resolutions are selectively varied only along the $y$ direction which is normal to the shearing motion of the top lid. The choices made for the various parameters for each set of grid aspect ratios, including the number of grid nodes, lid velocity, speed of sound, and the relaxation time $\tau=1.0/\omega_\nu$ are presented in Table~\ref{tab:2}.
\begin{table}[H]
\centering
\captionsetup{justification=centering}
\caption{Parameters used in the simulation of 3D lid-driven cubic cavity flow using 3DCCM-LBM.}
\begin{tabular}{c c | c c | c c c c}
\hline
 \mbox{Re} & $N_x \times N_y\times N_z$ & $r$ & $s$ & $U$ &  $c_s^2$  & $\tau$ & $\nu$ \\
\hline
\hline
\multirow{3}{4em}{100} & $80 \times 80\times 80$ & 1.0 & 1.0 & 0.1 & 0.3333   & 0.737   & 0.079\\
 &$80 \times 160\times 80$ & 0.5 & 1.0 & 0.05 & 0.14   & 0.782    & 0.039\\
 &$80 \times 240\times 80$ & 0.33 & 1.0 & 0.04 & 0.06   & 1.023    & 0.031  \\
\hline
\multirow{1}{4em}{400} & $80 \times 160\times 80$ & 0.5 & 1.0 & 0.1 & 0.3333   & 0.737   & 0.079\\
\hline
\multirow{1}{4em}{1000} & $80 \times 160\times 80$ & 0.5 & 1.0 & 0.1 & 0.3333   & 0.737   & 0.079 \\
\hline
\end{tabular}
\label{tab:2}
\end{table}
The no-slip boundary conditions are implemented using the half-way bounce back scheme; in particular, for representing the effect of the moving top lid, we impose the momentum-augmented corrections, which are parameterized by the grid aspect ratios as shown in Eq.~(\ref{eq:69}). Figure \ref{fig:vprofile} presents the steady state velocity profiles of the horizontal velocity component $u$ and the vertical velocity component $v$ along the vertical and horizontal geometric centerlines, respectively, in the mid $x-y$ plane at $z=0.5H$ at a fixed Reynolds number $\mbox{Re}=100$, and using the 3DCCM-LBM with three different grid aspect ratios as indicated in Table~\ref{tab:2} earlier. Here, we compared them with the benchmark numerical solution based on the method of differential quadrature to solve the incompressible NS equation in Shu \emph{et al}~\cite{shu2003numerical}. It is evident that our 3D central moment LBM based on the cuboid lattice is in very good agreement with the reference numerical data.
Moreover, Fig.~\ref{fig:Revelocity} shows the comparisons made for the velocity profiles at a fixed grid aspect ratios of $r= 0.5$ and $s=1.0$ for three different Reynolds numbers of $100$, $400$ and $1000$. Our 3DCCM-LBM is able to reproduce the reference results quite well for all the values of $\mbox{Re}$ considered. Thus, the use of the extended moment equilibria involving the correction terms in our LB algorithm effectively eliminates the anisotropy associated with the use of the cuboid lattice and computes solutions that are fully consistent with the 3D NS equations.
In addition, for visualizing the overall flow patterns, the streamlines along the mid- $x-y$ plane, $x-z$ plane and $y-z$ plane of the cubic cavity at $\mbox{Re}=100$, $400$ and $1000$ computed using $r=0.5$ and $s=1$ are presented in Fig.~\ref{fig:Restream}. The motion of the lid normal to the $y$ direction is seen to set up a major vortex in $x-y$ plane whose size and its center location appear to significantly change as the Reynolds number is varied. In particular, the eye of this vortex moves towards the center in the $x-y$ plane as $\mbox{Re}$ increases. Meanwhile, some secondary vortices start to form around the corners at higher $\mbox{Re}$ and grow in size with increase in the inertial effects as the Reynolds number is increased. The three-dimensionality of the resulting flow field is evident from observing the other $y-z$ and $x-z$ planes in these figures. These features are in agreement with the prior numerical solutions~\cite{ku1987pseudospectral,jiang1994large,shu2003numerical}.

\section{Demonstration of computational advantages of using cuboid lattice over cubic lattice: 3D anisotropic shear flows in a lid-driven shallow cuboid cavity} \label{sec:6}
\begin{figure}[H]
\centering
 \includegraphics[width=0.5\textwidth] {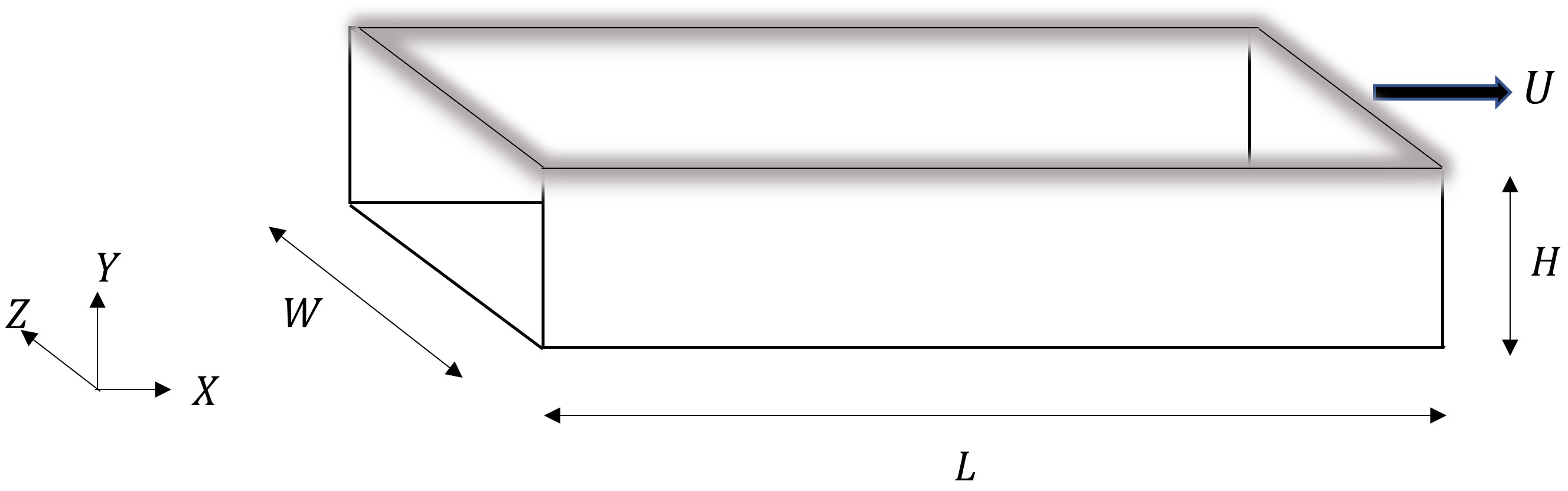}
 \caption{Schematic of the 3D shallow cuboid cavity of dimensions $L\times H \times W$.}
 \label{fig:schematicshallowcavity}
\end{figure}

The results from the above case studies show the ability of our 3DCCM-LBM in computing the flow fields accurately for a variety of standard benchmark flow problems. Now, we will demonstrate the advantages of using the cuboid lattice over the cubic lattice in more efficiently simulating a flow case study characterized by an anisotropic and inhomogeneous shear flow where their spatial gradients in one or more direction are larger than those in the other directions. In this regard, we consider a 3D cuboid cavity of span length $L$ in the
$x$ direction, height $H$ in the $y$ direction, and width $W$ in the $z$ direction enclosing a fluid of viscosity $\nu$. A flow is set up by the shearing motion of the lid located in the $x-z$ plane at $y=H$ at a velocity $U$ (see Fig.~\ref{fig:schematicshallowcavity}).
In particular, we choose the span-to-height ratio $L/H = 4$, and width-to-height ratio $W/H=2$ so that the cavity block is relatively shallow in nature, and the flow generated by this set up is inhomogeneous and anisotropic, with the spatial gradients along the $y$ direction that is normal to the direction of shear expected to be larger than those in the $x$ and $z$ directions. If $N_x\times N_y\times N_z$ represent the total number of grid nodes resolving the flow domain, the grid space steps in each of the directions are given by $\Delta x = L/N_x$, $\Delta y = H/N_y$ and $\Delta z = W/N_z$. In the case of using a cubic lattice, they are constrained by the uniform grid spacing requirement, i.e., $\Delta x = \Delta y = \Delta z$, which fixes the number of grid points in each coordinate direction according to the length dimension of the cuboid cavity in that direction, i.e., $N_x/L = N_y/H = N_z/W$. As an example, we consider a flow at a Reynolds number $\mbox{Re}=UL/\nu=100$ within this cavity block, and choose the number of grid nodes in the $y$ direction normal to the shear as $N_y = 64$. Based on the above considerations, for the cubic lattice, we obtain $N_x=256$ and $N_z=128$, and hence, the total number of grid nodes in this case is $256\times 64\times 128$. On the other hand, when the cuboid lattice is used $\Delta y = r\Delta x$ and $\Delta z = s\Delta x = (s/r)\Delta y$, so that $N_x = r(L/H)N_y$ and $N_z = (1/s)(W/L)N_x = (r/s)(W/H)N_y$. Taking $N_y=64$ again along the direction normal to the shearing motion of the lid, for the purpose of illustration, we consider the following three different possibilities for the grid aspect ratios along with the corresponding resolution requirements in the $x$ and $z$ directions: (i) $(r,s)= (0.406,1.3)$ with $(N_x,N_z)=(104,40)$, (ii) $(r,s)= (0.375,0.75)$ with $(N_x,N_z)=(96,64)$, and (iii) $(r,s)= (0.375,1.0)$ with $(N_x,N_z)=(96,48)$. Thus, the total number of grid nodes with using the cuboid lattice for the above three cases are $104\times 64\times 40$, $96\times 64\times 64$ and $96\times 64\times 48$, respectively. Clearly, the resolution requirements with using the cubic lattice involving the same grid size in all the directions are significantly greater than those with using the choices made above with the flexible cuboid lattice that can more naturally conform with the flow characteristics. For performing flow simulations, we choose $c_s^2=1/3$ for the cubic lattice and $c_s^2=0.04$ for the cuboid lattice.

\begin{figure}[H]
\centering
\advance\leftskip-1.7cm
    \subfloat[$x=0.5L$ and $z=0.5W$] {
        \includegraphics[width=.5\textwidth] {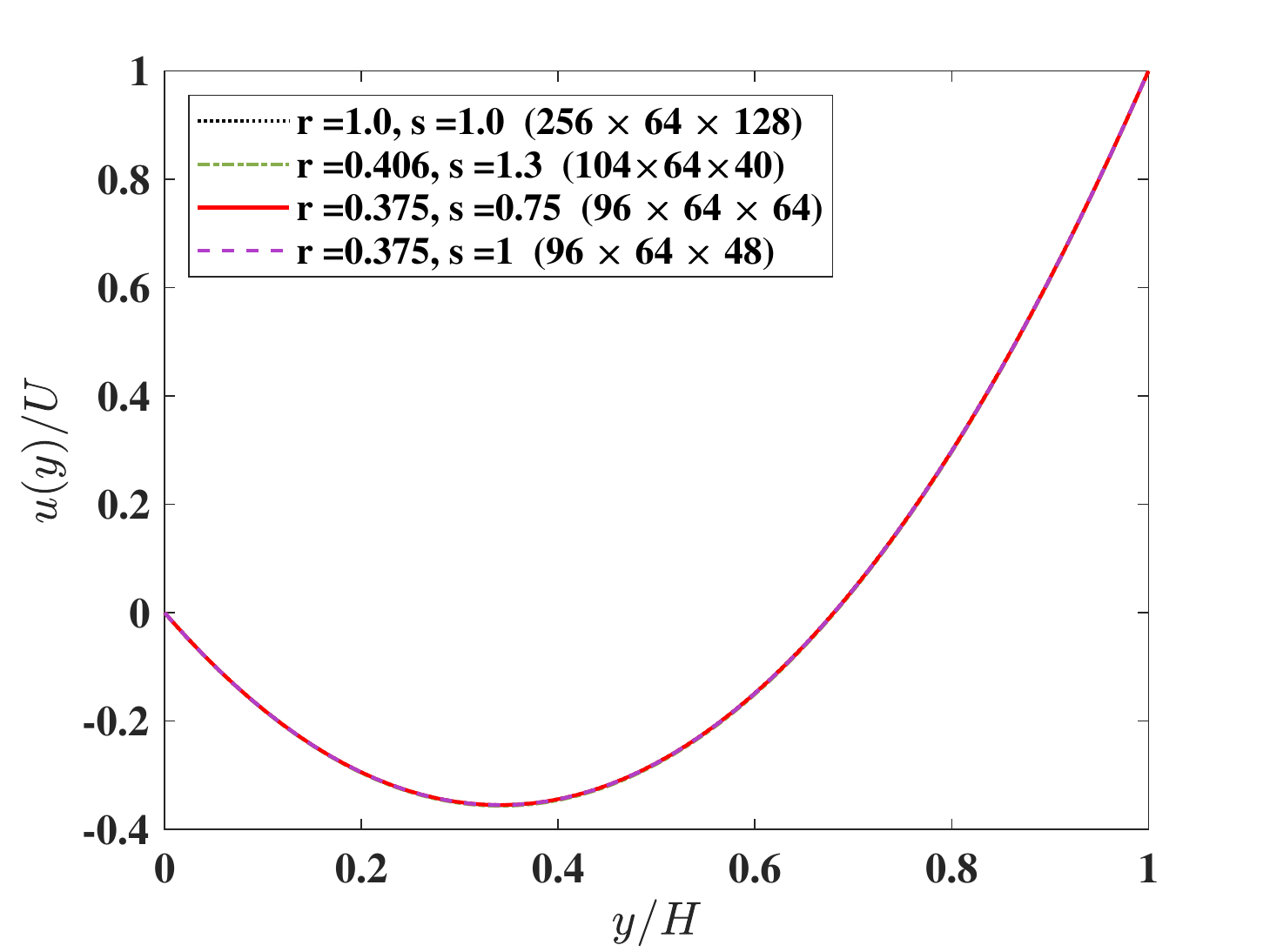}
        \label{fig:6a} } 
    \subfloat[$y=0.5H$ and $z=0.5W$] {
        \includegraphics[width=.5\textwidth] {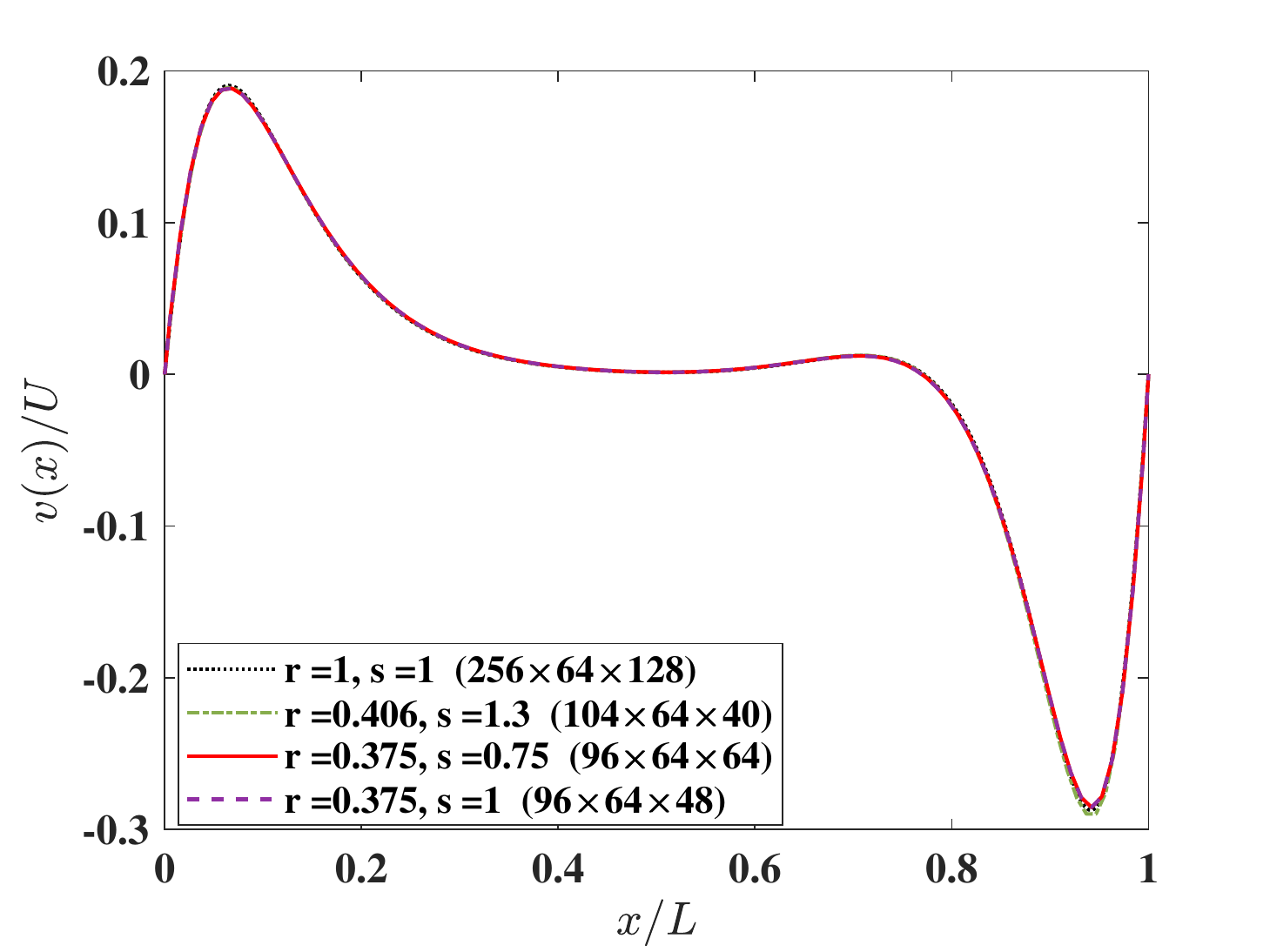}
        \label{fig:6b} } 
        \advance\leftskip0cm
    \caption{The velocity profiles along the centerlines of a 3D shallow lid driven cavity of aspect ratio $L/H=2$ and $L/W=4$ at a Reynolds number $\mbox{Re}=100$ computed using 3DCCM-LBM with the cubic lattice $(r=s=1)$ and a cuboid lattice with the following three different choices o fthe grid aspect ratios: $(r,s)= (0.406,1.3)$, $(r,s)= (0.375,0.75)$, and $(r,s)= (0.375,1.0)$. (a) $u$ component along the $y$ direction at $x=0.5L$ and $z=0.5W$, and (b) $v$ component along the $x$ direction at $y=0.5H$ and $z=0.5W$.}
    \label{fig:6}
\end{figure}

\begin{figure}[H]
\centering
\advance\rightskip-1.3cm
    \subfloat[Plane at $z=0.5W$ (cubic lattice)] {
        \includegraphics[width=.5\textwidth] {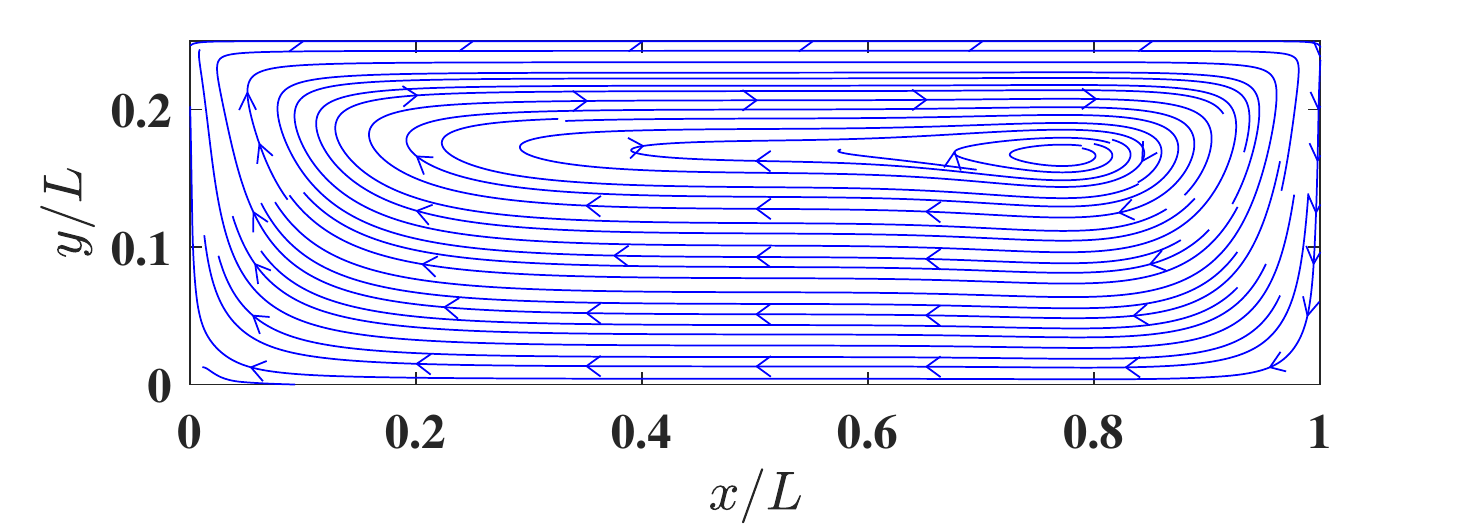}
        \label{fig:7a} } 
       \subfloat[Plane at $z=0.5W$ (cuboid lattice)] {
        \includegraphics[width=.5\textwidth] {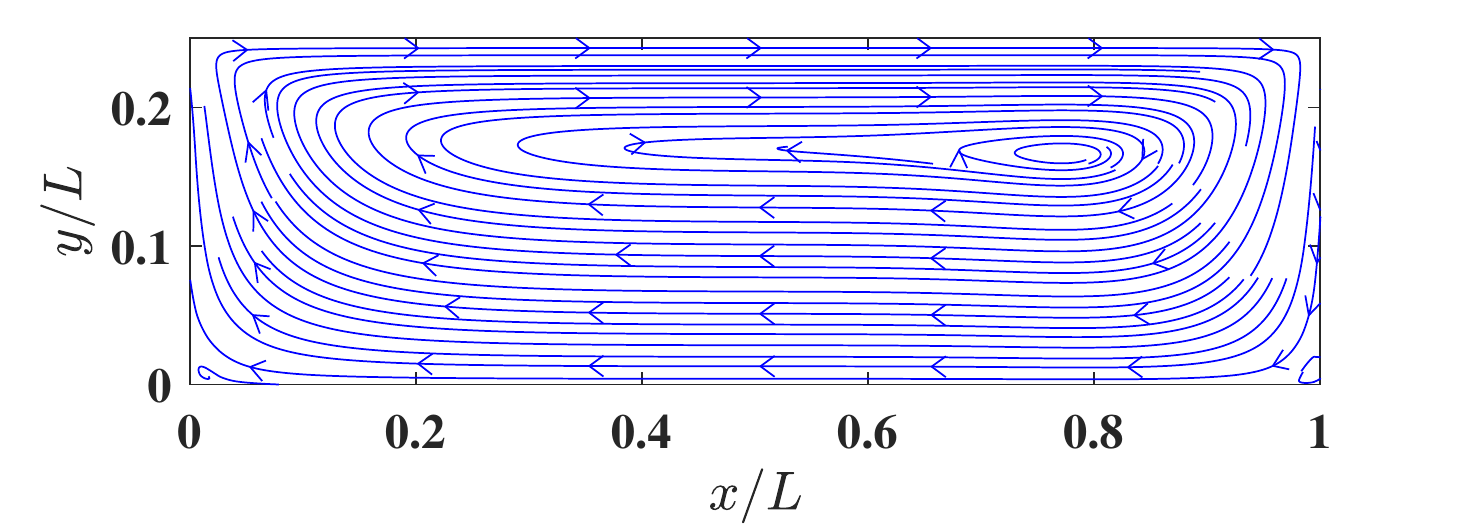}
        \label{fig:7b} } 
        \\
        \subfloat[Plane at $y=0.5H$ (cubic lattice)] {
        \includegraphics[width=.5\textwidth] {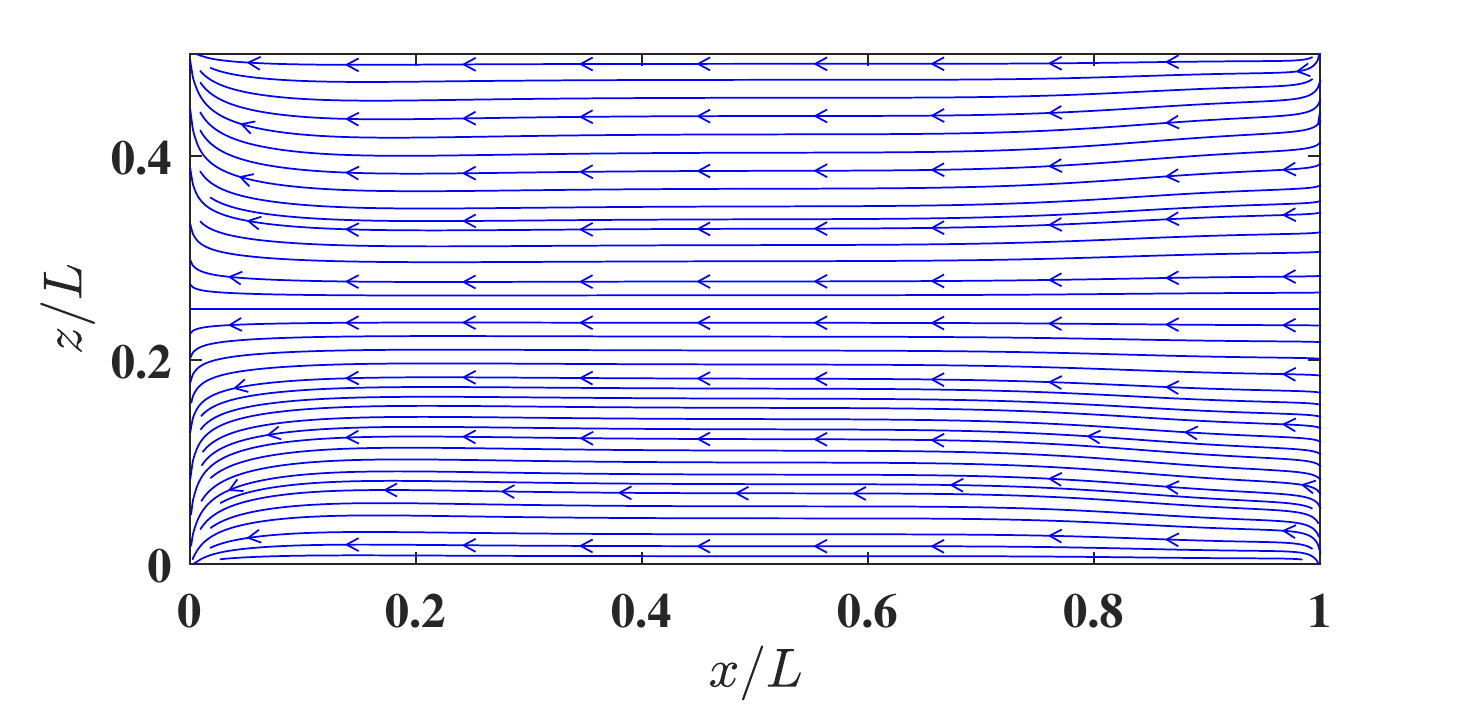}
        \label{fig:7c} } 
       \subfloat[Plane at $y=0.5H$ (cuboid lattice)] {
        \includegraphics[width=.5\textwidth] {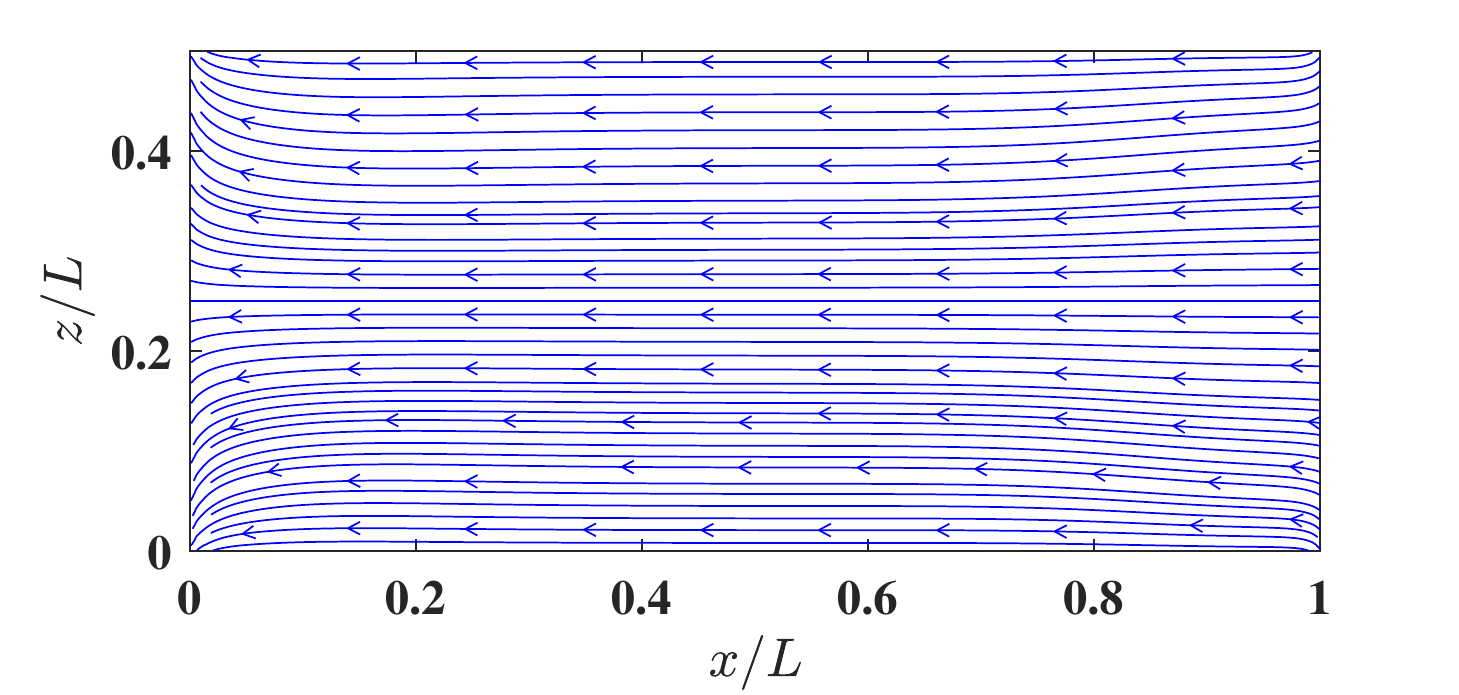}
        \label{fig:7d} } 
         \\
        \subfloat[Plane at $x=0.5L$ (cubic lattice)] {
        \includegraphics[width=.3\textwidth] {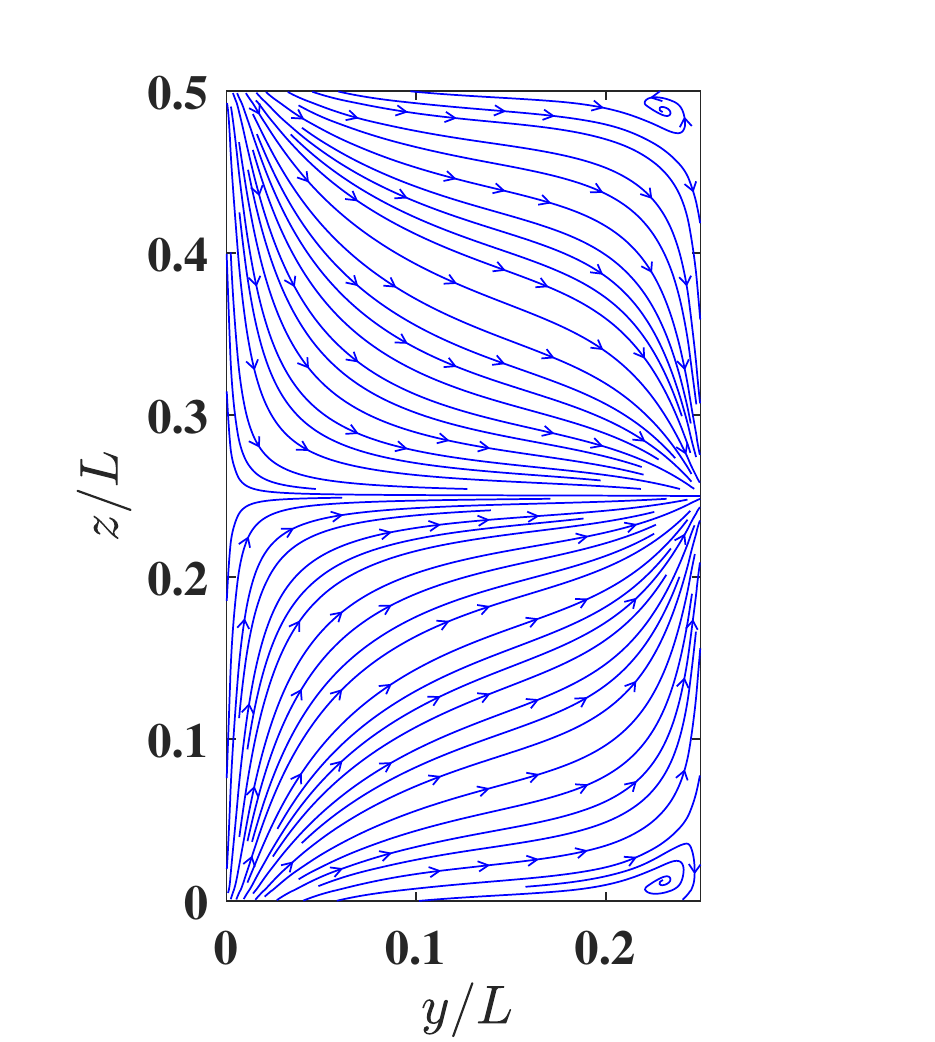}
        \label{fig:7e} } \hspace*{10em}
       \subfloat[Plane at $x=0.5L$ (cuboid lattice)] {
        \includegraphics[width=.3\textwidth] {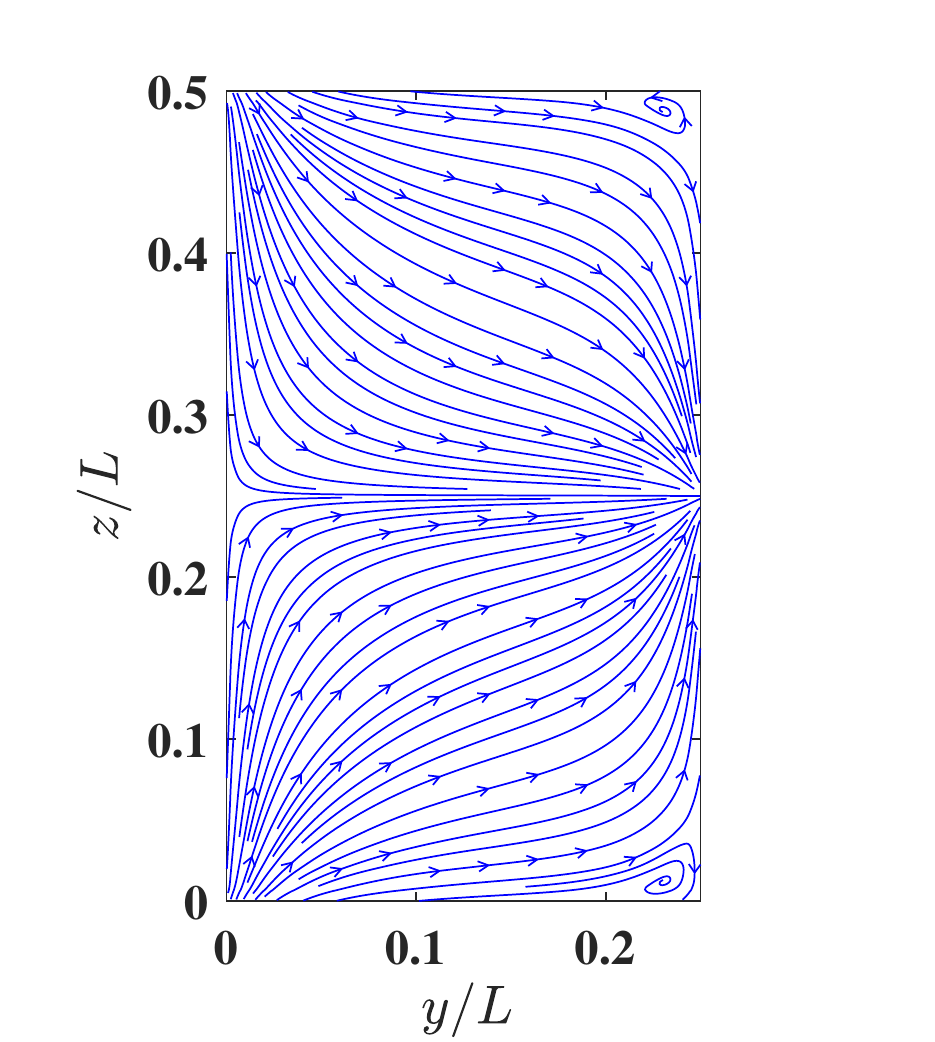}
        \label{fig:7f} } 
        \advance\leftskip0cm
    \caption{Comparison of streamline patterns along the mid-planes of a 3D shallow lid driven cavity of aspect ratio $L/H=2$ and $L/W=4$ at a Reynolds number $\mbox{Re}=100$ computed using 3DCCM-LBM with the cubic lattice ($(r,s)=(1.0,1.0)$) using a grid resolution of $256\times128\times64$ shown in sub-figures (a), (c) and (e)  and compared with the 3DCCM-LBM with the cuboid lattice ($(r,s)=(0.406,1.3)$) using a grid resolution of $104\times64\times40$ shown in sub-figures (b), (d) and (f).}
    \label{fig:7}
\end{figure}

Figure~\ref{fig:6} presents comparisons of the velocity profiles of the $u$ component along the $y$ direction at $x=0.5L$ and $z=0.5W$, and $v$ component along the $x$ direction at $y=0.5H$ and $z=0.5W$ obtained using the 3DCCM-LBM with the above three sets of grid aspect ratios for the cuboid lattice and the cubic lattice with $(r,s)=(1.0,1.0)$. Moreover, the streamlines along the three mid-planes of the 3D shallow cuboid cavity using both the cubic lattice and the cuboid lattice using the grid aspect ratios for the case (i) (corresponding to that with the lowest total number of grid points among the three cuboid lattice cases) are shown in Fig.~\ref{fig:7}. Clearly, the cuboid lattice results, while using significantly fewer total number of grid nodes, are in excellent agreement with those using the cubic lattice. In particular, to achieve practically similar accuracy for the numerical results, the flexibility accorded by the cuboid lattice yielded significant savings in the computer memory storage by a factor of $7.88$, $5.33$ and $7.11$ for cases (i), (ii), and (iii), respectively, and about similar reductions in the computational cost for the simulation turnaround time when compared to the cubic lattice. While for the specific flow configuration considered here a reduction in the computational resource requirements by a factor between 5 to 7.5 was achieved for the choices made for the grid aspect ratios, if the flow geometry happens to be further skewed in the different coordinate directions, e.g., when $H/L \ll 1$ and/or $H/W \ll 1$, additional improvements with using the cuboid lattice can be expected for simulating such anisotropic shear flow problems.

\section{Numerical stability test results of raw moment and central moment based LB algorithms on the D3Q27 cuboid lattice}\label{sec:7}
\begin{figure}[H]
\centering
 \includegraphics[width=0.5\textwidth] {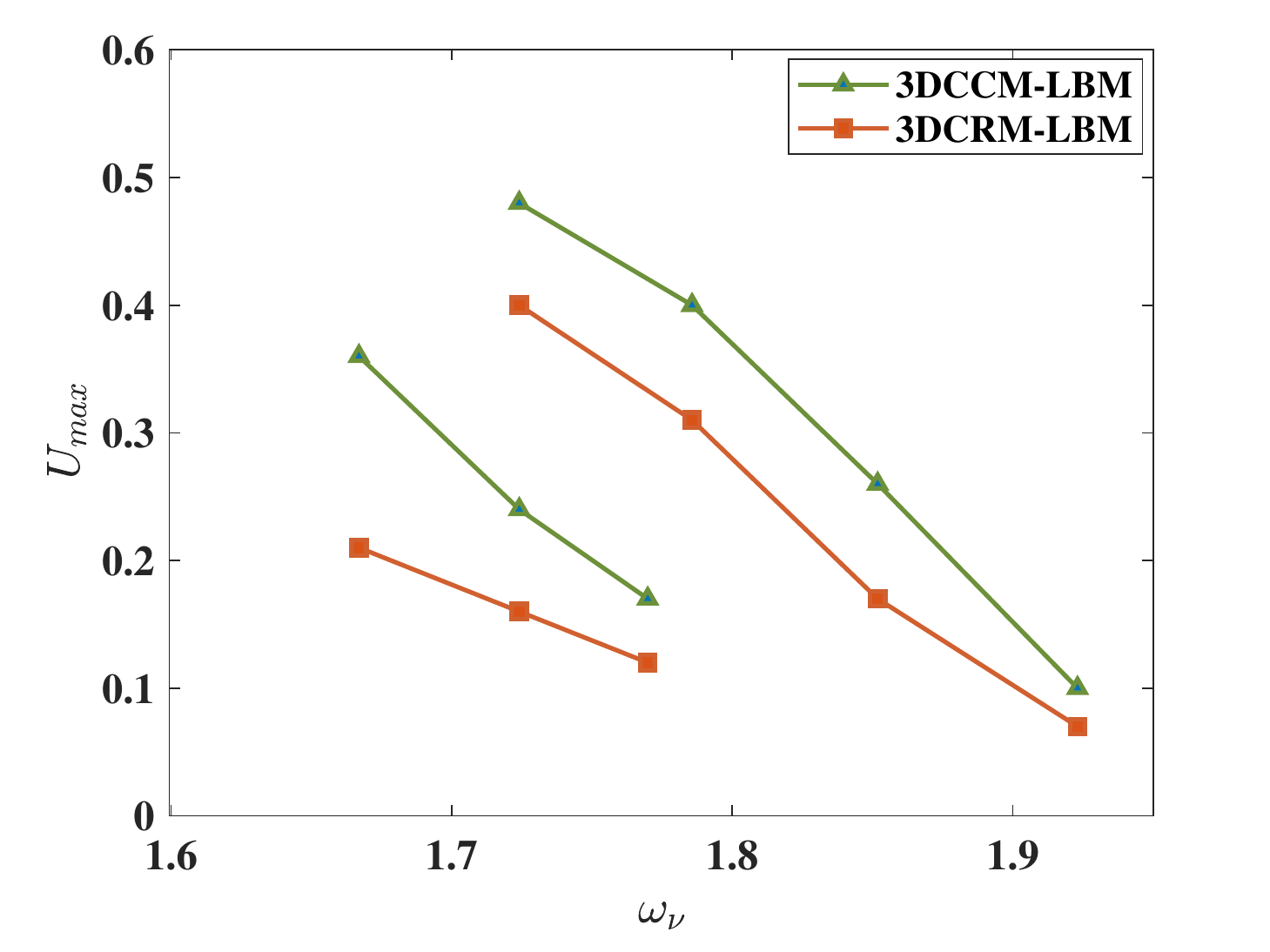}
 \caption{Numerical stability test results showing the maximum threshold velocity of the lid $U$ in a 3D lid driven cubic cavity flow at different values of the relaxation parameter corresponding to the shear viscosity for the 3DCCM-LBM (i.e., based on central moments) and 3DCRM-LBM (i.e., based on central moments) at grid aspect ratios of $r=0.75$, $s=1.0$ (upper two curves), and $r=0.5$, $s=1.0$ (lower two curves).}
    \label{fig:8}
\end{figure}

The algorithmic details given in Sec.~\ref{subsec:algorithmicdetails3DCCM-LBM} is general and readily extends to other collision models. For example, if instead of central moments, the raw moments are used in performing the relaxations under collision with the same corrections in the equilibria as before and omitting the steps involving the mappings between the central moments and raw moments (i.e., involving $\tensor{\mathcal{F}}$ and $\tensor{\mathcal{F}}^{-1}$), the algorithm reduces to a 3D cuboid raw moment LBM, which is referred to as the 3DCRM-LBM in what follows. Such a 3DCRM-LBM would be significantly better than other prior 3D cuboid LB raw moment based formulations, as it avoids the orthogonalization of the moment basis, uses better forms of the equilibria obtained from matching with those of the continuous Maxwell distribution function and the elimination of the non-GI cubic velocity errors, and with simpler expressions for the corrections and transport coefficients. However, it would be interesting to see how the resulting 3DCRM-LBM compares with the 3DCCM-LBM based on central moments that has been used as the method of choice in this work and validated in detail earlier. While the 3DCRM-LBM incurs a slightly lower computational cost than the 3DCCM-LBM because of the absence of the additional mappings indicated above, the 3DCCM-LBM, which executes the collision step in the local moving frame of reference, is expected to be more robust with better numerical stability characteristics in simulating shear flows at larger characteristic velocities or lower viscosities than the 3DCRM-LBM, as observed in the recent 2D studies involving square~\cite{ning2016numerical} and rectangular~\cite{yahia2021central} lattice sets. Let's verify if this is indeed observed in 3D simulations of the lid-driven cubic cavity flow using the cuboid lattice. In this regard, we first perform a stability test in which, for a chosen grid resolution (or the grid aspect ratio), and by varying the relaxation parameter $\omega_\nu$ controlling the shear viscosity, we determine the maximum possible lid velocity $U$, i.e., $U_{max}$ that maintains stable simulation for 100,000 steps in each case. We chose two different grid resolutions of $30\times 40 \times 30$ and $30\times 60\times 30$ corresponding to the grid ratios of $r=0.75$ and $s=1.0$ and $r=0.5$ and $s=1.0$, respectively. In both cases, we use $c_s^2=0.08$, which is the minimum of the optimal values for the two grid aspect ratios, in order to have the same reference speed of sound in calculating the Mach number. All the relaxation parameters, other than that for the shear viscosity, are set to $1.0$ for simplicity. The results are shown in Fig.~\ref{fig:8}. Clearly, the 3DCCM-LBM is able to support larger magnitudes of shear consistently than the 3DCRM-LBM for all the choices of the grid resolution used.

\begin{figure}[H]
\centering
 \includegraphics[width=0.5\textwidth] {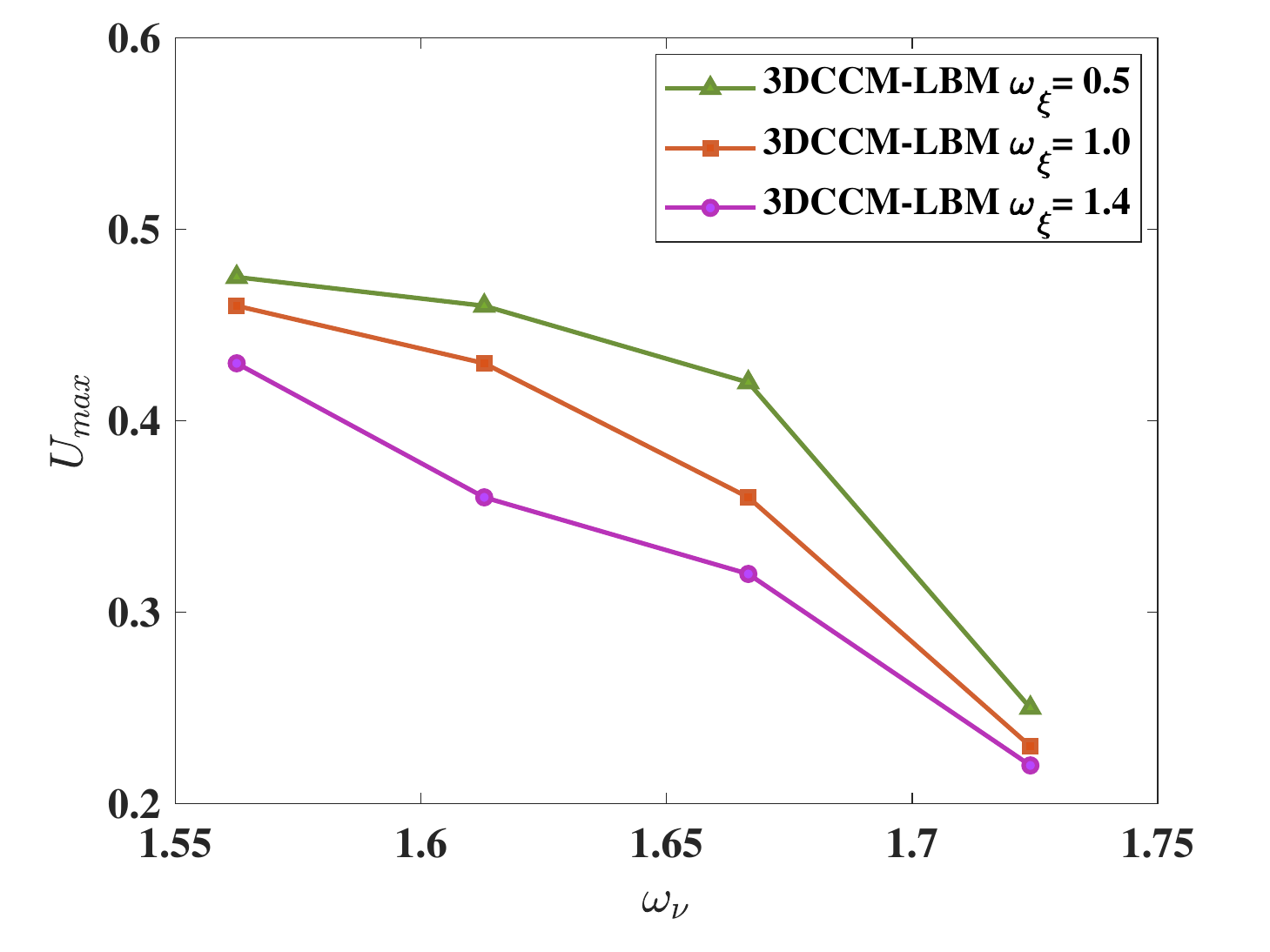}
 \caption{Numerical stability results of the 3DCCM-LBM showing the maximum threshold velocity of the lid $U$ in a 3D lid driven cubic cavity flow at different values of the relaxation parameter corresponding to the shear viscosity obtained for three different values of the relaxation parameter controlling the bulk viscosity, i.e., $\omega_\xi = 0.5, 1.0$, and $1.4$.}
    \label{fig:9}
\end{figure}

Moreover, it was pointed out earlier that in our 3D cuboid LB algorithms, the trace of the diagonal components of the second order moments is evolved separately from the other moments with its own relaxation parameter $\omega_\xi$ after applying $\tensor{B}$, and the individual components are subsequently recovered via applying $\tensor{B}^{-1}$, so that the bulk viscosity can be specified independently from shear viscosity. It is known that for the common LB schemes based on the cubic lattice sets that a higher value of the bulk viscosity (via decreasing $\omega_\xi$) enhances stability by suppressing the spurious pressure waves. We will now verify if this also holds true for the cuboid LB formulation based on central moments. Repeating the stability test mentioned above by choosing $\omega_{\xi}=0.5, 1.0$, and $1.4$ at $30\times 60 \times 30$ and $c_s^2=0.08$, Fig.~\ref{fig:9} shows the resulting maximum threshold velocity of the lid $U$ for numerical stability. This figure clarifies that the robustness of the 3DCCM-LBM is significantly improved by choosing lower $\omega_{\xi}$ or higher bulk viscosity.
In addition, we now perform a different type of stability study, where the grid resolution and lid velocity are fixed, and the lowest possible shear viscosity or the maximum Reynolds number that maintains stable simulations for 3DCCM-LBM and 3DCRM-LBM are determined. In this regard, we set $U=0.2$, $c_s^2=0.08$, $r=0.5$ and $s=1.0$, and three different choices of the grid resolutions $40\times 80\times 40$, $50\times 100\times 50$ and $90\times 180\times 90$ are used to determine the maximum Reynolds number. The central moment formulation is found to be more stable and allows larger Reynolds number than that due to the raw moment formulation, which again support the superior numerical characteristics of the 3DCCM-LBM in simulating shear flows on the cuboid lattice.
\begin{figure}[H]
\centering
 \includegraphics[width=0.5\textwidth] {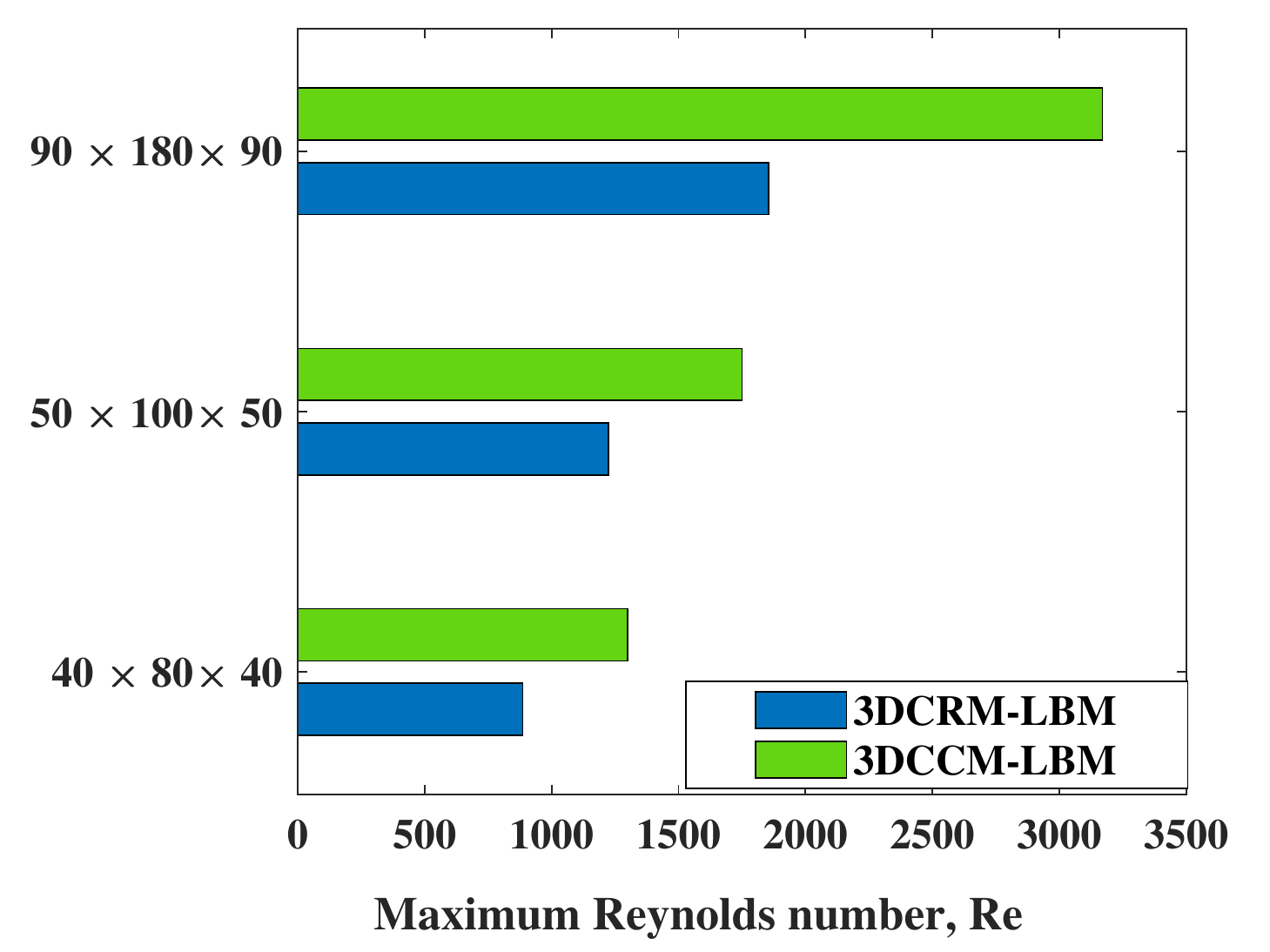}
 \caption{Comparison of the maximum Reynolds number for numerical stability of 3DCCM-LBM and 3DCRM-LBM at different grid resolutions with cuboid grid aspect ratios of $r=0.5$ and $s=1.0$ for a fixed lid velocity of $U=0.2$ and $c_s^2=0.08$.}
    \label{fig:10}
\end{figure}

\section{Summary and Conclusions} \label{sec:8}
In this paper, we have presented a new 3D LB algorithm based on central moments for the D3Q27 lattice using a cuboid grid, which is parameterized by two grid aspect ratios that are related to the ratios of the particle speeds with respect to that along a reference coordinate direction. The additional degrees of freedom introduced by the flexible specification of the cuboid lattice grid enables the method to naturally and more efficiently compute flows having different characteristic length scales in different directions. It is constructed to simulate the Navier-Stokes equations consistently via introducing counteracting corrections to the second order moment equilibria that eliminate the associated grid anisotropy and the non-Galilean invariant third order velocity errors obtained via a Chapman-Enskog analysis. Such corrections are related to the diagonal components of the velocity gradient tensor, which are computed locally using non-equilibrium moments in our approach and depend on the grid aspect ratios. The implementation is shown to be compact and modular, with an interpretation based on special matrices, admitting ready extension of the standard algorithm for the cubic lattice to the cuboid lattice via appropriate scaling of moments based on grid aspect ratios before and after collision step and equilibria corrections. The resulting formulation is general in that the same grid corrections developed for the D3Q27 lattice for recovering the correct viscous stress tensor is applicable for other lattice subsets, and a variety of collision models, including those based on the relaxation of raw moments, central moments and cumulants, as well as their special case involving the distribution functions. The cuboid central moment LBM was validated against a variety of benchmark flows, and when used in lieu of the corresponding raw moment formulation for simulating shear flows, we show that it results in significant improvements in numerical stability. Finally, we demonstrated that our cuboid LB approach is efficient in simulating anisotropic shear flow problems with significant savings in computational cost and memory storage when compared to that based on the cubic lattice. Future improvements are expected by extending the present formulation as a multiblock cuboid LB algorithm, which will be considered in a future study.

\section*{Acknowledgements}
Parts of this work were presented virtually at the SIAM Conference on Computational Science and Engineering (CSE-21) in March 2021 and at the International Conference on Mesoscopic Methods in Engineering Science (ICMMES 2021) in July 2021. The first author (EY) thanks the Department of Mechanical Engineering at the University of Colorado Denver for financial support. The second (WS) and third (KNP) authors would like to acknowledge the support of the US National Science Foundation (NSF) under Grant CBET-1705630. The third author (KNP) would like to also thank the NSF for support of the development of a computer cluster infrastructure through the project ``CC* Compute: Accelerating Science and Education by Campus and Grid Computing'' under Award 2019089.

\appendix

\section{Mapping pre-collision distribution functions to raw moments}\label{sec:appendix1}
The intermediate raw moments with respect to the moment basis for the D3Q27 cubic lattice computed from $\mathbf{m}= \tensor{P}\mathbf{f}$ are expressed in the following. While the full expressions are shown in this appendix and all the ones that follow for better clarity, in actual implementations they should be optimized by grouping together the common sub-expressions when possible.
\begin{align*}
&\Kps{000} = \f{0} + \f{1} + \f{2} + \f{3} + \f{4} + \f{5} + \f{6} + \f{7} +  \f{8} + \f{9} + \f{10} + \f{11} + \f{12} + \f{13} + \f{14} +\f{15} + \f{16} + \f{17} + \f{18} \\
&\quad\quad\quad\;\;\;+ \f{19} + \f{20} + \f{21} + \f{22} + \f{23} +\f{24} + \f{25} +\f{26},\\
&\Kps{100} = \f{1} -\f{2} + \f{7} - \f{8} + \f{9} - \f{10} + \f{11}-\f{12} + \f{13} - \f{14}+\f{19} - \f{20} + \f{21} - \f{22} + \f{23} - \f{24} + \f{25} -\f{26}, \\
&\Kps{010} = \f{3} - \f{4} + \f{7} + \f{8} - \f{9} - \f{10} + \f{15}-\f{16} + \f{17} - \f{18} + \f{19} + \f{20}- \f{21} - \f{22} + \f{23} +\f{24} - \f{25} -\f{26},\\
&\Kps{001}= \f{5} - \f{6} + \f{11} + \f{12} - \f{13} - \f{14} + \f{15}+ \f{16} - \f{17} - \f{18} + \f{19} + \f{20} + \f{21} + \f{22} - \f{23}-\f{24}-\f{25} - \f{26}, \\
&\Kps{110} = \f{7} - \f{8} - \f{9} + \f{10} + \f{19} - \f{20} - \f{21}+ \f{22} + \f{23} - \f{24}- \f{25}+ \f{26},\\
&\Kps{101} = \f{11} - \f{12} - \f{13} + \f{14} + \f{19} - \f{20} + \f{21}- \f{22} - \f{23} + \f{24}- \f{25}+ \f{26},\\
&\Kps{011} = \f{15} - \f{16} - \f{17} + \f{18} + \f{19} + \f{20} - \f{21}- \f{22} - \f{23} - \f{24}+ \f{25}+ \f{26},\\
&\Kps{200} = \f{1} + \f{2} +\f{7} +  \f{8} + \f{9} + \f{10} +\f{11} + \f{12} + \f{13} + \f{14}+\f{19} + \f{20} + \f{21} + \f{22} + \f{23} +\f{24} + \f{25} +\f{26},\\
&\Kps{020} = \f{3} + \f{4} +\f{7} +  \f{8} + \f{9} + \f{10} +\f{15} + \f{16} + \f{17} + \f{18}+\f{19} + \f{20} + \f{21} + \f{22} + \f{23} +\f{24} + \f{25} +\f{26},\\
&\Kps{002} = \f{5} + \f{6} +\f{11} + \f{12} + \f{13} + \f{14} +\f{15} + \f{16} + \f{17} + \f{18}+\f{19} + \f{20} + \f{21} + \f{22} + \f{23} +\f{24} + \f{25} +\f{26},\\
&\Kps{120} = \f{7} - \f{8} +\f{9} - \f{10}+\f{19} - \f{20} + \f{21} - \f{22} + \f{23} - \f{24} + \f{25} -\f{26},\\
&\Kps{102} = \f{11} - \f{12} +\f{13} - \f{14}+\f{19} - \f{20} + \f{21} - \f{22} + \f{23} - \f{24} + \f{25} -\f{26},\\
&\Kps{210} = \f{7} + \f{8} -\f{9} - \f{10}+ \f{19} + \f{20}- \f{21} - \f{22} + \f{23} +\f{24} - \f{25} -\f{26},\\
&\Kps{012} = \f{15} - \f{16} +\f{17} - \f{18}+ \f{19} + \f{20}- \f{21} - \f{22} + \f{23} +\f{24} - \f{25} -\f{26},\\
&\Kps{201} = \f{11} + \f{12} -\f{13} - \f{14}+\f{19} + \f{20} + \f{21} + \f{22} - \f{23}-\f{24}-\f{25} - \f{26},\\
&\Kps{021} = \f{15} + \f{16} -\f{17} - \f{18}+\f{19} + \f{20} + \f{21} + \f{22} - \f{23}-\f{24}-\f{25} - \f{26},\\
&\Kps{111} = \f{19} - \f{20} -\f{21} + \f{22}-\f{23} + \f{24}+\f{25} - \f{26},\\
&\Kps{220} = \f{7} +  \f{8} + \f{9} + \f{10} +\f{19} + \f{20} + \f{21} + \f{22} + \f{23} +\f{24} + \f{25} +\f{26},\\
&\Kps{202} = \f{11} + \f{12} + \f{13} + \f{14} +\f{19} + \f{20} + \f{21} + \f{22} + \f{23} +\f{24} + \f{25} +\f{26}, \\
&\Kps{022} = \f{15} + \f{16} + \f{17} + \f{18}+\f{19} + \f{20} + \f{21} + \f{22} + \f{23} +\f{24} + \f{25} +\f{26},\\
&\Kps{211} = \f{19} + \f{20} -\f{21} - \f{22}-\f{23} - \f{24}+\f{25} + \f{26},\\
&\Kps{121} = \f{19} - \f{20} +\f{21} - \f{22}-\f{23} + \f{24}-\f{25} + \f{26},\\
&\Kps{112} = \f{19} - \f{20} -\f{21} + \f{22}+\f{23} - \f{24}-\f{25} + \f{26},\\
&\Kps{122} = \f{19} - \f{20} + \f{21} - \f{22} + \f{23} - \f{24} + \f{25} -\f{26},\\
&\Kps{212} = \f{19} + \f{20}- \f{21} - \f{22} + \f{23} +\f{24} - \f{25} -\f{26},\\
&\Kps{221} = \f{19} + \f{20} + \f{21} + \f{22} - \f{23}-\f{24}-\f{25} - \f{26},\\
&\Kps{222} = \f{19} + \f{20} + \f{21} + \f{22} + \f{23} +\f{24} + \f{25} +\f{26}.
\end{align*}

\section{Scaling pre-collision raw moments by grid aspect ratios}\label{sec:appendix2}
In order to obtain the pre-collision raw moments for the cuboid lattice from those for the cubic lattice, we scale the elements of the latter, i.e., $k_{mnp}^\prime$  by the grid aspect ratios using the factor $r^ns^p$ for the moment of order $(m+n+p)$. Thus, the operations involving $\mathbf{m}\leftarrow\tensor{S}\mathbf{m}$ can be written as follows:
\begin{equation*}
\begin{matrix}
  \Kps{000}=\Kps{000} & \Kps{100} =\Kps{100} & \Kps{010}= r \Kps{010}, &  & & \\
  \Kps{001}=s \Kps{001} &  \Kps{200}=\Kps{200}     &  \Kps{102}= s^2  \Kps{102} &  \Kps{021}=r^2 s \Kps{021}& \Kps{022}=r^2 s^2\Kps{022} &  \Kps{122}= r^2 s^2  \Kps{122}, \\
  \Kps{110} =r \Kps{110}&  \Kps{020}= r^2 \Kps{020}&  \Kps{210}= r\Kps{210}     & \Kps{111}=r s \Kps{111}& \Kps{211}=r s \Kps{211} & \Kps{212}=r s^2\Kps{212}, \\
  \Kps{101}= s  \Kps{101}& \Kps{002}=s^2 \Kps{002}&   \Kps{012}=r s^2 \Kps{012}&  \Kps{220}=r^2 \Kps{220}& \Kps{121}=r^2 s\Kps{121}  &  \Kps{221} =r^2 s\Kps{221},  \\
  \Kps{011}=r s \Kps{011}& \Kps{120} =r^2\Kps{120}&   \Kps{201}=s \Kps{201}    &  \Kps{202}=s^2 \Kps{202}&\Kps{112} =r s^2 \Kps{112} & \Kps{222}= r^2 s^2 \Kps{222}.
 \end{matrix}
\end{equation*}

\section{Mapping pre-collision raw moments to central moments} \label{sec:appendix3}
By writing the central moments for all the components for the D3Q27 lattice in terms of the raw moments via the binomial expansions, represented by $\mathbf{m}^c=\tensor{\mathcal{F}}\mathbf{m}$, the following expressions are obtained:
\[\Ks{000} = \Kps{000},\]
\[\Ks{100} = \Kps{100} - \ux\Kps{000}, \]
\[\Ks{010} = \Kps{010} - \uy\Kps{000}, \]
\[\Ks{001} = \Kps{001} -\uz\Kps{000},\]
\[\Ks{110} = \Kps{110} - \ux\Kps{010}- \uy\Kps{100} + \ux\uy\Kps{000},\]
\[\Ks{101} = \Kps{101} - \ux\Kps{001} - \uz\Kps{100} + \ux\uz\Kps{000},\]
\[\Ks{011} = \Kps{011} -  \uy\Kps{001} - \uz\Kps{010} +  \uy\uz\Kps{000},\]
\[\Ks{200} = \Kps{200}  - 2\ux\Kps{100} + \uxx\Kps{000 },\]
\[\Ks{020} = \Kps{020}  - 2\uy\Kps{010} + \uyy\Kps{000} ,\]
\[\Ks{002} = \Kps{002}  - 2\uz \Kps{001} +\uzz\Kps{000} ,\]
\[\Ks{120} = \Kps{120} -\ux\Kps{020} - 2\uy \Kps{110}+ 2\ux\uy\Kps{010}+\uyy\Kps{100}- \ux\uyy\Kps{000},\]
\[\Ks{102} = \Kps{102} - \ux\Kps{002} - 2\uz \Kps{101} + 2\ux\uz \Kps{001}+ \uzz\Kps{100}  - \ux\uzz\Kps{000},\]
\[\Ks{210} = \Kps{210} -\uy\Kps{200}- 2\ux\Kps{110}  + \uxx\Kps{010}+ 2\ux\uy\Kps{100} - \uxx\uy\Kps{000},\]
\[\Ks{012} = \Kps{012} -\uy\Kps{002} - 2\uz\Kps{011}+ \uzz\Kps{010} + 2\uy\uz\Kps{001} - \uy\uzz\Kps{000},\]
\[\Ks{201} = \Kps{201}- \uz\Kps{200}  - 2\ux\Kps{101} + \uxx\Kps{001} + 2\ux\uz\Kps{100}- \uxx\uz\Kps{000}, \]
\[\Ks{021} = \Kps{021} - \uz\Kps{020}- 2\uy\Kps{011} + \uyy\Kps{001}+ 2\uy\uz\Kps{010} - \uyy\uz\Kps{000},\]
\[\Ks{111} = \Kps{111} - \ux\Kps{011} -\uy\Kps{101} - \uz\Kps{110} + \ux\uy\Kps{001} + \ux\uz\Kps{010} + \uy\uz\Kps{100} - \ux\uy\uz\Kps{000},\]
\[\Ks{220} = \Kps{220} - 2\uy \Kps{210}- 2\ux \Kps{120}+ \uxx\Kps{020}+ \uyy\Kps{200}+ 4\ux\uy \Kps{110} - 2\uxx\uy \Kps{010}  - 2\ux\uyy \Kps{100} +\uxx\uyy\Kps{000},\]
\[\Ks{202} =  \Kps{202}-2\uz\Kps{201} - 2\ux \Kps{102}+ \uxx\Kps{002} + \uzz\Kps{200}+ 4\ux\uz\Kps{101}- 2\uxx\uz\Kps{001}- 2\ux\uzz\Kps{100} \uxx\uzz\Kps{000} ,  \]
\[\Ks{022} = \Kps{022}- 2\uz \Kps{021}- 2\uy \Kps{012}+\uzz \Kps{020}+\uyy\Kps{002} + 4\uy\uz\Kps{011}- 2\uyy\uz \Kps{001} - 2\uy\uzz\Kps{010}+ \uyy\uzz\Kps{000},  \]
\vspace{-2em}
\begin{align*}
\Ks{211} &= \Kps{211} -2\ux \Kps{111}- \uy\Kps{201} -\uz\Kps{210} + \uxx\Ks{011} + 2\ux\uy \Kps{101}+ \uy\uz\Kps{200} + 2\ux\uz \Kps{110}  \\ & \; \; \; \;-\uxx\uy\Kps{001} -\uxx\uz\Kps{010}-2\ux\uy\uz\Kps{100} + \uxx\uy\uz\Kps{000},
\end{align*}
\vspace{-2em}
\begin{align*}
\Ks{121} &= \Kps{121} - 2\uy\Kps{111}  -\ux\Kps{021}- \uz\Kps{120}+ \ux\uz\Kps{020} + 2\ux\uy\Kps{011} + \uyy\Kps{101}  + 2\uy\uz\Kps{110}  \\& \; \; \; \;  - \ux\uyy\Kps{001} -2\ux\uy\uz\Kps{010}- \uyy\uz\Kps{100}  +\ux\uyy\uz\Kps{000},
\end{align*}
\vspace{-2em}
\begin{align*}
\Ks{112} &= \Kps{112}-2\uz \Kps{111}- \ux\Kps{012} -\uy\Kps{102}+ \ux\uy\Kps{002}+ 2\ux\uz\Kps{011} + 2\uy\uz \Kps{101}  + \uzz\Kps{110}\\  & \; \; \; \;-2\ux\uy\uz\Kps{001}- \ux\uzz\Kps{010} - \uy\uzz\Kps{100} +\ux\uy\uzz\Kps{000},
\end{align*}
\vspace{-2em}
\begin{align*}
\Ks{122} &= \Kps{122}- 2\uy\Kps{112} - 2\uz \Kps{121}-\ux\Kps{022} + 4\uy\uz \Kps{111}  + 2\ux\uz\Kps{021}+ 2\ux\uy\Kps{012}+ \uyy\Kps{102} \\& \; \; \; \; + \uzz\Kps{120}- \ux\uyy\Kps{002} -\ux\uzz\Kps{020}- 4\ux\uy\uz \Kps{011} - 2\uyy\uz\Kps{101} - 2\uy\uzz\Kps{110} \\& \; \; \; \; + 2\ux\uyy\uz\Kps{001}  + 2\ux\uy\uzz \Kps{010}+ \uyy\uzz\Kps{100} - \ux\uyy\uzz\Kps{000},
\end{align*}
\vspace{-2em}
\begin{align*}
\Ks{212} &= \Kps{212}- 2\ux\Kps{112}- 2\uz \Kps{211} -\uy\Kps{202} + 4\ux\uz \Kps{111} + 2\uy\uz \Kps{201}  + \uxx\Kps{012} + \uzz\Kps{210} \\ & \; \; \; \; + 2\ux\uy \Kps{102}- \uxx\uy\Kps{002}-\uy\uzz\Kps{200}- 2\uxx\uz\Kps{011}- 4\ux\uy\uz \Kps{101}  -2\ux\uzz\Kps{110} \\ & \; \; \; \;+ 2\uxx\uy\uz \Kps{001}+ \uxx\uzz\Kps{010} + 2\ux\uy\uzz \Kps{100} - \uxx\uy\uzz\Kps{000},
\end{align*}
\vspace{-2em}
\begin{align*}
\Ks{221} &=  \Kps{221}- 2\ux \Kps{121} - 2\uy\Kps{211}- \uz\Kps{220}+ 4\ux\uy \Kps{111} + \uxx\Kps{021} + \uyy\Kps{201}+ 2\uy\uz \Kps{210}\\ & \; \; \; \; + 2\ux\uz \Kps{120} - \uxx\uz\Kps{020} - \uyy\uz\Kps{200}  - 2\uxx\uy\Kps{011}-2\ux\uyy \Kps{101}- 4\ux\uy\uz \Kps{110}\\ & \; \; \; \;+\uxx\uyy\Kps{001}+ 2\uxx\uy\uz \Kps{010}+ 2 \ux\uyy\uz\Kps{100}- \uxx\uyy\uz\Kps{000},
\end{align*}
\vspace{-2em}
\begin{align*}
\Ks{222} &=\Kps{222} - 2\uz\Kps{221}- 2\uy\Kps{212}- 2 \ux\Kps{122}+ 4\ux\uy\Kps{112}+ 4\ux\uz \Kps{121}\\& \; \; \; \;+ 4\uy\uz\Kps{211}+\uxx\Kps{022}+ \uyy\Kps{202}+ \uzz\Kps{220}- 8\ux\uy\uz\Kps{111}- 2\uxx\uz\Kps{021}\\& \; \; \; \;- 2\uyy\uz\Kps{201}- 2\uxx\uy \Kps{012}  - 2\uy\uzz\Kps{210} - 2\ux\uyy\Kps{102} - 2\ux\uzz\Kps{120}+\uxx\uyy\Kps{002}\\&  \; \; \; \;+ \uxx\uzz\Kps{020}+\uyy\uzz\Kps{200} + 4\uxx\uy\uz\Kps{011} + 4\ux\uyy\uz\Kps{101}+ 4\ux\uy\uzz\Kps{110}- 2\uxx\uyy\uz\Kps{001}\\ & \; \; \; \; - 2\uxx\uy\uzz\Kps{010} - 2\ux\uyy\uzz\Kps{100} +\uxx\uyy\uzz\Kps{000}.
\end{align*}

\section{Compute post-collision central moments via relaxations under collision using extended equilibria for cuboid lattice} \label{sec:appendix4}
First, reflecting the original moment basis in Eq.~(\ref{eq:5}), we apply $\tensor{B}$ and combine the selected pre-collision central moments of 2nd, 3rd and 4th orders as follows:
\begin{eqnarray}\label{eq:combinedprecollisioncentramoments}
&& \;\;k_{2s} = k_{200}+ k_{020} + k_{002},\quad k_{2d1} = k_{200}- k_{020},\quad\quad k_{2d2} = k_{200}- k_{002}\nonumber\\
&& k_{3s1} =  k_{120} + k_{102}, \quad k_{3m1} = k_{120} - k_{102},\nonumber\\
&& k_{3s2} = k_{210} + k_{012}, \quad k_{3m2} = k_{210} - k_{012},\nonumber\\
&& k_{3s3} =  k_{201} + k_{021}, \quad k_{3m3} = k_{201} - k_{021},\nonumber\\
&&k_{4s}  = k_{220} + k_{202} + k_{022}, \quad k_{4d1}=k_{220} + k_{202}- k_{022}, \quad\quad k_{4d2} = k_{220}- k_{202}.
\end{eqnarray}

Next, from Sec.~\ref{sec:2}, we can write the effective equilibrium central moments as follows:
\begin{align*}
&k_{000}^{eq}=\rho, \quad\;\; k_{100}^{eq}=k_{010}^{eq}=k_{001}^{eq}=0,\quad\;\; k_{110}^{eq}=k_{101}^{eq}=k_{011}^{eq}=0,
\end{align*}
\begin{eqnarray}\label{eq:63}
k_{2s}^{eq} &=& k_{200}^{eq}+ k_{020}^{eq}+ k_{002}^{eq} = 3 c_s^2\rho + \Big( \theta_{bx}\partial_x u_x+ \theta_{by} \partial_y u_y + \theta_{bz} \partial_z u_z +  \lambda_{bx} \partial_x \rho  + \lambda_{by} \partial_y \rho+ \lambda_{bz} \partial_z \rho \Big) \Delta t,\nonumber\\
k_{2d1}^{eq} &=& k_{200}^{eq}- k_{020}^{eq} = \Big( \theta_{sx}\partial_x u_x- \theta_{sy} \partial_y u_y + \lambda_{sx} \partial_x \rho  + \lambda_{sy} \partial_y \rho \Big) \Delta t,\nonumber\\
k_{2d2}^{eq} &=& k_{200}^{eq}- k_{002}^{eq} = \Big( \theta_{sx}\partial_x u_x- \theta_{sz} \partial_z u_z + \lambda_{sx} \partial_x \rho  + \lambda_{sz} \partial_z \rho \Big) \Delta t,\nonumber
 \end{eqnarray}
\begin{align}\label{eq:64}
\begin{matrix}
  k_{3s1}^{eq}=k_{120}^{eq}+ k_{102}^{eq}=0, &  k_{3m1}^{eq}=k_{120}^{eq}- k_{102}^{eq}=0, &\\
  k_{3s2}^{eq}=k_{210}^{eq}+ k_{012}^{eq}=0,& k_{3m2}^{eq}=k_{210}^{eq}- k_{012}^{eq}=0, &  \\
  k_{3s3}^{eq}=k_{201}^{eq}+ k_{021}^{eq}=0,& k_{3m3}^{eq}=k_{201}^{eq}- k_{021}^{eq}=0, &  k_{111}^{eq}=0,\\
  k_{211}^{eq}=k_{121}^{eq}= k_{112}^{eq}=0, &k_{122}^{eq}=k_{212}^{eq}= k_{221}^{eq}=0,&  k_{222}^{eq}=c_s^6\rho, \\
  k_{4s}^{eq}=k_{220}^{eq}+ k_{202}^{eq}+ k_{022}^{eq}= 3 c_s^4\rho, & k_{4d1}^{eq}=k_{220}^{eq}+ k_{202}^{eq}- k_{022}^{eq}= c_s^4\rho, &  k_{4d2}^{eq}=k_{220}^{eq}- k_{202}^{eq}=0.\\
 \end{matrix}
\end{align}
Note that in Eq.~\eqref{eq:64}, we have extended the combinations of the diagonal parts of the second order moments by including the corrections due to grid anisotropy for the cuboid grid and non-GI velocity errors due to aliasing on the D3Q27 lattice using the results from Sec.~\ref{subsec:correctionsviaCEanalysis}. Moreover, in a slight economy for the notations, in the following, we will introduce the subscript `b' for the coefficients related to corrections associated with bulk viscosity and the subscript `s' for those other components influencing the shear viscosity (rather than the numerical subscripts used in Sec.~\ref{subsec:correctionsviaCEanalysis}):
\begin{align*}
\begin{matrix}
  \theta_{sx}= -\rho\left(3 u_x^2 + 3 c_s^2 - 1 \right) \left( \dfrac{1}{\omega_{\nu}}- \dfrac{1}{2}\right), &\quad \quad \lambda_{sx}= - \left(3 c_s^2 - 1 \right) \left(\dfrac{1}{\omega_{\nu}}- \dfrac{1}{2}\right) u_x, \\
  \theta_{bx}= -\rho\left(3 u_x^2 + 3 c_s^2 - 1 \right) \left( \dfrac{1}{\omega_{\xi}}- \dfrac{1}{2}\right),&\quad \quad \lambda_{bx}= - \left(3 c_s^2 - 1 \right) \left(\dfrac{1}{\omega_{\xi}}- \dfrac{1}{2}\right) u_x, \\
  \theta_{sy}= -\rho \left( 3 u_y^2 +3 c_s^2 - r^2  \right)]\left(\dfrac{1}{\omega_{\nu}}- \dfrac{1}{2}\right), &\quad \quad \lambda_{sy}= + (3 c_s^2 - r^2) \left(\dfrac{1}{\omega_{\nu}}- \dfrac{1}{2}\right) u_y, \\
  \theta_{by}= -\rho \left( 3 u_y^2 + 3 c_s^2 - r^2 \right)\left(\dfrac{1}{\omega_{\xi}}- \dfrac{1}{2}\right),&\quad \quad \lambda_{by}= - (3 c_s^2 - r^2) \left(\dfrac{1}{\omega_{\xi}}- \dfrac{1}{2}\right) u_y,\\
  \theta_{sz}= -\rho \left( 3 u_z^2 +3 c_s^2 - s^2 \right)\left(\dfrac{1}{\omega_{\nu}}- \dfrac{1}{2}\right), &\quad\quad \lambda_{sz}= + (3 c_s^2 - s^2) \left(\dfrac{1}{\omega_{\nu}}- \dfrac{1}{2}\right) u_z,\\
  \theta_{bz}= -\rho \left( 3 u_z^2 + 3 c_s^2 - s^2 \right)\left(\dfrac{1}{\omega_{\xi}}- \dfrac{1}{2}\right),  & \quad \quad\lambda_{bz}= - (3 c_s^2 - s^2)\left(\dfrac{1}{\omega_{\xi}}- \dfrac{1}{2}\right) u_z.\\ \\
\end{matrix}
\end{align*}
Here, $\omega_{\xi}$ is the relaxation parameter associated with the trace of the second order moments $k_{2s}$ and $\omega_{\nu}$ for the other two diagonal components $k_{2d1}$ and $k_{2d2}$ defined in Eq.~(\ref{eq:combinedprecollisioncentramoments}) (as well as the off-diagonal components $k_{110}$, $k_{101}$ and $k_{011}$). As we will show later that this would enable independent specification of the bulk viscosity from the shear viscosity. For the associated equilibria in Eq.~\eqref{eq:64}, the density gradients $\partial_x \rho$, $\partial_y \rho$ and $\partial_z \rho$ will be obtained from a second order (isotropic) finite difference scheme, while the diagonal components of the velocity gradient tensor $\partial_x u_x$, $\partial_y u_y$ and $\partial_z u_z$ will be computed locally from non-equilibrium moments (using the results in Eqs.~\eqref{eq:44}-\eqref{eq:49} derived in Sec.~\ref{subsec:correctionsviaCEanalysis}) as follows:

Defining
\begin{eqnarray*}
&&A = \frac{1}{2}(3 c_s^2 - 1)u_x, \quad\quad B= \frac{1}{2}(3 c_s^2 - r^2)u_y, \quad\quad C= \frac{1}{2}(3 c_s^2 - s^2)u_z,\\
&&e_{sr1} = -A \;\partial_x \rho +B \;\partial_y \rho,\quad\quad e_{sr2} = -A \;\partial_x \rho +C \;\partial_z \rho,\quad\quad e_{br}= -A \;\partial_x \rho -B \;\partial_y \rho-C \;\partial_z \rho
\end{eqnarray*}
\begin{eqnarray*}
  R_{b} &=& \left({k_{200}} + {k_{020}} + {k_{002}}\right)- 3c_s^2\rho +e_{br}= k_{2s}- 3c_s^2\rho +e_{br},  \\
  R_{s1} &=& \left({k_{200}} - {k_{020}}\right)+ e_{sr1}=k_{2d1}+ e_{sr1}, \\
  R_{s2} &=& \left({k_{200}} - {k_{002}}\right)+ e_{sr2}=k_{2d2}+ e_{sr2}.
  \end{eqnarray*}
  \begin{eqnarray*}
  &C_{sx}= \left[-\dfrac{2c_s^2 }{\omega_{\nu}}+\dfrac{\left(3 c_s^2 - 1 \right)}{2}+ \dfrac{3u_x^2}{2} \right] \rho ,&C_{bx}=\left[-\dfrac{2 c_s^2}{\omega_{\xi}} +\dfrac{\left(3 c_s^2 - 1 \right)}{2}+ \dfrac{3 u_x^2}{2} \right] \rho,\\
  &C_{sy}= \left[\dfrac{2 c_s^2}{\omega_{\nu}} -\dfrac{\left(3 c_s^2 - r^2 \right)}{2}- \dfrac{3u_y^2}{2} \right] \rho ,&C_{by} = \left[-\dfrac{2 c_s^2}{\omega_{\xi}} + \dfrac{\left(3 c_s^2 - r^2 \right)}{2} + \dfrac{3u_y^2}{2}\right] \rho, \\
  &C_{sz} = \left[\dfrac{2 c_s^2}{\omega_{\nu}} -\dfrac{\left(3 c_s^2 - s^2 \right)}{2}- \dfrac{3u_z^2}{2} \right] \rho,& C_{bz} = \left[-\dfrac{2 c_s^2}{\omega_{\xi}} + \dfrac{\left(3 c_s^2 - s^2 \right)}{2}+ \dfrac{3 u_z^2}{2} \right] \rho,
  \end{eqnarray*}
we can then write the local expressions for the velocity gradients as
\begin{eqnarray*}\label{eq:65}
&\partial_x u_x = {\cfrac{\left[- C_{sz}C_{by}R_{s1} - C_{sy} \left( C_{bz} R_{s2}  - C_{sz} R_{b} \right)\right]}{\left[- C_{sx}C_{sz}C_{by} - C_{sy} \left( C_{sx} C_{bz}  - C_{sz}C_{bx} \right)\right]}},\\
&\partial_y u_y = \cfrac{1}{C_{sy}} \left[ R_{s1}- C_{sx}\; \partial_x u_x \right],\;\;\;\;\;\;
\partial_z u_z = \cfrac{1}{C_{sz}} \left[ R_{s2} - C_{sx}\; \partial_x u_x \right].
\end{eqnarray*}

Based on these quantities, we can then prescribe the post-collision central moments via relaxations under collision as follows:
\begin{eqnarray*}
\tilde{k}_{000} & = & k_{000},\\
\tilde{k}_{100} & = & k_{100}+\omega_1(k_{100}^{eq}-k_{100})+(1-\omega_1/2)\sigma_{100}\Delta t,\\
\tilde{k}_{010} & = & k_{010}+\omega_1(k_{010}^{eq}-k_{010})+(1-\omega_1/2)\sigma_{010}\Delta t,\\
\tilde{k}_{001} & = & k_{001}+\omega_1(k_{001}^{eq}-k_{001})+(1-\omega_1/2)\sigma_{001}\Delta t,\\
\tilde{k}_{110} & = & k_{110}+\omega_\nu(k_{110}^{eq}-k_{110}),\\
\tilde{k}_{101} & = & k_{101}+\omega_\nu(k_{101}^{eq}-k_{101}),\\
\tilde{k}_{011} & = & k_{011}+\omega_\nu(k_{011}^{eq}-k_{011}),\\
\tilde{k}_{2d1} & = & k_{2d1}+\omega_\nu(k_{2d1}^{eq}-k_{2d1}),\\
\tilde{k}_{2d2} & = & k_{2d2}+\omega_\nu(k_{2d2}^{eq}-k_{2d2}),\\
\tilde{k}_{2s} & = & k_{2s}+\omega_\xi(k_{2s}^{eq}-k_{2s}),\\
\tilde{k}_{3s1} & = & k_{3s1}+\omega_{3s1}(k_{3s1}^{eq}-k_{3s1}),\\
\tilde{k}_{3s2} & = & k_{3s2}+\omega_{3s2}(k_{3s2}^{eq}-k_{3s2}),\\
\tilde{k}_{3s3} & = & k_{3s3}+\omega_{3s3}(k_{3s3}^{eq}-k_{3s3}),\\
\tilde{k}_{3m1} & = & k_{3m1}+\omega_{3m1}(k_{3m1}^{eq}-k_{3m1}),\\
\tilde{k}_{3m2} & = & k_{3m2}+\omega_{3m2}(k_{3m2}^{eq}-k_{3m2}),\\
\tilde{k}_{3m3} & = & k_{3m3}+\omega_{3m3}(k_{3m3}^{eq}-k_{3m3}),\\
\tilde{k}_{111} & = & k_{111}+\omega_{111}(k_{111}^{eq}-k_{111}),\\
\tilde{k}_{211} & = & k_{211}+\omega_{211}(k_{211}^{eq}-k_{211}),\\
\tilde{k}_{121} & = & k_{121}+\omega_{121}(k_{121}^{eq}-k_{121}),\\
\tilde{k}_{112} & = & k_{112}+\omega_{112}(k_{112}^{eq}-k_{112}),\\
\tilde{k}_{4s} & = & k_{4s}+\omega_{4s}(k_{4s}^{eq}-k_{4s}),\\
\tilde{k}_{4d1} & = & k_{4d1}+\omega_{4d1}(k_{4d1}^{eq}-k_{4d1}),\\
\tilde{k}_{4d2} & = & k_{4d2}+\omega_{4d2}(k_{4d2}^{eq}-k_{4d2}),\\
\tilde{k}_{122} & = & k_{122}+\omega_{122}(k_{122}^{eq}-k_{122}),\\
\tilde{k}_{212} & = & k_{212}+\omega_{212}(k_{212}^{eq}-k_{212}),\\
\tilde{k}_{221} & = & k_{221}+\omega_{221}(k_{221}^{eq}-k_{221}),\\
\tilde{k}_{222} & = & k_{222}+\omega_{222}(k_{222}^{eq}-k_{222}).
\end{eqnarray*}
Here, the relaxation parameters for the second order moments $\omega_\xi$ and $\omega_\nu$ independently specify the bulk viscosity $\xi$ and shear viscosity $\nu$, respectively, and are related to one another via
\begin{eqnarray*}
& \nu = c_s^2\left(\dfrac{1}{\omega_{\nu}}- \dfrac{1}{2}\right)\Delta t, \quad  \xi = \dfrac{2c_s^2}{3}\left(\dfrac{1}{\omega_{\xi}}- \dfrac{1}{2}\right)\Delta t,
\end{eqnarray*}
so that, in conjunction the moment equilibria corrections, the 3DCCM-LBM simulates the 3D NS equations. The rest of the relaxation parameters for the evolution of other central moments under collision are set to unity in this work. That is, $\omega_1=\omega_{3s1}=\omega_{3s2}=\omega_{3s3}=\omega_{3m1}=\omega_{3m2}=\omega_{3m3}=\omega_{111}=\omega_{211}=\omega_{121}=\omega_{11 2}=\omega_{4s}=\omega_{4d1}=\omega_{4d2}=\omega_{122}=\omega_{212}=\omega_{221}=\omega_{222}=1.0$. Finally, we segregate the post-collision values of those selectively combined central moments in the above, which is represented by $\tensor{B}^{-1}$, as
\begin{eqnarray*}
  &\tilde{k}_{200} = \dfrac{1}{3}\left(\tilde{k}_{2s} + \tilde{k}_{2d1}+ \tilde{k}_{2d2}\right) ,\quad\;\;\;\;
  \tilde{k}_{120} = \dfrac{1}{2}\left(\tilde{k}_{3s1} + \tilde{k}_{3m1}\right),\quad\;\;\;\;
  \tilde{k}_{102} = \dfrac{1}{2}\left(\tilde{k}_{3s1} - \tilde{k}_{3m1}\right)\\
  &\tilde{k}_{020} = \dfrac{1}{3}\left(\tilde{k}_{2s} -2 \tilde{k}_{2d1}+ \tilde{k}_{2d2}\right) ,\quad\;\;\;\;
  \tilde{k}_{210} = \dfrac{1}{2}\left(\tilde{k}_{3s2} + \tilde{k}_{3m2}\right)\quad\;\;\;\;
  \tilde{k}_{012} = \dfrac{1}{2}\left(\tilde{k}_{3s2} - \tilde{k}_{3m2}\right) ,\\
  &\tilde{k}_{002} = \dfrac{1}{3}\left(\tilde{k}_{2s} + \tilde{k}_{2d1}-2 \tilde{k}_{2d2}\right) , \quad\;\;\;\;
  \tilde{k}_{201} = \dfrac{1}{2}\left(\tilde{k}_{3s3} + \tilde{k}_{3m3}\right) , \quad\;\;\;\;
  \tilde{k}_{021} = \dfrac{1}{2}\left(\tilde{k}_{3s3} - \tilde{k}_{3m3}\right) , \\[4pt]
  &\tilde{k}_{220} = \dfrac{1}{4}\left(\tilde{k}_{4s} + \tilde{k}_{4d1}+ 2 \tilde{k}_{4d2}\right) ,\\
  &\tilde{k}_{202} = \dfrac{1}{4}\left(\tilde{k}_{4s} + \tilde{k}_{4d1}- 2 \tilde{k}_{4d2}\right) ,\\
  &\tilde{k}_{022} = \dfrac{1}{2}\left(\tilde{k}_{4s} - \tilde{k}_{4d1}\right).
\end{eqnarray*}
As a result, at the end of this computation, all the linearly independent post-collision central moments $\tilde{k}_{mnp}$ of the D3Q27 lattice have been computed.

\section{Mapping post-collision central moments to raw moments} \label{sec:appendix5}
The post-collision raw moments can be obtained from the central moments via $\tilde{\mathbf{m}}=\tensor{\mathcal{F}}^{-1}\tilde{\mathbf{m}}^c$, where the inverse of the frame transformation matrix, i.e., $\tensor{\mathcal{F}}^{-1}$ is given by $\tensor{\mathcal{F}}^{-1}=\tensor{\mathcal{F}}(-u_x,-u_y,-u_z)$. As noted in our previous work~\cite{yahia2021central}, this follows from the property of the binomial transforms in their generating function representation. In other words, by using the expressions presented in~\ref{sec:appendix3} and performing the exchanges $k_{mnp}\leftrightarrow k_{mnp}^\prime$ and, $u_x \leftrightarrow -u_x$, $u_y \leftrightarrow -u_y$ and $u_z \leftrightarrow -u_z$, and then applying $\tilde{(\cdot)}$ over the moments as they represent the post-collision states, we obtain the required mapping relations which are listed as follows:
\[\dKps{000} = \dKs{000},\]
\[\dKps{100} = \dKs{100} + \ux\dKs{000}, \]
\[\dKps{010} = \dKs{010} + \uy\dKs{000}, \]
\[\dKps{001} = \dKs{001} +\uz\dKs{000},\]
\[\dKps{110} = \dKs{110} + \ux\dKs{010}+ \uy\dKs{100} + \ux\uy\dKs{000},\]
\[\dKps{101} = \dKs{101} + \ux\dKs{001} + \uz\dKs{100} + \ux\uz\dKs{000},\]
\[\dKps{011} = \dKs{011} +  \uy\dKs{001} +  \uz\dKs{010} +  \uy\uz\dKs{000},\]
\[\dKps{200} = \dKs{200}  + 2\ux\dKs{100} + \uxx\dKs{000 },\]
\[\dKps{020} = \dKs{020}  + 2\uy\dKs{010} + \uyy\dKs{000} ,\]
\[\dKps{002} = \dKs{002}  + 2\uz \dKs{001} +\uzz\dKs{000} ,\]
\[\dKps{120} = \dKs{120} +\ux\dKs{020} + 2\uy \dKs{110}+ 2\ux\uy\dKs{010}+\uyy\dKs{100}+ \ux\uyy\dKs{000},\]
\[\dKps{102} = \dKs{102} + \ux\dKs{002} + 2\uz \dKs{101} + 2\ux\uz \dKs{001}+ \uzz\dKs{100}  + \ux\uzz\dKs{000},\]
\[\dKps{210} = \dKs{210} +\uy\dKs{200}+ 2\ux\dKs{110}  + \uxx\dKs{010}+ 2\ux\uy\dKs{100} + \uxx\uy\dKs{000},\]
\[\dKps{012} = \dKs{012} +\uy\dKs{002} + 2\uz\dKs{011}+ \uzz\dKs{010} + 2\uy\uz\dKs{001} + \uy\uzz\dKs{000},\]
\[\dKps{201} = \dKs{201}+ \uz\dKs{200}  + 2\ux\dKs{101} + \uxx\dKs{001} + 2\ux\uz\dKs{100}+ \uxx\uz\dKs{000}, \]
\[\dKps{021} = \dKs{021} + \uz\dKs{020} + 2\uy\dKs{011} + \uyy\dKs{001}+ 2\uy\uz\dKs{010} + \uyy\uz\dKs{000},\]
\[\dKps{111} = \dKs{111} + \ux\dKs{011} +\uy\dKs{101} + \uz\dKs{110} + \ux\uy\dKs{001} + \ux\uz\dKs{010} + \uy\uz\dKs{100} + \ux\uy\uz\dKs{000},\]
\[\dKps{220} = \dKs{220} + 2\uy \dKs{210}+ 2\ux \dKs{120}+ \uxx\dKs{020}+ \uyy\dKs{200}+ 4\ux\uy \dKs{110} + 2\uxx\uy \dKs{010}  + 2\ux\uyy \dKs{100} +\uxx\uyy\dKs{000},\]
\[\dKps{202} =  \dKs{202}+2\uz\dKs{201} + 2\ux \dKs{102}+ \uxx\dKs{002} + \uzz\dKs{200}+ 4\ux\uz\dKs{101}+ 2\uxx\uz\dKs{001}+ 2\ux\uzz\dKs{100}+ \uxx\uzz\dKs{000} ,  \]
\[\dKps{022} = \dKs{022}+ 2\uz \dKs{021}+ 2\uy \dKs{012}+\uzz \dKs{020}+\uyy\dKs{002} + 4\uy\uz\dKs{011}+ 2\uyy\uz \dKs{001} + 2\uy\uzz\dKs{010}+ \uyy\uzz\dKs{000},  \]
\vspace{-2em}
\begin{align*}
\dKps{211} &= \dKs{211} +2\ux \dKs{111}+ \uy\dKs{201} +\uz\dKs{210} + \uxx\Ks{011} + 2\ux\uy \dKs{101}+ \uy\uz\dKs{200} + 2\ux\uz \dKs{110}  \\ & \; \; \; \;+\uxx\uy\dKs{001} +\uxx\uz\dKs{010}+2\ux\uy\uz\dKs{100} + \uxx\uy\uz\dKs{000},
\end{align*}
\vspace{-2em}
\begin{align*}
\dKps{121} &= \dKs{121} + 2\uy\dKs{111}  +\ux\dKs{021}+ \uz\dKs{120}+ \ux\uz\dKs{020} + 2\ux\uy\dKs{011} + \uyy\dKs{101}  + 2\uy\uz\dKs{110}  \\& \; \; \; \;  - \ux\uyy\dKs{001} +2\ux\uy\uz\dKs{010}+ \uyy\uz\dKs{100}  +\ux\uyy\uz\dKs{000},
\end{align*}
\vspace{-2em}
\begin{align*}
\dKps{112} &= \dKs{112}+2\uz \dKs{111}+ \ux\dKs{012} +\uy\dKs{102}+ \ux\uy\dKs{002}+ 2\ux\uz\dKs{011} + 2\uy\uz \dKs{101}  + \uzz\dKs{110}\\  & \; \; \; \;+2\ux\uy\uz\dKs{001}+ \ux\uzz\dKs{010} + \uy\uzz\dKs{100} +\ux\uy\uzz\dKs{000},
\end{align*}
\vspace{-2em}
\begin{align*}
\dKps{122} &= \dKs{122}+ 2\uy\dKs{112} + 2\uz \dKs{121}+ \ux\dKs{022} + 4\uy\uz \dKs{111}  + 2\ux\uz\dKs{021}+ 2\ux\uy\dKs{012}+ \uyy\dKs{102} \\& \; \; \; \; + \uzz\dKs{120}+ \ux\uyy\dKs{002} +\ux\uzz\dKs{020}+ 4\ux\uy\uz \dKs{011} + 2\uyy\uz\dKs{101} + 2\uy\uzz\dKs{110} \\& \; \; \; \; + 2\ux\uyy\uz\dKs{001}  + 2\ux\uy\uzz \dKs{010}+ \uyy\uzz\dKs{100} + \ux\uyy\uzz\dKs{000},
\end{align*}
\vspace{-2em}
\begin{align*}
\dKps{212} &= \dKs{212}+ 2\ux\dKs{112}+ 2\uz \dKs{211} +\uy\dKs{202} + 4\ux\uz \dKs{111} + 2\uy\uz \dKs{201}  + \uxx\dKs{012} + \uzz\dKs{210} \\ & \; \; \; \; + 2\ux\uy \dKs{102}+ \uxx\uy\dKs{002}+\uy\uzz\dKs{200}+ 2\uxx\uz\dKs{011}+ 4\ux\uy\uz \dKs{101}  +2\ux\uzz\dKs{110} \\ & \; \; \; \;+ 2\uxx\uy\uz \dKs{001}+ \uxx\uzz\dKs{010} + 2\ux\uy\uzz \dKs{100} + \uxx\uy\uzz\dKs{000},
\end{align*}
\vspace{-2em}
\begin{align*}
\dKps{221} &=  \dKs{221}+ 2\ux \dKs{121} + 2\uy\dKs{211}+ \uz\dKs{220}+ 4\ux\uy \dKs{111} + \uxx\dKs{021} + \uyy\dKs{201}+ 2\uy\uz \dKs{210}\\ & \; \; \; \; + 2\ux\uz \dKs{120} + \uxx\uz\dKs{020} + \uyy\uz\dKs{200}  + 2\uxx\uy\dKs{011}+2\ux\uyy \dKs{101}+ 4\ux\uy\uz \dKs{110}\\ & \; \; \; \;+\uxx\uyy\dKs{001}+ 2\uxx\uy\uz \dKs{010}+ 2 \ux\uyy\uz\dKs{100}+ \uxx\uyy\uz\dKs{000},
\end{align*}
\vspace{-2em}
\begin{align*}
\dKps{222} &=\dKs{222} + 2\uz\dKs{221}+ 2\uy\dKs{212}+ 2 \ux\dKs{122}+ 4\ux\uy\dKs{112}+ 4\ux\uz \dKs{121}\\& \; \; \; \;+ 4\uy\uz\dKs{211}+\uxx\dKs{022}+ \uyy\dKs{202}+ \uzz\dKs{220}+ 8\ux\uy\uz\dKs{111}+ 2\uxx\uz\dKs{021}\\& \; \; \; \;+ 2\uyy\uz\dKs{201}+ 2\uxx\uy \dKs{012}  + 2\uy\uzz\dKs{210} + 2\ux\uyy\dKs{102} + 2\ux\uzz\dKs{120}+\uxx\uyy\dKs{002}\\&  \; \; \; \;+ \uxx\uzz\dKs{020}+\uyy\uzz\dKs{200} + 4\uxx\uy\uz\dKs{011} + 4\ux\uyy\uz\dKs{101}+ 4\ux\uy\uzz\dKs{110}+ 2\uxx\uyy\uz\dKs{001}\\ & \; \; \; \; + 2\uxx\uy\uzz\dKs{010}+ 2\ux\uyy\uzz\dKs{100} +\uxx\uyy\uzz\dKs{000}.
\end{align*}

\section{Inverse scaling post-collision raw moments by grid aspect ratios}\label{sec:appendix6}
The inverse scaling of the post-collision raw moments, i.e., $\tilde{\mathbf{m}}\leftarrow\tensor{S}^{-1}\tilde{\mathbf{m}}$ involves dividing the post-collision raw moments $\tilde{k}_{mnp}^\prime$ of the order $(m+n+p)$ by $r^ns^p$. That is,
\begin{align*}
  &\dKps{000}=\dKps{000}, \quad\quad\;\;\;\;\;\;\;\;\; \dKps{100} =\dKps{100}, \quad\;\;\;\;\;\;\;\;\quad\;\; \dKps{010}= r^{-1} \dKps{010},  \nonumber\\
  &\dKps{001}=s^{-1} \dKps{001},  \quad\;\;\;\;\;\;\; \dKps{200}=\dKps{200}, \quad\;\;\;\;\;\;\;\;\quad\;\; \dKps{102}= s^{-2}\dKps{102},  \quad\;\;\;\;\;\;\; \dKps{021}=r^{-2} s^{-1} \dKps{021},\nonumber\\
  &\dKps{110} =r^{-1} \dKps{110}, \quad \;\;\;\;\;\;\; \dKps{020}= r^{-2} \dKps{020},\quad\;\; \quad\;\; \dKps{210}= r^{-1}\dKps{210}, \quad\;\;\;\;\;\;\;  \dKps{111}=r^{-1} s^{-1} \dKps{111},\nonumber\\
  &\dKps{101}= s^{-1}  \dKps{101},\quad \;\;\;\;\;\;\; \dKps{002}=s^{-2} \dKps{002},\quad\;\;\;\quad\;\dKps{220}=r^{-2} \dKps{220}, \quad\;\quad \;\;\;  \dKps{012}=r^{-1} s^{-2} \dKps{012},\nonumber\\
  &\dKps{011}=r^{-1} s^{-1}\dKps{011}, \;\;\;\;\;  \dKps{120} =r^{-2}\dKps{120},\quad\;\;\;\quad\;  \dKps{201}=s^{-1} \dKps{201},\quad\;\;\;\;\;\;\;  \dKps{202}=s^{-2} \dKps{202},\nonumber\\
  &\dKps{022} =r^{-2} s^{-2}\dKps{022} ,\quad\;\; \dKps{211}=r^{-1} s^{-1} \dKps{211},\quad\;\; \dKps{121}=r^{-2} s^{-1}\dKps{121},\quad\;\; \dKps{112} =r^{-1} s^{-2} \dKps{112} ,\nonumber\\
&\dKps{122}= r^{-2} s^{-2}  \dKps{122},\quad \;\; \dKps{212}=r^{-1} s^{-2}\dKps{212}, \quad\;\;
\dKps{221} =r^{-2} s^{-1}\dKps{221} ,\quad\;\; \dKps{222}= r^{-2} s^{-2} \dKps{222}.
\end{align*}

\section{Mapping post-collision raw moments to distribution functions} \label{sec:appendix7}
The post-collision distribution functions can be obtained from the raw moments via $\tilde{\mathbf{f}}=\tensor{P}^{-1}\tilde{\mathbf{m}}$, where $\tensor{P}^{-1}$ is the inverse of the simpler moment basis for the cubic lattice presented in Eq.~(\ref{eq:momentbasiscubiclattice}). Thus, we get
\begin{align*}
& \df{0} = \dKps{000} - \dKps{200} - \dKps{020}-\dKps{002} + \dKps{220}+\dKps{202} + \dKps{022}- \dKps{222},\\
& \df{1} = \frac{1}{2}  \left(\dKps{100} + \dKps{200} -\dKps{120}-\dKps{102} - \dKps{220} - \dKps{202} +\dKps{122}+ \dKps{222}\right),\\
& \df{2} = \frac{1}{2}  \left(-\dKps{100} + \dKps{200} +\dKps{120}+\dKps{102} - \dKps{220} - \dKps{202} -\dKps{122}+ \dKps{222}\right),\\
& \df{3} = \frac{1}{2}  \left( \dKps{010} + \dKps{020}- \dKps{210}- \dKps{012} - \dKps{220}- \dKps{022} +\dKps{212} + \dKps{222}\right),\\
& \df{4} = -\frac{1}{2}  \left(\dKps{010} + \dKps{020}+ \dKps{210}+ \dKps{012} - \dKps{220}- \dKps{022} -\dKps{212} +\dKps{222}\right),\\
& \df{5} = \frac{1}{2}  \left(\dKps{001} + \dKps{002}- \dKps{201}- \dKps{021} - \dKps{202} - \dKps{022} + \dKps{221}+ \dKps{222}\right),\\
& \df{6} = \frac{1}{2}  \left(-\dKps{001}+\dKps{002} + \dKps{201}+ \dKps{021} - \dKps{202}- \dKps{022} - \dKps{221}+ \dKps{222}\right),\\
& \df{7} = \frac{1}{4}  \left(\dKps{110} + \dKps{120} +\dKps{210}+ \dKps{220}- \dKps{112} - \dKps{122} - \dKps{212} - \dKps{222}\right),\\
& \df{8} = \frac{1}{4}  \left(-\dKps{110} - \dKps{120}+\dKps{210}+ \dKps{220}+ \dKps{112} + \dKps{122} - \dKps{212} - \dKps{222}\right),\\
& \df{9} = \frac{1}{4}  \left(-\dKps{110} + \dKps{120} -\dKps{210}+ \dKps{220}+\dKps{112}- \dKps{122} + \dKps{212} - \dKps{222}\right),\\
& \df{10} = \frac{1}{4}  \left(\dKps{110} - \dKps{120} -\dKps{210}+ \dKps{220}-\dKps{112}+ \dKps{122} + \dKps{212} - \dKps{222}\right),\\
& \df{11} = \frac{1}{4}  \left(\dKps{101} + \dKps{102} +\dKps{201}+ \dKps{202}- \dKps{121} - \dKps{122} - \dKps{221} - \dKps{222}\right),\\
& \df{12} = \frac{1}{4}  \left(-\dKps{101} - \dKps{102} +\dKps{201}+ \dKps{202}+ \dKps{121} + \dKps{122} - \dKps{221} - \dKps{222}\right),\\
& \df{13} = \frac{1}{4}  \left( -\dKps{101} +\dKps{102}-\dKps{201} + \dKps{202}+ \dKps{121} - \dKps{122} + \dKps{221} - \dKps{222}\right),\\
& \df{14} = \frac{1}{4}  \left(\dKps{101} -\dKps{102}-\dKps{201} + \dKps{202}- \dKps{121} + \dKps{122} + \dKps{221} - \dKps{222}\right),\\
& \df{15} = \frac{1}{4}  \left(\dKps{011} +\dKps{012} + \dKps{021} +\dKps{022}- \dKps{211} - \dKps{212} - \dKps{221} - \dKps{222}\right),\\
& \df{16} =\frac{1}{4}  \left(-\dKps{011} -\dKps{012} + \dKps{021} +\dKps{022}+ \dKps{211} + \dKps{212} - \dKps{221} - \dKps{222}\right),\\
& \df{17} = \frac{1}{4}  \left(-\dKps{011}+ \dKps{012} -\dKps{021} + \dKps{022}+ \dKps{211} - \dKps{212} + \dKps{221} - \dKps{222}\right),\\
& \df{18} = \frac{1}{4}  \left(\dKps{011}- \dKps{012} -\dKps{021} + \dKps{022}- \dKps{211} + \dKps{212} + \dKps{221} - \dKps{222}\right),\\
& \df{19} = \frac{1}{8} \left(\dKps{111} + \dKps{211}+ \dKps{121} + \dKps{112}+ \dKps{122} + \dKps{212} + \dKps{221} + \dKps{222}\right),\\
& \df{20}= \frac{1}{8}  \left(-\dKps{111} + \dKps{211}- \dKps{121} - \dKps{112}- \dKps{122} + \dKps{212} + \dKps{221} + \dKps{222}\right),\\
& \df{21} = \frac{1}{8}  \left(- \dKps{111}-\dKps{211}+\dKps{121} - \dKps{112} + \dKps{122} - \dKps{212} + \dKps{221} + \dKps{222}\right),\\
& \df{22} = \frac{1}{8}  \left(\dKps{111}-\dKps{211}-\dKps{121} + \dKps{112} - \dKps{122} - \dKps{212} + \dKps{221} + \dKps{222}\right),\\
& \df{23} = \frac{1}{8}  \left(-\dKps{111}-\dKps{211}- \dKps{121}+ \dKps{112} + \dKps{122} + \dKps{212} - \dKps{221} + \dKps{222}\right),\\
& \df{24}= \frac{1}{8}  \left(\dKps{111}-\dKps{211}+ \dKps{121}- \dKps{112} - \dKps{122} + \dKps{212} - \dKps{221} + \dKps{222}\right),\\
& \df{25} = \frac{1}{8}  \left(\dKps{111} + \dKps{211}- \dKps{121}- \dKps{112} + \dKps{122} - \dKps{212} - \dKps{221} + \dKps{222}\right),\\
& \df{26} = \frac{1}{8}  \left(-\dKps{111} + \dKps{211}+ \dKps{121}+ \dKps{112} - \dKps{122} - \dKps{212} - \dKps{221} + \dKps{222}\right).
\end{align*}


\end{document}